# Mémoire sur la mécanique quantique et l'approche ondulatoire

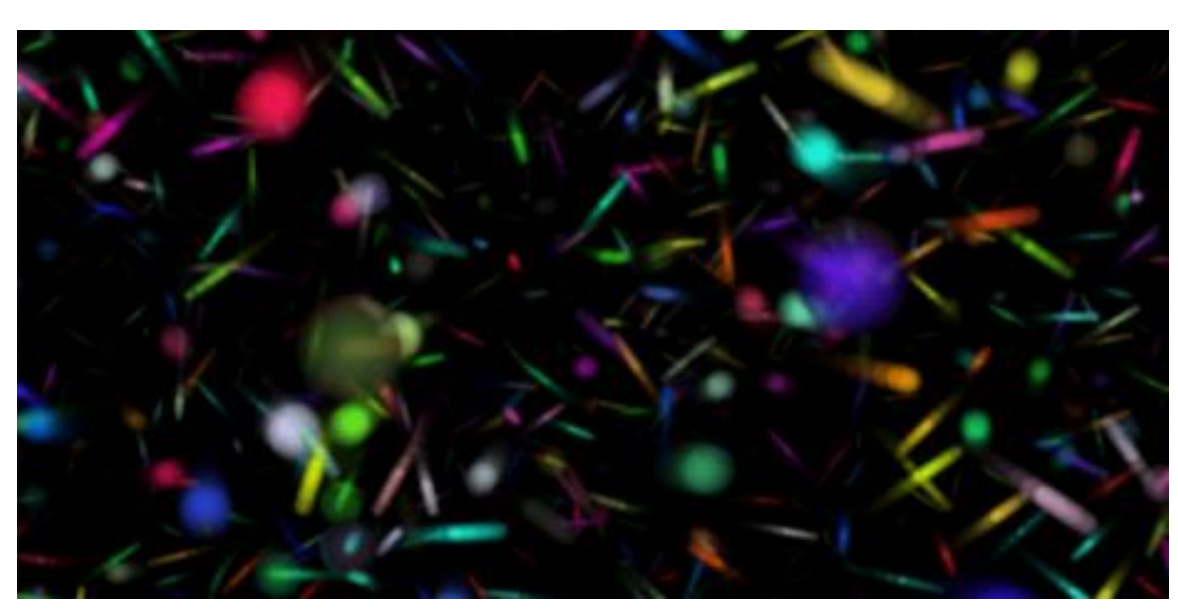

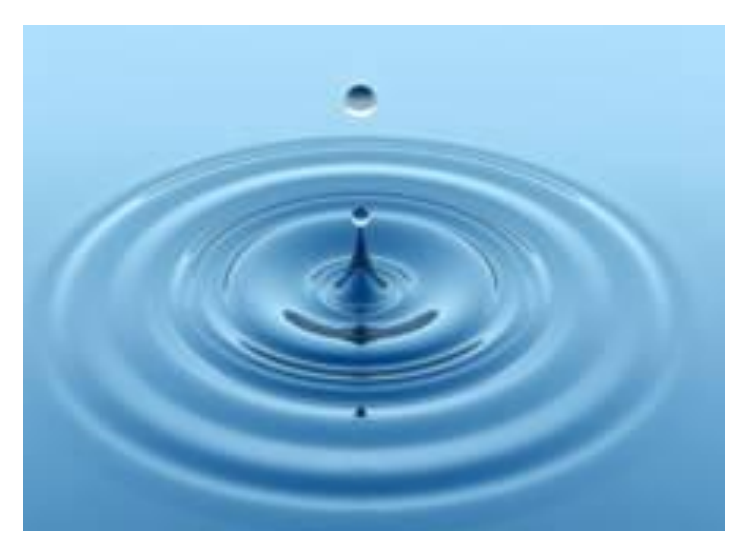

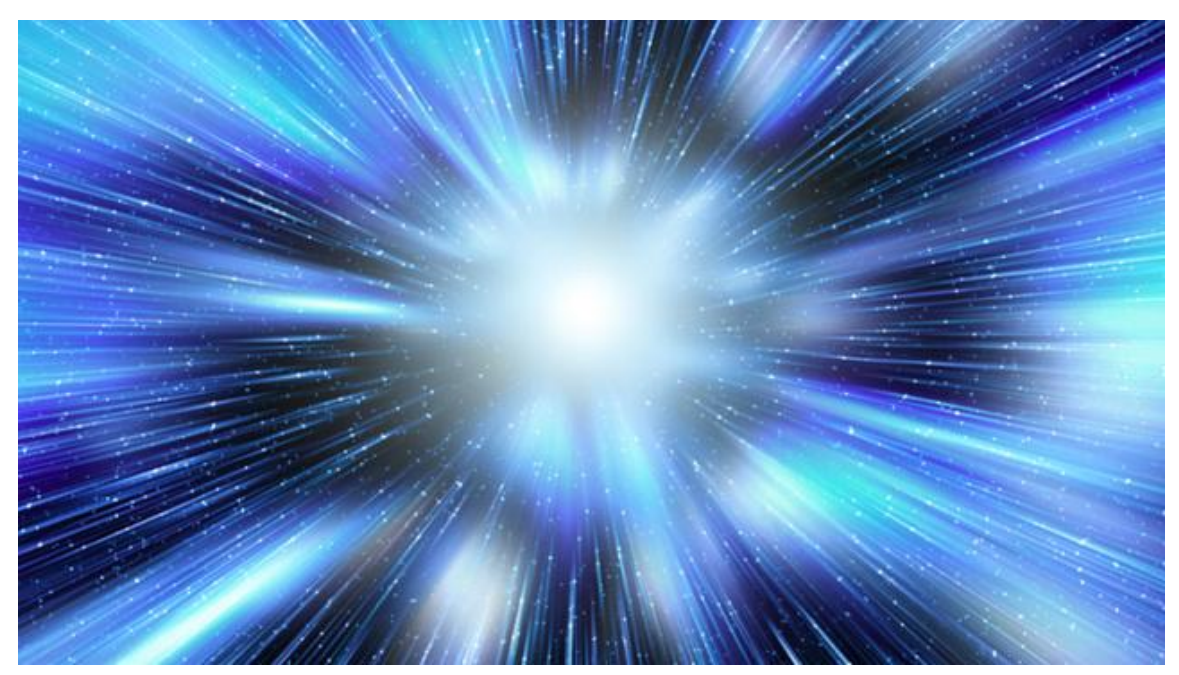

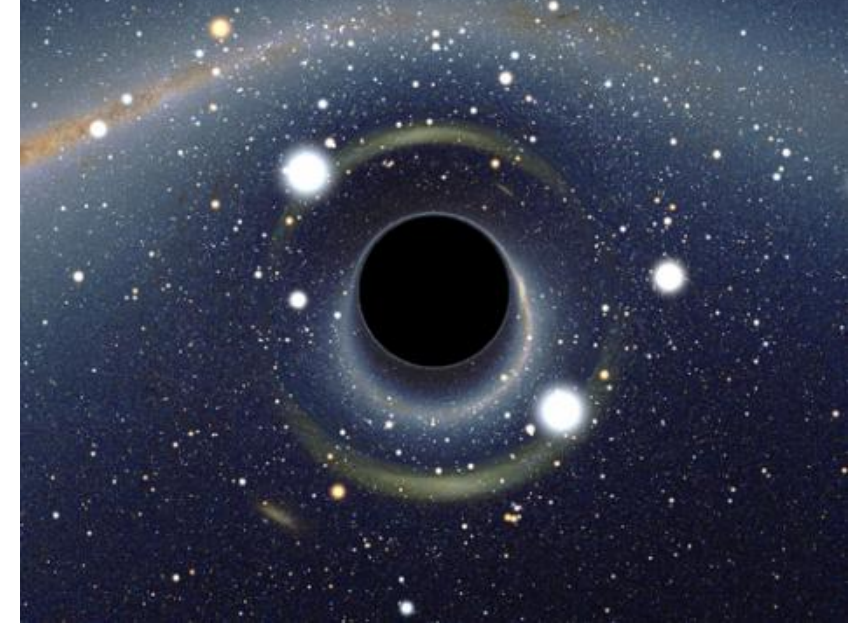

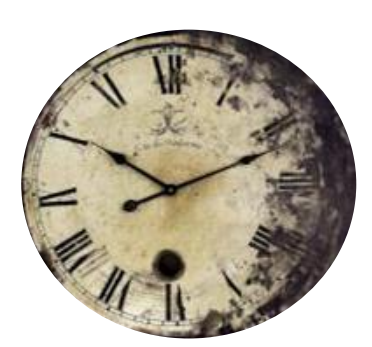

Olivier Rousselle

Mémoire ENS Ulm M2 philosophie des sciences – 2018 / 2019

Tuteur : Christian Bracco, Observatoire de Paris

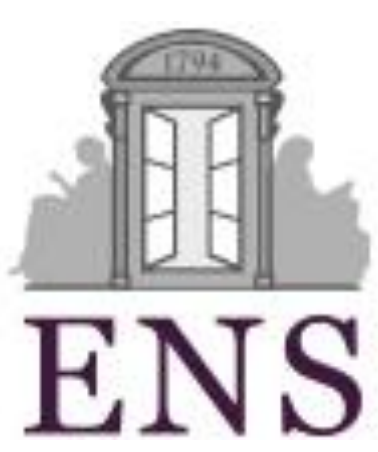

## Abstract

The Copenhagen interpretation has been the subject of much criticism, notably by De Broglie and Einstein ("God doesn't play dice"), because it contradicts the principles of causality and realism.

The aim of this essay is to study the wave mechanics as an alternative to traditional quantum mechanics, in the continuity of the ideas of Louis de Broglie: the pilot wave theory of De Broglie (where each particle is associated with a wave which guides it), De Broglie-Bohm theory, stochastic electrodynamics (where the stochastic character of particles is caused by the energy field of the fluctuating vacuum), and the analogies between quantum mechanics and hydrodynamics.

# Introduction

Qu'est-ce que le temps, qu'est-ce l'espace ? Qu'est-ce que la matière ? Ces questions n'ont cessé d'obséder les hommes au cours des siècles derniers.

Les plus grands esprits ont permis de faire avancer ces questions, avec un besoin profond de théoriser le monde, c'est à dire d'y trouver une unité face à la diversité des expériences sensibles. L'unification est au cœur de la démarche scientifique, et la physique révèle que des phénomènes en apparence très différents ont une origine commune: les mouvements terrestres et les mouvements célestes (Newton); l'électricité et le magnétisme (Maxwell); l'espace et le temps (Einstein)...

Au cours des siècles derniers, la science nous apprend que le monde visible, la notion de matérialité, sont peut-être illusoires, et que les entités ont une essence ondulatoire. Cela a été montré pour la lumière, puis pour les particules de matière.

Le but de cet essai est d'étudier l'approche ondulatoire de la matière selon les idées des physiciens Louis de Broglie, David Bohm et d'autres savants, et aussi d'analyser son évolution au cours du XX$^{e}$s. et sa portée contemporaine.

Les articles analysés ont une approche à la fois philosophique (questionnement des fondements), physique (raisonnements basés sur l'expérience) et mathématique (démarches rigoureuses). Les concepts seront illustrés à l'aide de schémas, et d'équations assez simples.

Nous mettrons également en valeur à travers nos développements le lien entre la philosophie et la physique. En effet, la philosophie fournit des outils et méthodes dont la physique a besoin : analyse conceptuelle, capacité à détecter les incohérences dans les arguments standards, rechercher des explications alternatives,...

Nous allons étudier, dans un premier temps, les théories physiques ayant conduit au développement de la mécanique ondulatoire, du XVII$^{e}$ jusqu'au début du XX$^{e}$s.

Nous analyserons l'évolution des idées sur la lumière, et en particulier l'opposition de la théorie corpusculaire prônée par Newton, et la théorie ondulatoire prônée par

Huygens, Young,…

Pour développer la mécanique ondulatoire, De Broglie s'est appuyé sur la théorie de la relativité restreinte, développée au début du XX$^{e}$s. par Einstein, Poincaré, avec les concepts d'espace et de temps relatifs, de contraction et dilatation des longueurs et des durées,… que nous tâcherons d'appréhender.

Enfin, dans cette partie préliminaire, nous verrons comment a évolué la notion de matière au cours des siècles, avec les propriétés fondamentales de charge, de masse, et quels étaient les constituants élémentaires connus au début du XX$^{e}$s.

Dans un deuxième temps, nous verrons, à l'aide d'articles historiques, le développement des idées ondulatoires au XX$^{e}$s.

Nous analyserons la thèse de doctorat de Louis de Broglie, *Recherches sur la théorie des Quanta (1924),* dans laquelle il montre que les particules de matière (électrons) possèdent un caractère ondulatoire. C'est la dualité onde-corpuscule appliquée à la matière, et le début de la mécanique ondulatoire. Ces idées révolutionnaires vaudront à De Broglie le prix Nobel en 1929, il « a levé un coin du grand voile » selon Einstein.

Ensuite, nous étudierons, avec l'article *Interpretation of quantum mechanics by the double solution theory,* la théorie de l'onde pilote de De Broglie (1926-1927) qui découle de sa thèse et vise à donner une interprétation à la dualité onde-corpuscule. Dans cette vision, chaque particule est associée à une onde qui la guide, elle est à l'origine des phénomènes quantiques comme les interférences des électrons.

A partir des idées de Broglie et son postulat d'équivalence onde-matière, Erwin Schrödinger formula sa célèbre équation d'onde en 1926. Un nouvel outil mathématique est introduit alors, la fonction d'onde $\psi$, qui caractérise la particule mais reste une notion abstraite. En 1926, Born montra que cette fonction d'onde peut s'interpréter en termes de probabilité de présence de la particule. Le congrès de Solvay de 1927 marque la naissance de la mécanique quantique et l'interprétation de Copenhague où les probabilités, l'indéterminisme, le principe de superposition, jouent un rôle majeur. Selon la vision de Bohr, Heisenberg, Pauli,… il faut renoncer à vouloir comprendre la nature et se cantonner à l'observation.

L'interprétation de Copenhague a fait l'objet de nombreuses critiques, notamment par De Broglie et Einstein. Elle entrait en contradiction avec deux principes auxquels ils tenaient : la causalité (tout phénomène a une cause, il n'y a pas d'indéterminisme) et le réalisme (il existe une réalité indépendante de l'observation; la physique doit décrire la nature, et non uniquement notre relation à l'expérience). Ce qu'Einstein résuma par ses célèbres phrases : « Dieu ne joue pas aux dés » et « J'aime penser que la lune est là, même si je ne la regarde pas ». Nous détaillerons la critique de De Broglie à travers son article *La physique quantique restera-t-elle indéterministe ?* (1952)

La théorie de l'onde pilote de De Broglie fut abandonnée durant des dizaines d'années du fait de la remarquable efficacité de l'approche de Copenhague. Elle fut cependant reprise par David Bohm en 1952 pour donner ce que l'on appelle aujourd'hui la théorie de De Broglie-Bohm (également appelée « Bohmian mechanics»). Cette théorie déterministe est plus rigoureuse plus complète, plus simple que l'onde pilote de De Broglie, elle introduit des variables cachées, et est fondée sur la non-localité. Nous étudierons cette théorie à travers l'article fondateur de Bohm, *A suggested interpretation of the quantum theory in terms of hidden variables (1952).* « Je vis l'impossible se réaliser » indiqua le physicien John Bell à la publication de cet article. Elle est encore considérée de nos jours comme une alternative pertinente à l'interprétation de Copenhague, dans un souci de préservation du réalisme et de la causalité, et réalise les mêmes prédictions. L'étude du livre de M. & A. Gondran, *Mécanique quantique : Et si Einstein et de Broglie avaient raison ? (2014),* vise à approfondir cette interprétation à travers des expériences cruciales (fentes de Young, expérience de Stern et Gerlach, EPR), et à en étudier les limites.

L'interprétation de De Broglie-Bohm est cependant méconnue des physiciens et ignorée du grand public, alors qu'elle ne manque pas de pertinence. Nous verrons des faits modernes qui confirment l'intérêt de l'onde pilote, et les analogies de la mécanique quantique avec la physique statistique et l'hydrodynamique. En effet, en 2006, des chercheurs français ont montré qu'une gouttelette (analogue à la particule), dans un fluide en vibration, est associée à une onde (analogue à l'onde pilote) et cette expérience permet de reproduire la quasi-totalité des résultats quantiques (interférences, effet tunnel,…). Nous verrons également une interprétation moderne de la théorie de De Broglie-Bohm, l'électrodynamique stochastique, où le caractère indéterminé des particules est provoqué par le champ d'énergie du vide fluctuant.

Enfin, dans une troisième partie d'ouverture, nous proposerons des idées pour se représenter différemment la notion de particule, de masse, de mouvement, dans la continuité de l'approche ondulatoire.

Selon De Broglie, la physique a un besoin urgent de pouvoir définir une structure des particules et notamment de pouvoir introduire un rayon de l'électron. C'est ainsi que nous proposerons un modèle ondulatoire de la particule, dotée d'un rayon, où la masse, la charge et le spin ont une origine électromagnétique. Nous tenterons de dériver les lois de la mécanique quantique et de la relativité restreinte avec cette approche ondulatoire.

Le saint-graal de la physique moderne est une théorie qui engloberait la mécanique quantique et la relativité générale. Deux voix peuvent être empruntées : soit rendre la relativité générale non réaliste et non déterministe; soit rendre la mécanique quantique réaliste et déterministe. C'est cette deuxième voix que nous suivrons, celle qui est empruntée par les physiciens étudiés dans cet essai. Les idées de De Broglie ont une portée importante et sont toujours d'actualité de nos jours...

« La philosophie est écrite dans ce vaste livre qui se trouve ouvert en permanence sous nos yeux, que j'appelle univers » Galilée

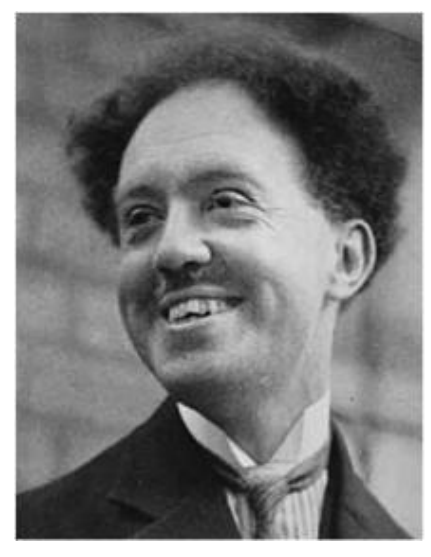

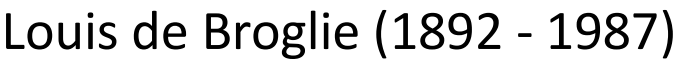

Louis de Broglie (1892 - 1987)

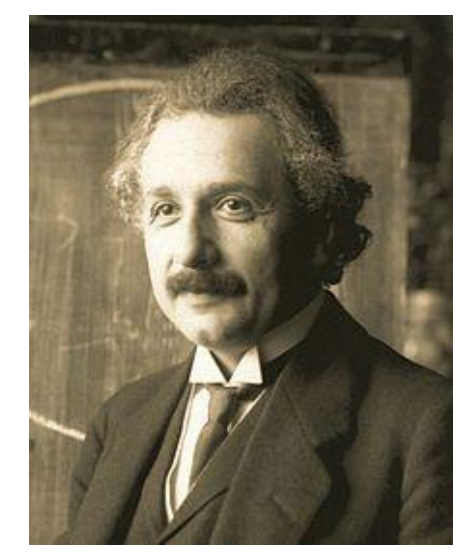

Albert Einstein (1879 - 1955)

# I. Historique de la physique fondamentale du XVII$^{e}$ au début du XX$^{e}$s.

## 1) Évolution des idées sur la lumière : approches corpusculaire et ondulatoire

Dans les premières descriptions mythiques du monde, la lumière est une sorte de «brume claire» opposée à la «brume sombre» des ténèbres et de la nuit. Les Grecs s'interrogent sur la nature physique du monde, ils s'intéressent à la vision et sont les premiers à comprendre les ténèbres comme étant une absence de lumière. Au X$^{e}$s., dans la science arabe, la lumière désignait l'éclat brillant observable par nos yeux. (*Bracco, 2004*)

Au XVII$^{e}$s., dans le contexte de révolution scientifique initiée par Galilée, différentes théories sur la lumière furent élaborées par Descartes, Newton, Huygens,... Les travaux de Galilée et Newton ont étendu la vision des Grecs, avec notamment la physique d'Aristote et l'ordre mathématique idéal de Platon.

La nécessaire théorisation de la lumière devait se fonder sur des faits expérimentaux découverts alors : la lumière se propage en ligne droite; la lumière blanche que l'on voit est un mélange de toutes les couleurs du spectre visible par l'œil humain; les couleurs sont des propriétés intrinsèques du rayon de lumière, chacun avec un degré propre de réfrangibilité (indice de réfraction); on observe une déviation du rayon lumineux quand il passe d'un milieu à un autre (réfraction).

Deux approches de la lumière s'opposaient pour théoriser ces phénomènes :

- L'approche corpusculaire portée par le savant britannique Isaac Newton : la lumière se comporte comme des balles de fusil. Les particules ne touchent pas tous les points de l’espace qu’elles traversent, et toute interaction répond au principe du tout ou rien (soit on heurte une particule, soit on l’évite). Cette approche permet de bien rendre compte de la propagation rectiligne de la lumière, de la diversité des couleurs, la réflexion de la lumière sur un miroir,... (*Newton, 1704*)

- L’approche ondulatoire portée par le savant néerlandais Christiaan Huygens : la lumière est une perturbation d’un milieu que l’on appelle alors éther, de façon analogue aux vagues sur l’eau ou à une corde qui vibre. Les ondes ne se concentrent pas sur des points ou des trajectoires linéaires, elles se propagent et balayent l’espace; deux ondes se croisent et continuent leur chemin comme s’il ne s’était rien passé. Cette approche permet de rendre compte des lois de Snell-Descartes, de la double réfraction dans les cristaux, mais ne permet pas encore d’expliquer la propagation rectiligne de la lumière et la diffraction. (*Huygens, 1690*)

Au début du XIX$^{e}$s., le physicien Thomas Young élabora l’expérience de la double fente (bien que certains historiens mettent en doute cette réalisation). Il mit alors en évidence le phénomène d’interférences (alternance de bandes sombres et claires), ce qui constitue une preuve supplémentaire du caractère ondulatoire de la lumière. (*Young, 1802; Worral, 1976*)

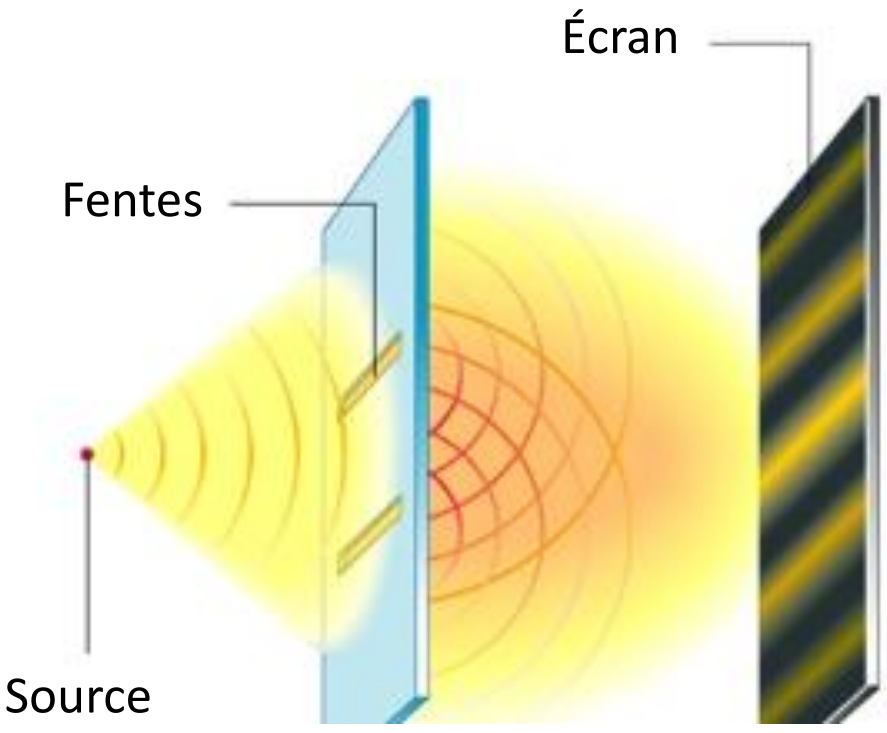

Fig. 1.1 : Fentes de Young et figure d’interférences

La théorie ondulatoire sera ensuite développée et mathématisée par Augustin Fresnel, qui donnera son nom au principe de Huygens-Fresnel : la lumière se propage par front d’onde, et chaque particule touchée par un front lumineux est à son tour à l’origine d’un nouveau front.

Fresnel put expliquer les interférences avec la notion d’onde de phase : lumière + lumière = lumière quand les deux ondes sont en phase; lumière + lumière = ombre quand les deux ondes sont en opposition de phase.

Il découvrit également la nature transversale des ondes lumineuses et expliqua ainsi les phénomènes de polarisation.

En 1905, Albert Einstein émet l’idée que la lumière se comporte comme une particule, il introduit la notion de quantas d’énergie $h\nu$.

Le rayonnement se propage sous forme un « nombre fini de quantas », Einstein pose alors les bases de ce qui allait devenir la mécanique quantique. Cette vision de la lumière lui permit d’expliquer l’effet photoélectrique : la lumière se propage, «frappe» un électron du matériau; si l’énergie du photon est suffisante, l’électron sera expulsé, comme une boule de billard.

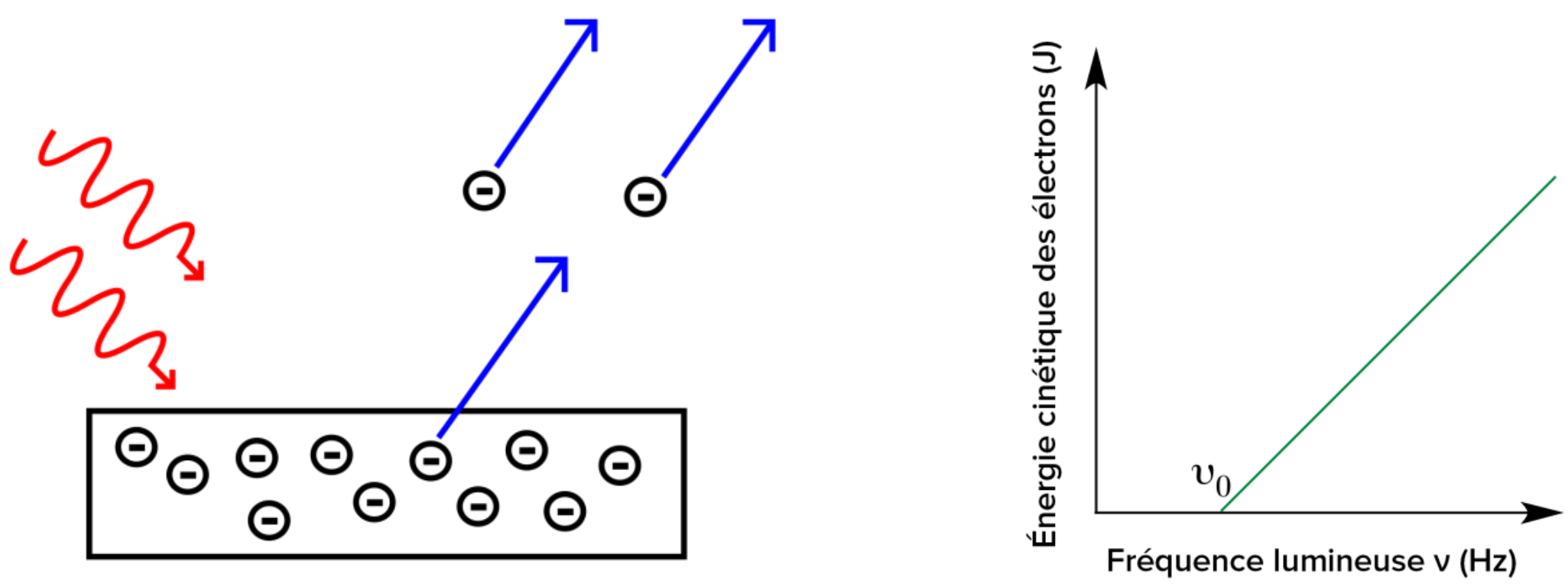

Fig. 1.2 : Illustration de l’effet photoélectrique

Le caractère corpusculaire de la lumière permet également d’expliquer l’effet Compton (*Compton, 1923*). C’est avec cet effet que les quantas de lumière se doteront d’une quantité de mouvement $\frac{h\nu}{c}$.

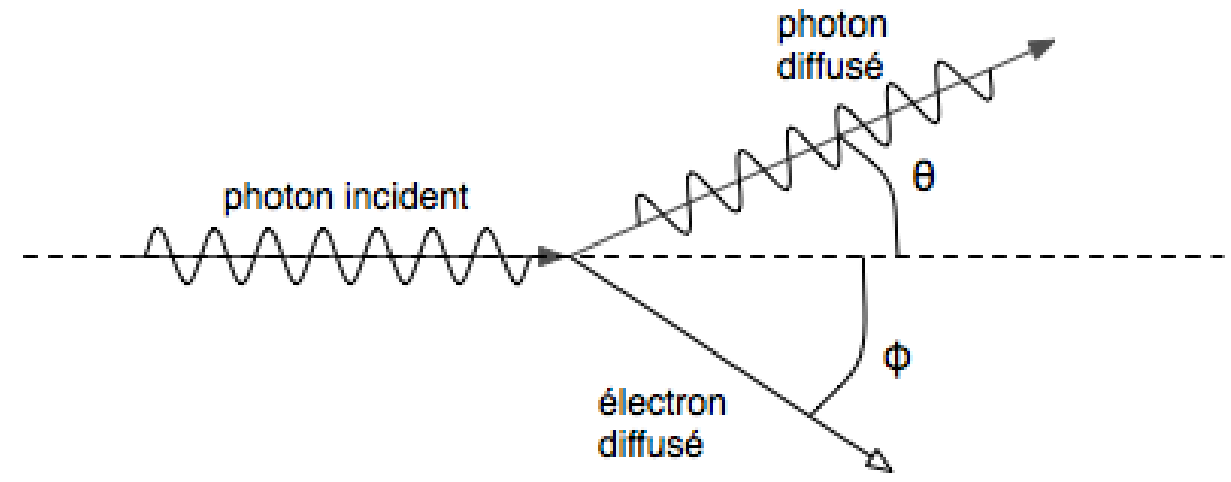

Fig. 1.3 : Effet Compton : collision d'un photon avec un électron au repos

Ces deux approches ne sont cependant pas nécessairement incompatibles, il est possible que la lumière puisse revêtir un caractère dualiste (à la fois onde et corpuscule).

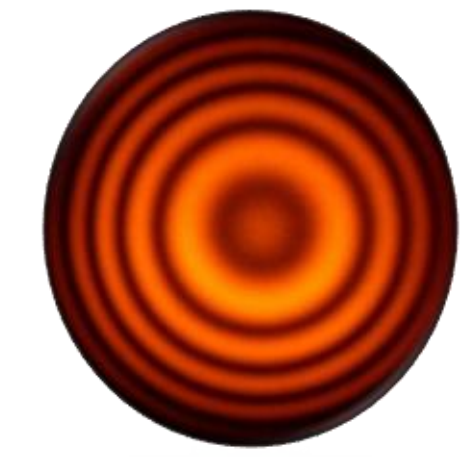

Fig. 1.4 : Anneaux de Newton

Cette dualité a été pensée pour la première fois par Newton, pour expliquer la formation d'une figure d'interférences formée d'anneaux concentriques (voir figure) lorsque l'on place une lentille sur une surface plane. (*Newton, 1704*)

De nos jours, on parle de dualité onde-corpuscule : la lumière se présente dans certaines limites comme une onde (avec une fréquence), ou une particule (avec une impulsion, une énergie). Ces deux faces de la lumière sont reliées par la relation suivante :

$$E = h\nu = \frac{h}{T} \qquad (1.1)$$

avec $E$ l'énergie du photon, $\nu$ la fréquence de l'onde, $T$ la période et $h$ la constante de Planck.

## 2) Historique de la notion de matière : masse, charge, atomes, électron,...

Nous décrivons ici une autre composante importante du monde physique : la matière, qui se distingue de la lumière par le fait qu'elle possède une masse. De plus, elle peut être chargée électriquement.

### La masse

Selon les philosophes grecs, dont Aristote, le poids est une qualité intrinsèque de la matière, laquelle est attirée vers le bas car c'est son lieu de repos. Dans cette vision des choses, des corps plus lourds tombent plus rapidement que des corps légers.

Au XVIe s., Galilée a formulé le principe de la chute libre (en faisant abstraction de la résistance de l'air), selon lequel tous les corps chutent à la même vitesse quel que soit leur masse.

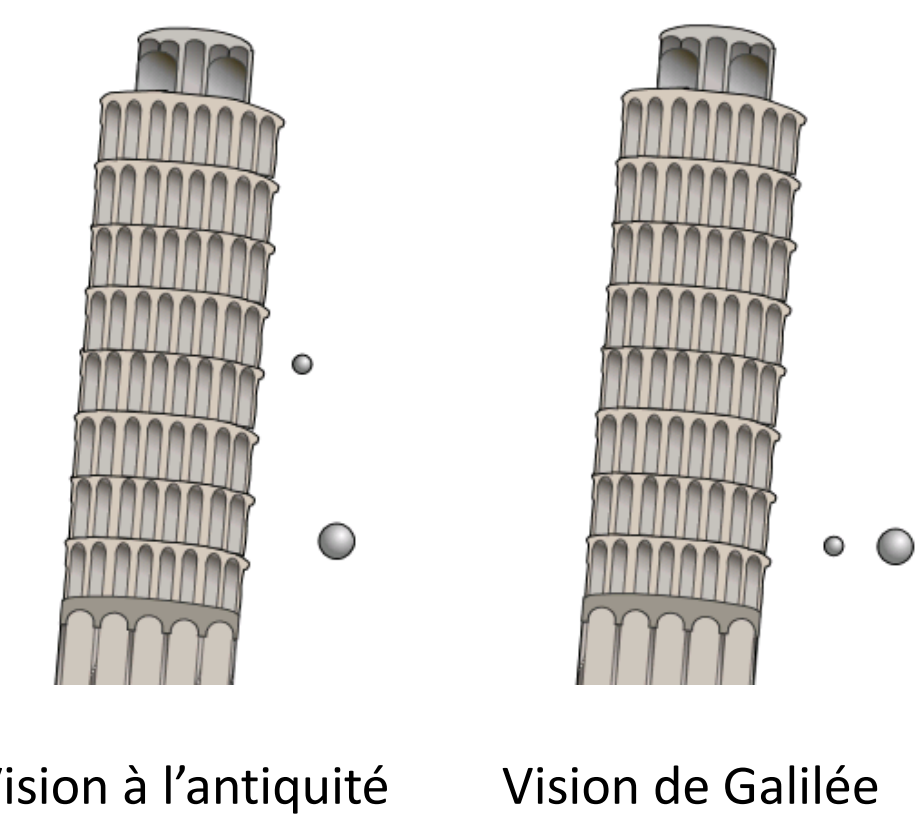

Vision à l'antiquité        Vision de Galilée

Fig. 1.5 : Illustration de la chute libre

Dans ses lois de la dynamique, Newton introduit deux types de masse (*Newton, 1687*) :

- La masse inerte, qui se manifeste par l'inertie des corps. Elle est la quantification de la résistance d'un corps à l'accélération, ce qui s'exprime par la 2e loi de Newton :

$$\vec{F} = m_i \vec{a} \qquad (1.2).$$

En introduisant la quantité de mouvement $\vec{p} = m\vec{V}$, on a : $\vec{F} = \frac{d\vec{p}}{dt}$. Une force a pour effet de modifier la valeur et la direction de la vitesse. En l'absence de force, le corps reste dans un mouvement rectiligne uniforme (inertie).

- La masse grave ou pesante, qui se manifeste par l'attraction universelle des corps. Newton l'a introduite dans la loi universelle de la gravitation : $\vec{F} = G\frac{m_1 m_2}{r^2}\vec{u}$.

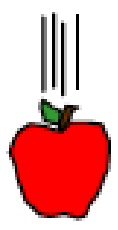

Dans la théorie de Newton, la force gravitationnelle se propage instantanément entre les corps massifs. Cependant, cette interaction à distance ne le satisfait guère.

Newton indique que la masse inerte et la masse pesante sont vraisemblablement égales, ce qui portera plus tard le nom de principe d'équivalence.

Ajoutons également que selon la vision newtonienne, les particules vivent dans un espace et un temps bien séparés. Newton considérait l'espace comme absolu, universel; ses lois traitent l'espace et le temps séparément, à l'aide de la géométrie euclidienne. L'espace existe comme une entité permanente, indépendamment de la matière qu'il contient, il est absolu et infini; Il s'oppose alors à Leibniz, pour qui l'espace est relatif.

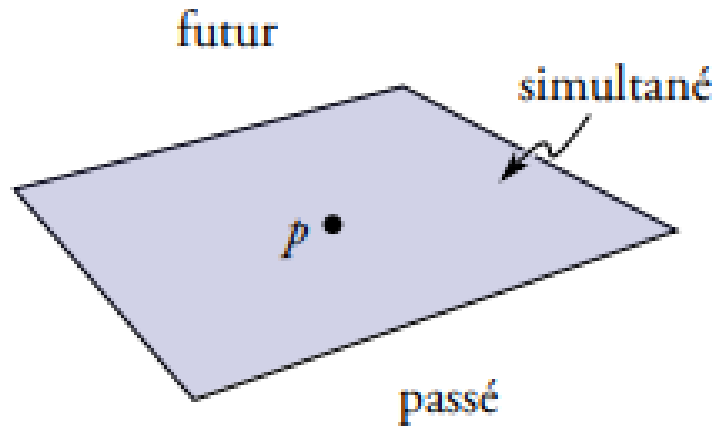

Fig. 1.6 : Conception newtonienne de l'espace et du temps

La conception d'espace a beaucoup évolué au cours des siècles, comme nous le verrons, sous l'influence des travaux de philosophes et physiciens.

Le principe d'équivalence est toujours valable de nos jours (à une précision relative de $10^{-14}$), tout en restant mystérieux. Comment ces deux quantités (masse pesante et masse inerte), si différentes de par leur définition, sont d'après l'expérience caractérisées par un seul et même nombre ? « Il doit y avoir une raison plus profonde pour cela » (*Einstein, 1952*).

## La charge

Certains constituants de la matière sont chargés électriquement. C'est le cas des entités telles que l'électron, les ions,...

La charge électrique (électricité signifiant «ambre» en grec ancien) est découverte par les Grecs. Ils remarquent en effet que si l'on frotte de l'ambre sur de la fourrure, elle devient «chargée» et peut alors attirer des objets tels que des cheveux.

Au XVIII^e^s., Benjamin Franklin met en évidence deux types de charge électrique : positive ou négative. Deux charges de même signe se repoussent, alors que deux charges opposées s'attirent.

La valeur de la force électrostatique est établie en 1785 par Charles-Augustin Coulomb, à l'aide d'une balance électrique (*Coulomb, 1785*) et des travaux de Cavendish (qui établit la loi en $1/r^2$) :

$$\vec{F} = k\frac{q_1 q_2}{r^2}\vec{u} \qquad (1.3).$$

L'intensité de la force est proportionnelle aux valeurs de charges et varie en raison inverse du carré de la distance r entre les deux charges.

L'unité usuelle de la charge est le coulomb (C), c'est une quantité qui se conserve lors des réactions.

L'expérience de la goutte d'huile (*Millikan, 1909*) a permis de mettre en évidence que la charge électrique est quantifiée : toute charge Q est une multiple entier de la charge élémentaire, e.

$$Q = n.e, \text{ avec } n \in \mathbb{N} \text{ et } e \approx 1{,}609.10^{-19}\, C \qquad (1.4).$$

## Le concept d'atome

En Grèce antique, on considère que la matière est composée de «grains» élémentaires, appelés alors atomes (signifiant «insécables»).

Au fil des siècles, la notion de matière et d'atome évolue. Au XVII$^{e}$s., le chimiste français Antoine Lavoisier énonce la loi de conservation de la masse, ce qui marque la naissance de la chimie moderne : « Rien ne se perd, rien ne se crée, tout se transforme » (*Lavoisier, 1789*).

Au XIX$^{e}$s., le chimiste russe Dmitri Mendeleïev classe les atomes par masse croissante, et remarque qu'il y a une périodicité dans leurs propriétés chimiques (*Mendeleïev, 1869*).

La notion d'atome a été définitivement acceptée au début du XX$^{e}$s., notamment avec les expériences menées par Jean Perrin. Ces expériences font appel au nombre d'Avogadro ($N_A$) dans l'interprétation théorique qui en était faite sur la base de l'hypothèse atomique (*Perrin, 1913*).

Au début du XIX$^{e}$s., Joseph John Thomson et ses collègues réalisent des expériences avec des tubes de Crookes (voir photo).

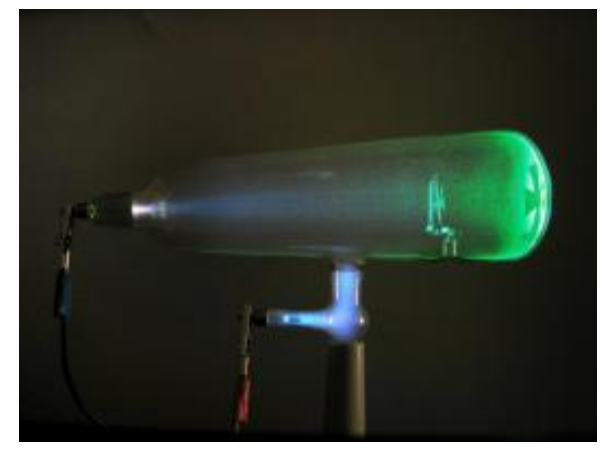

Fig. 1.7 : Tube de Crookes

Ils découvrent que les rayons cathodiques (en vert sur la photo) contenus dans ces tubes sont des particules individualisées; elles sont chargées négativement et ont environ un millième de la masse de l'hydrogène.

Le nom d'«électron» est proposé alors, il est la première particule élémentaire mise en évidence. (*Thomson, 1897)*

Ernest Rutherford découvre avec son expérience de la feuille d'or que l'atome est constitué d'un petit noyau comprenant toute la charge positive et presque toute la masse de l'atome (*Rutherford, 1911)*.

Niels Bohr explique les raies spectrales de l'atome d'hydrogène, en utilisant de nouveau la quantification. Bohr montre que les atomes peuvent avoir seulement des valeurs d'énergie discrètes et que les passages discontinus d'une énergie à une autre sont liés à l'émission ou l'absorption d'un quantum d'énergie $h\nu$. (*Bohr, 1913*)

En *1919, Rutherford* prouve que le noyau atomique est constitué de protons (chargés positivement) et d'hypothétiques neutrons (neutres).

Nous pouvons résumer ci-dessous les différents constituants de l'atome découverts au début du XX$^{e}$s. :

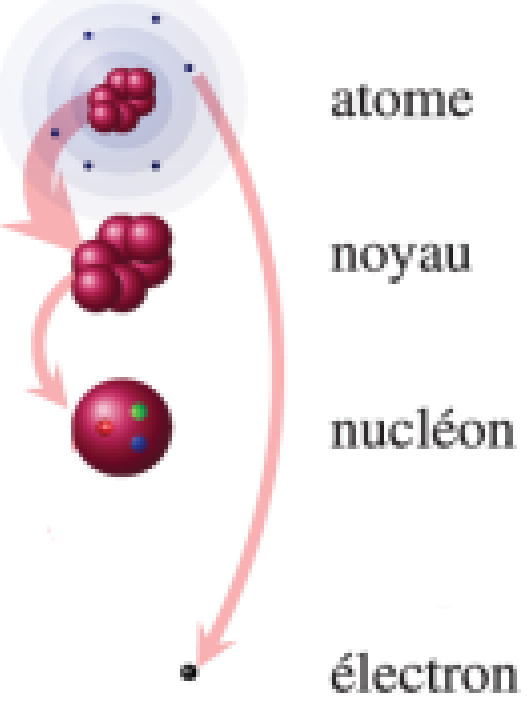

Fig. 1.8 : Structure du monde microscopique

## 3) L'électromagnétisme et la relativité restreinte

Nous allons décrire ici brièvement deux théories fondamentales de la physique contemporaine : l'électromagnétisme, théorisé par James Clerk Maxwell en 1864, et la relativité restreinte, découverte par Einstein en 1905.

### L'électromagnétisme

Au cours des siècles, les physiciens ont découvert que les particules de matière sont associées à des champs : champ gravitationnel, champ électrique, champ magnétique,...

Ce concept de champ a été introduit au XIX$^{e}$s., par Michael Faraday. Dans les expériences qu'il entreprend en laboratoire (comme avec la limaille de fer), il observe que les particules électriques paraissent créer dans l'espace environnant un état qui produit un certain ordre. Cet état préexiste à l'interaction entre deux corps; lorsqu'un second corps est introduit, la force « concrétise » cette énergie qui préexistait. Ces états de l'espace, appelés champs, pourraient selon lui expliquer les mystérieuses interactions électromagnétiques. Pour représenter ces idées, il inventa des schémas sur lesquels sont représentés des lignes de champ, comme illustré ci-dessous dans le cas du champ électrique.

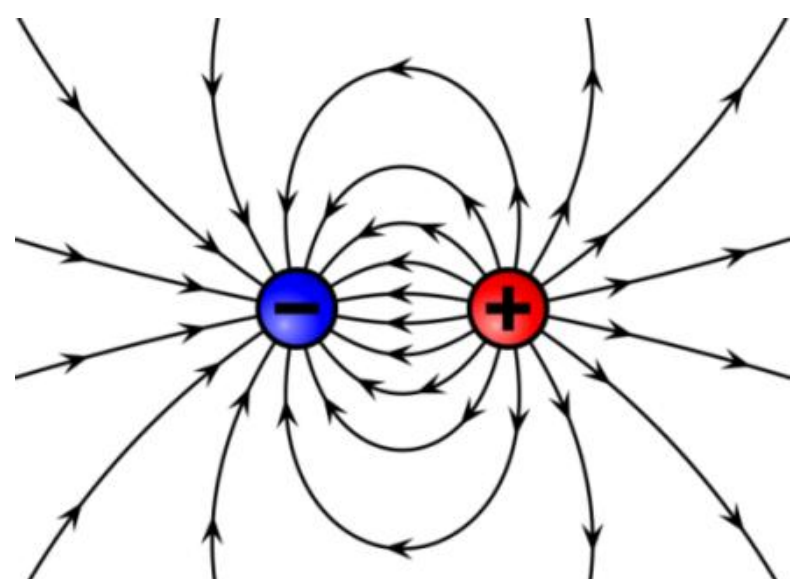

Fig. 1.9 : Champ électrique entre charges de signe opposé

En 1852, il dévoile l'existence du champ magnétique en décrivant les lignes de force le long desquelles s'oriente la limaille de fer (voir ci-dessous).

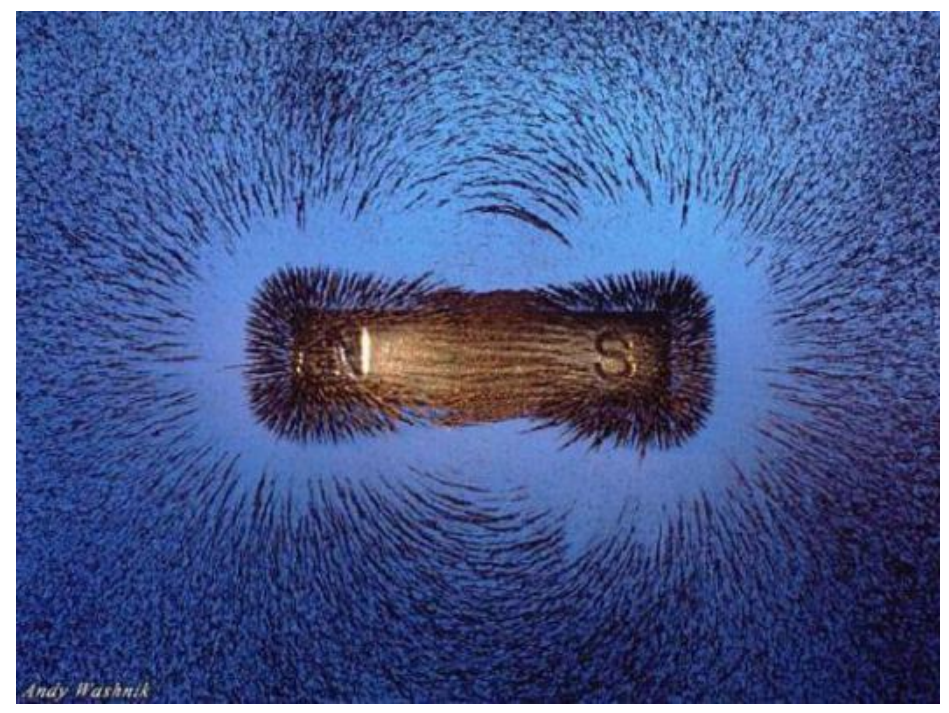

Fig. 1.10 : Lignes de champ magnétique

Faraday, dans le prolongement d'autres physiciens comme Ørsted, découvre le lien étroit entre champ électrique et champ magnétique.

James Clerk Maxwell formalise ensuite mathématiquement les idées de Faraday, les lois de champs, il unifie de façon rigoureuse l'électricité et le magnétisme, et fonde le courant de déplacement (*Maxwell, 1865*). Il fonde alors la théorie de l'électromagnétisme, décrit par les équations suivantes :

$$\begin{gathered}\vec{\nabla}\vec{E} = \frac{\rho}{\varepsilon_0} \\ \vec{\nabla}\vec{B} = 0 \\ \vec{\nabla} \wedge \vec{E} = -\frac{\partial \vec{B}}{\partial t} \\ \vec{\nabla} \wedge \vec{B} = \mu_0 \vec{J} + \mu_0 \varepsilon_0 \frac{\partial \vec{E}}{\partial t}.\end{gathered} \qquad (1.5)$$

avec $\rho$ la densité de charges, $\vec{J}$ le vecteur courant, $\varepsilon_0$la permittivité du vide, $\mu_0$ la perméabilité du vide.

Dans cette théorie, Maxwell introduit une nouvelle substance, le champ électromagnétique, qui avait pour support hypothétique l'éther luminifère de Huygens. Maxwell calcule la vitesse de propagation des ondes électromagnétiques dans l'éther, et obtient qu'elle coïncide avec la vitesse de la lumière :

$$c = \frac{1}{\sqrt{\varepsilon_0 \mu_0}} \approx 300.000\ km/s \quad (1.6).$$

La lumière est donc une onde électromagnétique transversale, associée à un champ électrique et un champ magnétique. Nous la représentons ci-dessous :

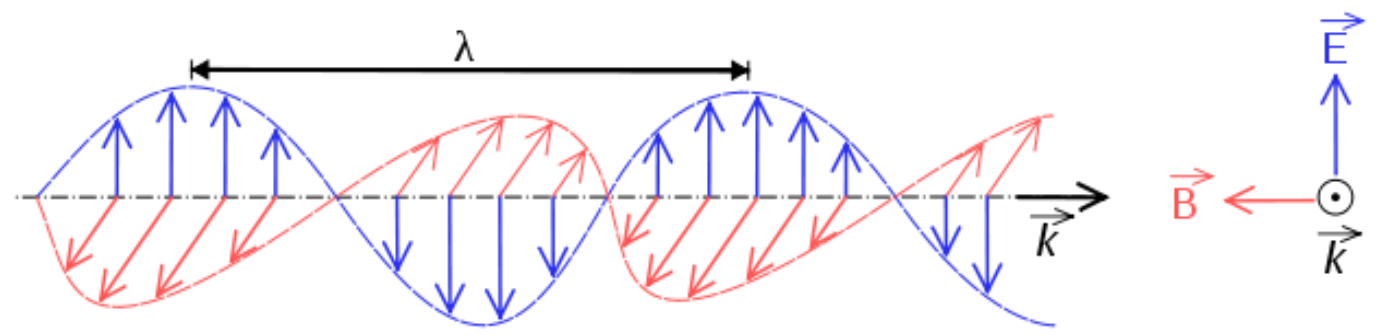

Fig. 1.11 : Onde électromagnétique

Le champ magnétique est toujours perpendiculaire au champ électrique, et le vecteur de Poynting correspond à la direction de propagation de l'onde électromagnétique. Il est donné par :

$$\vec{\pi} = \frac{\vec{E} \wedge \vec{B}}{\mu_0} \text{ , avec } \wedge \text{ l'opération de produit vectoriel} \quad (1.7).$$

## Le mouvement et la relativité restreinte

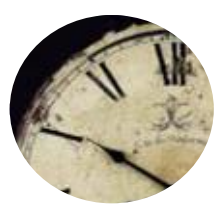

Le mouvement en physique a été théorisé pour la première fois par Galilée au XVI$^{e}$s. Le mouvement devient une propriété relative, il se définit par rapport à un référentiel. Naquirent alors les notions de vitesse instantanée et d'accélération, qui se développent entre le XVI$^{e}$ et XVII$^{e}$s.

Les transformations de Galilée sont décrites dans le cadre de la mécanique newtonienne de la façon suivante :

$$\begin{aligned} t' &= t \\ x' &= x - V * t \\ y' &= y \\ z' &= z. \end{aligned} \qquad (1.8)$$

Transformations de Galilée

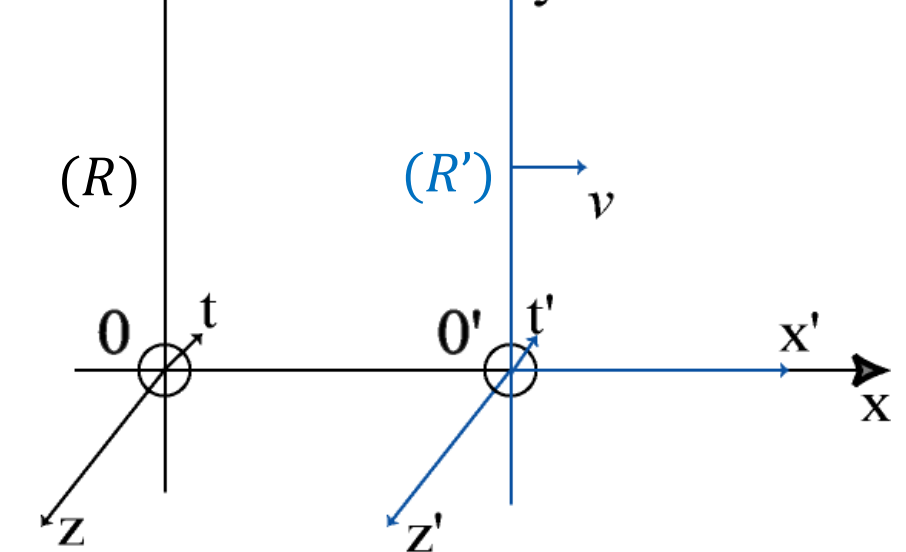

Selon la vision de Newton, le temps et l'espace sont absolus, universels. L'espace newtonien est un réceptacle inerte de la matière.

Avec la découverte de l'électromagnétisme par Maxwell, un problème se posa avec l'éther, support hypothétique des ondes électromagnétiques. Si l'on calcule la vitesse de la lumière par rapport à un référentiel mobile dans l'éther, on observe que la lumière se propage avec une vitesse différente et que les équations de Maxwell changent de forme. Cela signifie que les référentiels inertiels ne sont plus équivalents, et qu'il existe un référentiel privilégié (celui de l'éther).

A la fin du XIX^e^s., *Albert A. Michelson et Edward Morley* conduisent une expérience visant à mesurer la vitesse de la terre par rapport à l'éther, à l'aide un interféromètre (*Michelson & Morley, 1887*). Ils s'attendaient à obtenir une différence de vitesse de la lumière dans deux directions perpendiculaires. Toutefois, la vitesse de la lumière mesurée était la même, la terre restait «immobile» dans l'éther. Cela conduit les physiciens à mettre en doute l'existence de l'éther ou à formuler des hypothèses *ad hoc* (comme la contraction des bras de l'interféromètre dans le sens du mouvement).

En 1905, Einstein élabora la relativité restreinte sur la base notamment des travaux réalisés en premier ordre en $V/c$ par *Hendrik Lorentz (1895)* et *Henri Poincaré (1900)*.

La relativité repose sur deux postulats (*Einstein, 1905*) :

- Les lois physiques ont la même forme dans tout référentiel en mouvement uniforme (référentiel inertiel).
- La vitesse de la lumière dans le vide est la même dans n'importe quel référentiel inertiel. Elle est égale à c.

Cette théorie marque une profonde rupture avec la conception de Newton : l'espace et le temps sont ici relatifs, et les interactions ne sont plus instantanées. L'éther est alors une notion arbitraire qui n'est pas utile à l'expression de la relativité restreinte.

A partir des postulats de départ de la relativité restreinte, le physicien Lorentz déduit les transformations qui généralisent les transformations de Galilée. Nous pouvons les représenter de cette façon-ci (*Einstein, 1956*) :

$$\begin{aligned} ct' &= \gamma(ct - \beta x) \\ x' &= \gamma(x - \beta ct) \\ y' &= y \\ z' &= z \end{aligned} \qquad (1.9)$$

avec $\beta = \frac{V}{c}$ et $\gamma = \frac{1}{\sqrt{1-\frac{V^2}{c^2}}}$.

Transformations de Lorentz

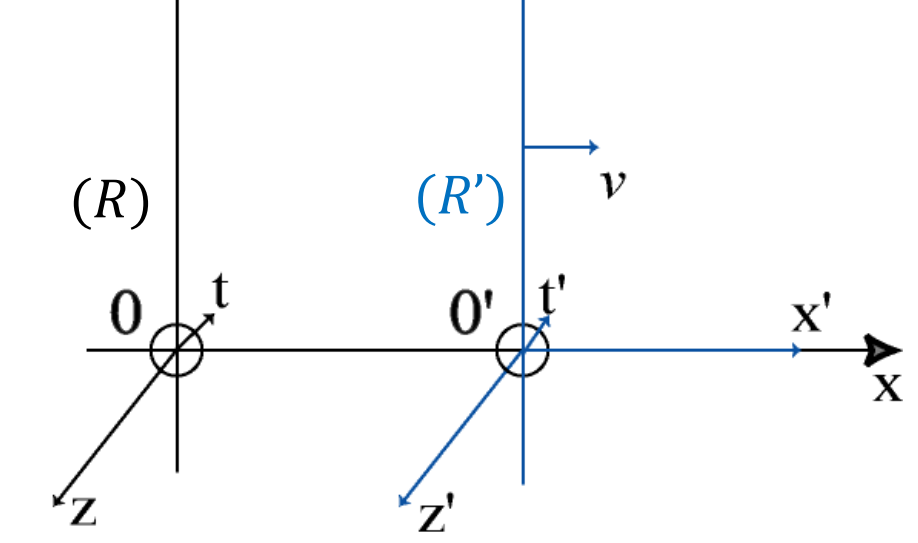

Cette transformation diffère de la transformation de Galilée de par la contraction des longueurs et la dilatation des durées lors du mouvement (facteur $\gamma$ ). A faible vitesse (contraction faible des objets), $V/c \approx 0$, et l'on retrouve alors les transformations de Galilée décrites précédemment.

Dès lors, l'espace et le temps ne sont plus des entités séparées, Hermann Minkowski les fusionne en une même réalité quadridimensionnelle appelée « espace-temps ».

Il généralise la notion de vecteur (3 dimensions) à celle de quadrivecteur (4 dimensions). Par exemple, le quadrivecteur énergie impulsion comporte l'énergie comme composante temporelle et la quantité de mouvement comme composante spatiale :

$$p^{\mu} = mu^{\mu} = \begin{pmatrix} E/c \\ p_x \\ p_y \\ p_z \end{pmatrix} \quad (1.10).$$

En relativité restreinte, l'énergie est donnée par :

$$E = mc^2 = \frac{m_0 c^2}{\sqrt{1 - \frac{V^2}{c^2}}} \quad (1.11)$$

Formule écrite par Einstein (1912)

$$E^2 = p^2c^2 + {m_0}^2c^4.$$

En relativité restreinte, la masse est considérée comme une forme d'énergie, appelée énergie de masse. La charge électrique, quant à elle, est un invariant relativiste : n'importe quelle particule, quelle que soit sa vitesse, garde toujours sa charge q.

L'espace newtonien tridimensionnel est remplacé par un espace quadridimensionnel, l'espace de Minkowski (*Einstein, 1956*). On dessine alors des diagrammes d'espace-temps où les cônes de lumière jouent un rôle prépondérant. Seules les trajectoires des particules à l'intérieur du cône sont autorisées (vitesse inférieure à celle de la lumière).

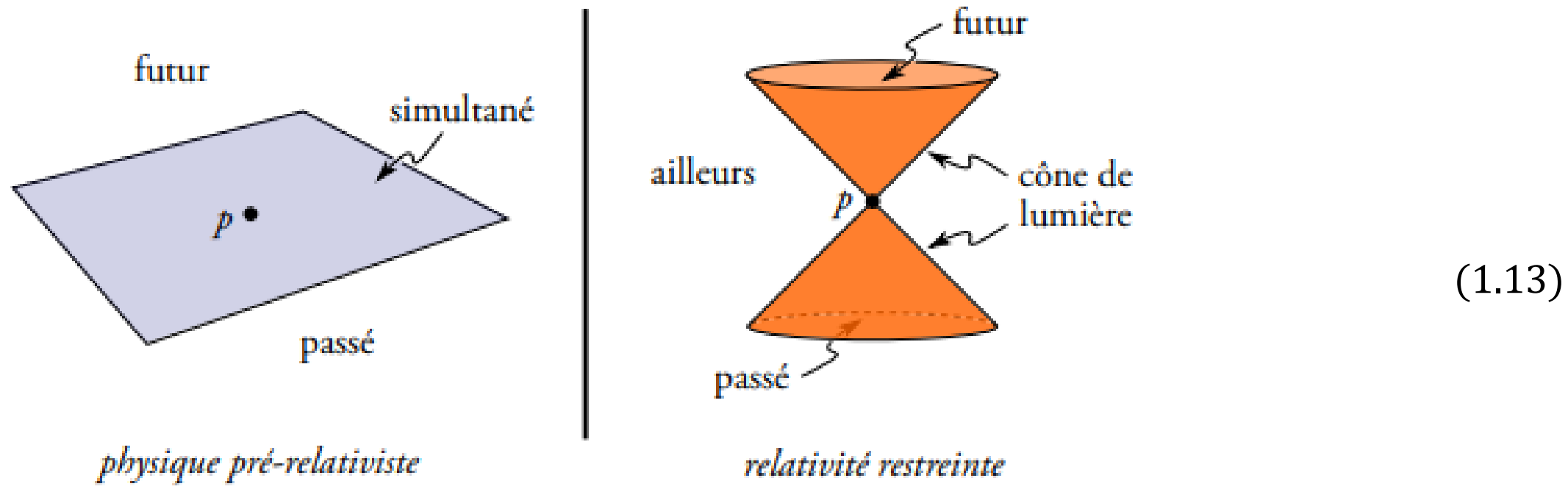

(1.13)

Fig. 1.12 : Conceptions de l'espace

Deux événements sont reliés par une distance spatio-temporelle (aussi appelé intervalle d'espace-temps) :

$$ds^2 = c^2 dt^2 - dl^2 \qquad (1.12).$$

Cet intervalle d'espace-temps est un invariant relativiste (invariant par changement de référentiel).

Les équations de la relativité conduisent à des phénomènes surprenants comme la contraction des longueurs et le ralentissement du temps.

Considérons deux référentiels : le référentiel $(R)$ où l'objet est immobile (longueur propre $L_0$, temps propre $t_0$), et un autre référentiel $(R')$ est translation à la vitesse $v$ par rapport à $(R)$.

On a :

$$L = \frac{L_0}{\gamma} = \sqrt{1 - \frac{V^2}{c^2}} L_0 \leq L_0 \qquad \rightarrow \text{ Contraction des longueurs}$$

$$t = \gamma . t_0 = \frac{t_0}{\sqrt{1 - \frac{V^2}{c^2}}} \geq t_0 \qquad \rightarrow \text{ Dilatation des durées}$$

En 1915, Einstein publie la relativité générale, généralisation de la relativité restreinte pour les mouvements non inertiels. C'est une théorie de la gravitation, où l'espace-temps est courbé par la matière et la matière est influencée par la courbure de l'espace-temps :

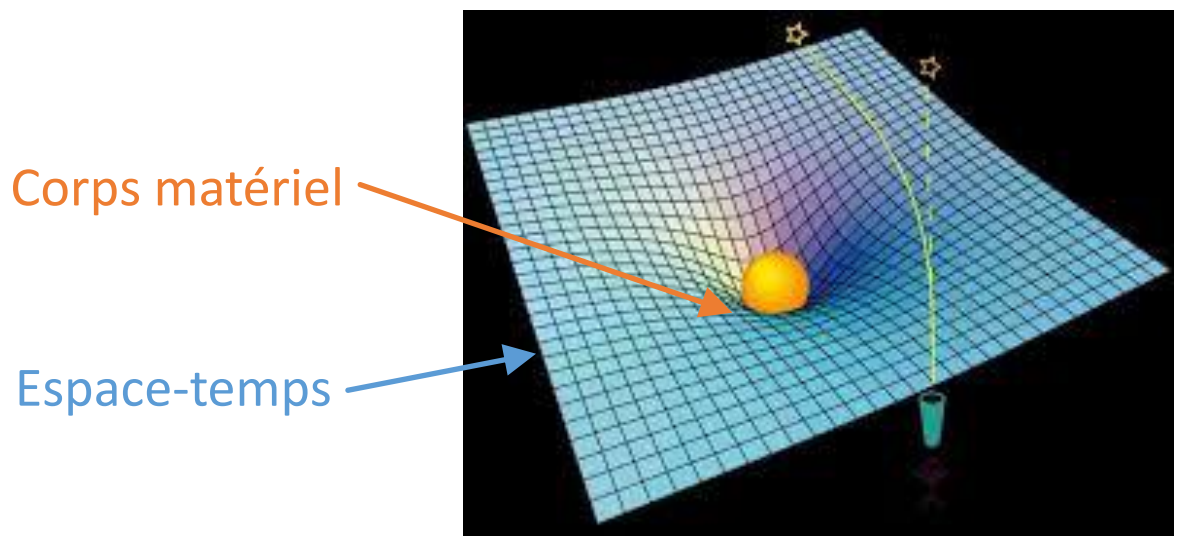

$$\underbrace{R_{\mu\nu} - \frac{1}{2} R g_{\mu\nu}}_{\text{Courbure}} = \frac{8\pi G}{c^4} \underbrace{T_{\mu\nu}}_{\text{Tenseur énergie-impulsion}} \tag{1.14}$$

Fig. 1.13 : Conception de l'espace-temps en relativité générale

Cette théorie se base sur le principe d'équivalence et la géométrie différentielle de Riemann.

Elle se fonde également sur la covariance généralisée, selon laquelle les lois de la nature doivent être covariantes relativement à toutes les transformations des coordonnées.

Les avancées majeures de la science du XX$^{e}$s. ont été fortement influencées par la philosophie. Par exemple, les idées de Mach, Kant et d'autres philosophes ont nourri les découvertes d'Einstein avec notamment la relativité.

Nous verrons qu'il existe un lien étroit entre philosophie et physique avec Louis de Broglie et sa mécanique ondulatoire.

## II. Du caractère ondulatoire de la matière

### 1) La dualité onde-corpuscule et la théorie de la double solution (De Broglie)

« Louis de Broglie a levé un coin du grand voile » Albert Einstein

De Broglie est un physicien français né en 1892 à Dieppe, il est issu de la maison noble De Broglie.

Son frère aîné, Maurice De Broglie, travaillait sur les rayons X, et il l'initie aux travaux de la physique moderne. Pendant la guerre, Louis de Broglie est affecté à la radiotélégraphie militaire et travaille à l'émetteur de la tour Eiffel. Après la guerre, il revient travailler au laboratoire de son frère, reprend ses études en science, et réfléchit longuement aux problèmes des ondes et des corpuscules. Les lectures qu'il suit (Collège de France, comptes rendus des congrès de Solvay,...) le mirent en contact avec de nombreuses branches de la physique, et notamment le problème des quanta. Il savait qu'Hamilton avait remarqué l'analogie entre les lois de la mécanique et la propagation des rayons lumineux, et se demandait si cette analogie aurait un contenu physique beaucoup plus profond. Après diverses études préliminaires, il publie sa thèse 1924 que nous analyserons dans la partie suivante.

Nous pourrons nous aider également dans cette partie de la thèse réalisée par Adrien Vila-Valls en 2012, intitulée *Louis de Broglie et la diffusion de la mécanique quantique en France (1925-1960)*.

De Broglie faisait partie de ces personnes qui manient à la fois l'art de philosopher et de faire de la science. Il cachait, sous son caractère réservé, un intérêt universel pour toutes les activités humaines. Penseur essentiellement solitaire, en grande partie autodidacte, il avait une vision originale de la physique théorique.

Il ne s'intéressait qu'aux questions fondamentales, qu'à la recherche de la connaissance pure et désintéressée, se considérait comme un théoricien, préoccupé par l'obtention d'un tableau d'ensemble qui permettent de comprendre de manière cohérente les phénomènes empiriques de la physique. La représentation concrète et imagée des phénomènes physiques est pour lui un des outils heuristiques les plus puissants du théoricien; « c'est elle qui révèle le sens véritable des formules, qui nous fait pénétrer dans la réalité physique profonde » indique De Broglie.

Thèse de doctorat de Louis De Broglie : *Recherches sur la théorie des Quanta, 1924*

Directeur de thèse : Paul Langevin; thèse soutenue à la Sorbonne

Comme nous avons vu précédemment, la lumière revêt une double nature : elle est soit une onde dans certaines limites, soit un corpuscule (on parle de dualité onde-corpuscule).

Cela est bien résumé par De Broglie au début de sa thèse de doctorat :

« L'histoire des théories optiques montre que la pensée scientifique a longtemps hésité entre une conception dynamique et une conception ondulatoire de la lumière : ces deux représentations sont donc sans doute moins en opposition qu'on ne l'avait supposé. »

Inspiré par la dualité onde-corpuscule de la lumière, Louis de Broglie proposa de la généraliser à toutes les particules de matière microscopiques (comme l'électron).

Tout d'abord, dans sa thèse, De Broglie étudie l'onde de phase associée à une particule, comme un électron. Il se fonde sur les formules de la relativité, dont $E = mc^2$.

De Broglie formule l'hypothèse qu'à chaque morceau d'énergie, de masse propre $m_0$ (masse dans le référentiel où il est au repos), est lié un phénomène périodique de fréquence $\nu_0$ tel que : $h\nu_0 = m_0c^2$

L'énergie de l'électron est répandue dans tout l'espace avec une très forte condensation dans une région de très petites dimensions dont les propriétés sont mal connues. La fréquence d'une onde plane monochromatique se transforme comme

$$\nu = \frac{\nu_0}{\sqrt{1-\beta^2}} \qquad (2.1).$$

Tandis qu'une fréquence d'horloge se transforme comme $\nu = \nu_0\sqrt{1-\beta^2}$ (2.2).

De Broglie pointe là une contradiction, et propose de la résoudre en introduisant une vitesse de groupe et une vitesse de phase :

- Fréquence de groupe $\nu = \nu_0\sqrt{1-\beta^2}$ et vitesse de groupe $V = \beta c\ (< c)$ du phénomène périodique lié au mobile.
- Fréquence de phase $\nu' = \frac{\nu_0}{\sqrt{1-\beta^2}}$ et vitesse de phase $V' = c/\beta\ (> c)$ de l'onde de phase associée. Cette onde ne transporte pas d'énergie et sa fréquence vérifie la relation $E = h\nu'$.

Cela signifie que la phase du phénomène périodique est invariante par changement de référentiel / par transformation de Lorentz : $\nu_0 t_0 = \nu'\left(t' - \frac{\beta x}{c}\right)$ (2.3).

De plus, $V.V' = c^2$.

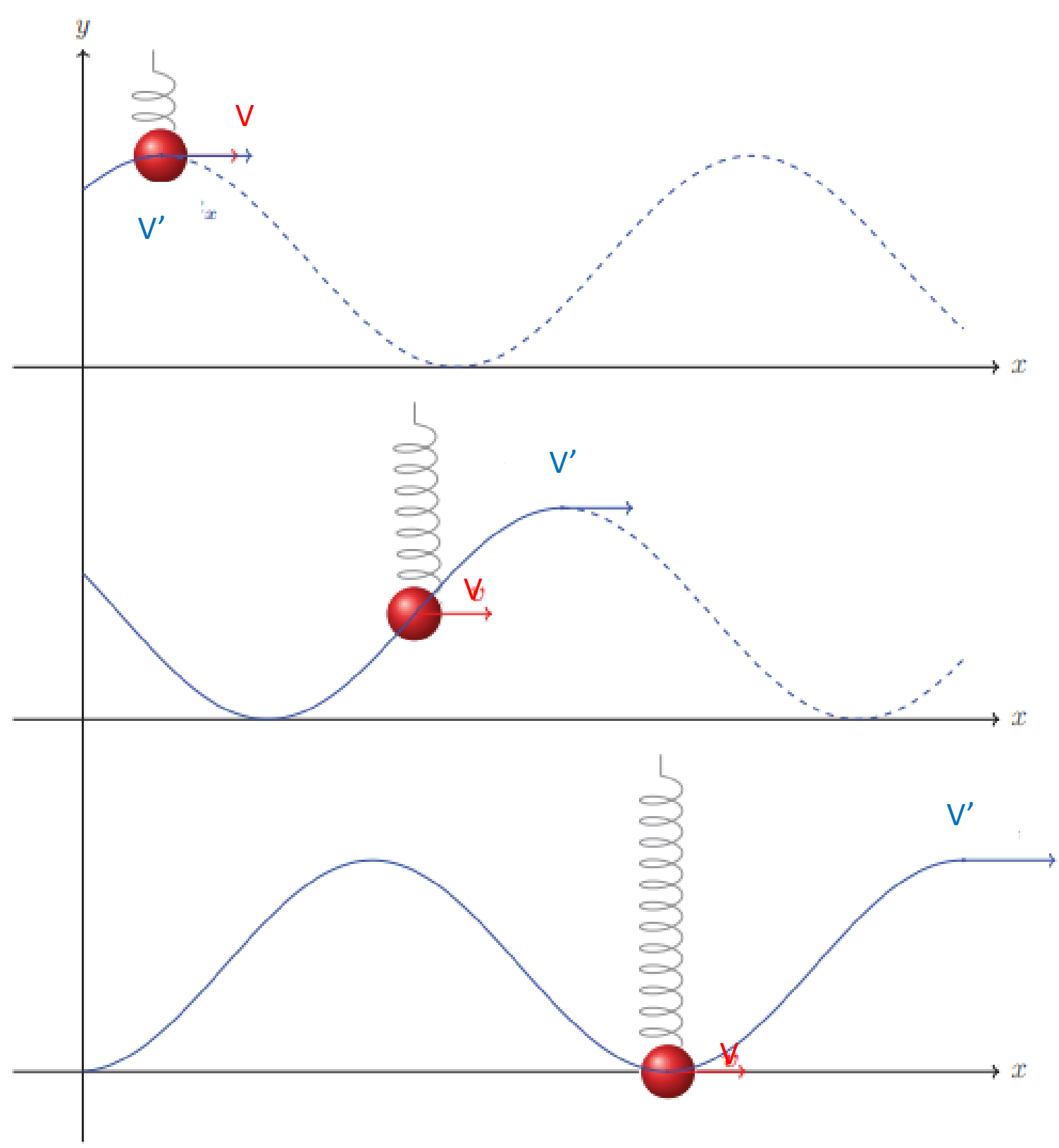

Fig. 2.1 : Représentation du corps se déplaçant à la vitesse V et de l'onde de phase se déplaçant à la vitesse V'. Les deux oscillations sont constamment en phase.

Source : *Besson, 2018*

De Broglie indique : « La particule peut être assimilée à une petite horloge possédant une vibration interne qui est constamment en phase avec celle de l'onde ».

Comment une particule assimilée à une petite horloge peut-elle se déplacer dans son onde de façon à ce que sa phase interne reste constamment égale à celle de l'onde ?

La variation de la phase d'une onde plane monochromatique se déplaçant suivant l'axe des $x$ s'exprime comme suit :

$$d\varphi = 2\pi\left(\nu dt - \frac{dx}{\lambda}\right) = 2\pi\left(\frac{\nu_0}{\sqrt{1-\beta^2}}dt - \frac{dx}{\lambda}\right) = \frac{2\pi}{h}\left(\frac{m_0c^2}{\sqrt{1-\beta^2}}dt - h\frac{dx}{\lambda}\right) \qquad (2.4).$$

La variation de la phase interne d'une particule se déplaçant le long de l'axe des $x$ s'exprime quant à elle comme suit :

$$d\varphi_i = 2\pi\nu_0\sqrt{1-\beta^2}dt = \frac{2\pi}{h}m_0c^2\sqrt{1-\beta^2}dt$$

$$d\varphi = d\varphi_i \;\; ; \;\; dx = V.dt$$

$$\rightarrow \frac{m_0c^2}{\sqrt{1-\beta^2}} - m_0c^2\sqrt{1-\beta^2} = \frac{h}{\lambda}\frac{dx}{dt} = \frac{hV}{\lambda}.$$

De plus, $\frac{m_0c^2}{\sqrt{1-\beta^2}} - m_0c^2\sqrt{1-\beta^2} = \frac{m_0V^2}{\sqrt{1-\beta^2}}$

$$\rightarrow p = \frac{m_0V}{\sqrt{1-\beta^2}} = \frac{h}{\lambda} \qquad (2.5).$$

De par ces raisonnements, De Broglie obtient les deux relations fondamentales de la mécanique ondulatoire :

$$E = h\nu \;\; ; \;\; p = \frac{h}{\lambda} \qquad (2.6).$$

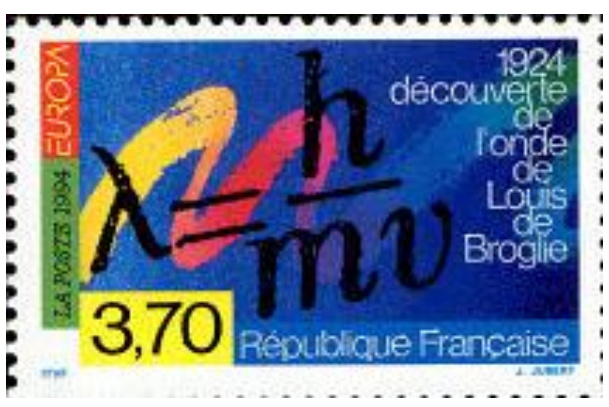

Fig. 2.2 : Timbre poste, datant de 1994

Le caractère ondulatoire de la matière n'est observable que pour les objets microscopiques (comme l'électron), à l'inverse des particules macroscopiques.

En *1923*, *Arthur Compton* découvre l'effet Compton (que nous avons décrit précédemment), où il introduit la longueur d'onde de l'électron : $\lambda_c = \frac{h}{mc}$.

En introduisant la fréquence, $\nu_c = \frac{c}{\lambda_c}$, la formule relativiste $E = mc^2$, on a: $E = h\nu_c$.

La fréquence de Compton de l'électron correspond donc à celle découverte par De Broglie relativement à l'énergie.

Dans une seconde partie, De Broglie développe ses idées en se fondant sur les principes de moindre action (Maupertuis, Hamilton) et le principe de Fermat (propagation des ondes). Il vise à répondre à la question : « Quand un mobile se déplace dans un champ de force d'un mouvement varié, comment se propage son onde de phase ? ».

En 1744, Maupertuis a énoncé le principe de moindre action en ces mots : « L'action est proportionnelle au produit de la masse par la vitesse et par l'espace. Lorsqu'il arrive quelque changement dans la Nature, la quantité d'Action employée pour ce changement est toujours la plus petite qu'il soit possible. » Ce qui se traduit mathématiquement par la formule (due à Euler) :

$$\delta \int_A^B mV dl = 0 \Leftrightarrow \delta \int_A^B \frac{m_0 \beta c}{\sqrt{1-\beta^2}} dl = 0 \qquad (2.7)$$

ou de façon générale $\delta \int_A^B \sum_i p_i dq^i = 0$.

Un corps prend la direction qui lui permet de dépenser le moins d'action dans l'immédiat (ou d'acquérir le plus d'énergie dans l'immédiat).

Les travaux de Maupertuis furent repris par Hamilton pour fonder la mécanique hamiltonienne.

On a les relations suivantes, utilisées par De Broglie dans sa thèse :

Action : $S = \int_{t_1}^{t_2} Ldt$ , $avec\ L = E_c - E_p$ le lagrangien

$$\delta S = 0 \rightarrow \delta \int_{t_1}^{t_2} Ldt = 0$$

$$\rightarrow \frac{d}{dt}\left(\frac{\partial L}{\partial \dot{q}_i}\right) = \frac{\partial L}{\partial q_i} \quad \text{(équation d'Euler-Lagrange)} \quad (2.8).$$

Un siècle avant, Fermat avait avancé un principe similaire pour les rayons lumineux de l'optique géométrique (trajectoire telle que la durée du parcours soit minimale).

Le principe de Fermat permet de déterminer les rayons d'une onde de fréquence $\nu$ :

$$\delta \int_A^B \frac{\nu}{V} dl = \delta \int_A^B \frac{m_0 c^2}{h\sqrt{1-\beta^2}} \frac{1}{c/\beta} dl = \delta \int_A^B \frac{m_0 \beta c}{h\sqrt{1-\beta^2}} dl = 0 \quad (2.9).$$

De Broglie montre qu'il est possible de relier le principe de moindre action et le principe de Fermat, en utilisant la quadri-impulsion du corps pour le 1er principe, et le quadrivecteur d'onde pour le 2e.

$$p_0 = E = \gamma m_0 c^2 = \hbar\omega = \hbar k_0, \text{ avec } \hbar = h/2\pi$$

$$\vec{p} = \gamma m_0 \vec{V} = \hbar\vec{k}. \qquad (2.10)$$

Généralisation : $p_i = \hbar k_i,\ i = 0,1,2,3.$

Louis de Broglie obtient là une formule relativiste, faisant intervenir des quadrivecteurs (avec 1 composante temporelle et 3 composantes spatiales).

Citons ici De Broglie : « Le principe de Fermat appliqué à l'onde de phase est identique au principe de Maupertuis appliqué au mobile : les trajectoires dynamiquement possibles du mobile sont identiques aux rayons possibles de l'onde ». Le mobile suit la trajectoire fixée par le principe de Fermat appliqué à l'onde. La nouvelle dynamique du point matériel libre est donc à l'ancienne dynamique ce que l'optique ondulatoire est à l'optique géométrique.

Enfin, De Broglie se penche sur l'étude du modèle quantique de l'atome.

Bohr avait montré que l'électron autour du noyau positif ne peut décrire que certaines trajectoires circulaires stables, les autres étant irréalisables dans la nature. On peut associer des numéros à ces orbites ($n = 1,2,3,4, \ldots$).

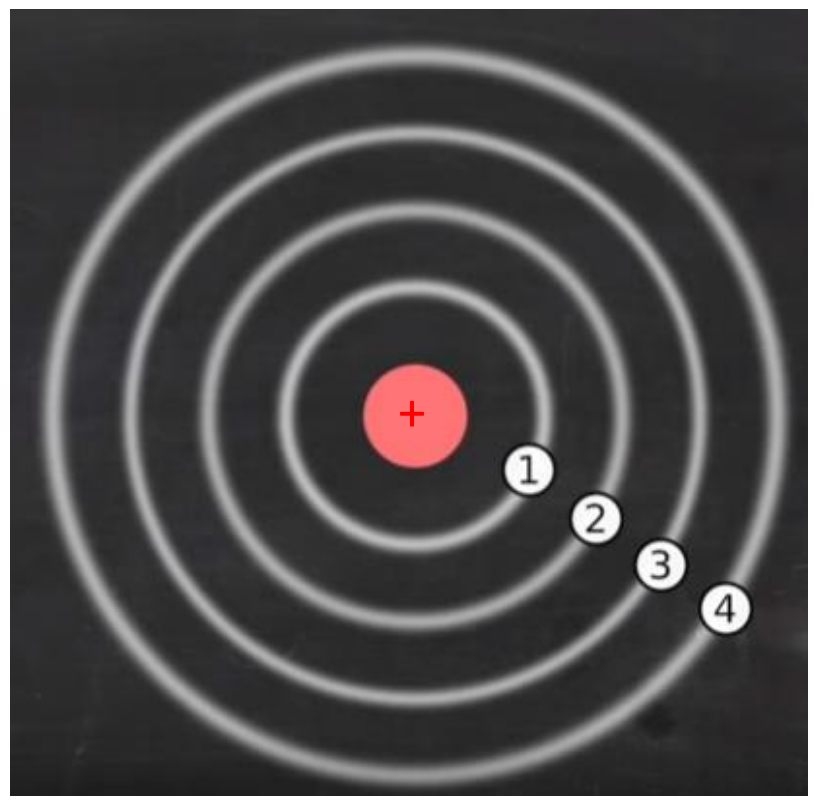

Fig. 2.3 : Modèle de Bohr : noyau positif au centre et orbites discrètes des électrons

De Broglie se demande comment comprendre ces états discrets et devine une ressemblance avec les ondes stationnaires. Il conçoit un électron tournant autour du noyau atomique comme étant lié à une onde.

Partant du principe de Fermat pour l'onde de phase de l'électron, il montre que les orbites stables sont celles dont la longueur $l$ vérifie $l = n.\lambda$ (condition de résonnance), avec $\lambda$ la longueur d'onde de l'onde de phase et $n \in \mathbb{N}$.

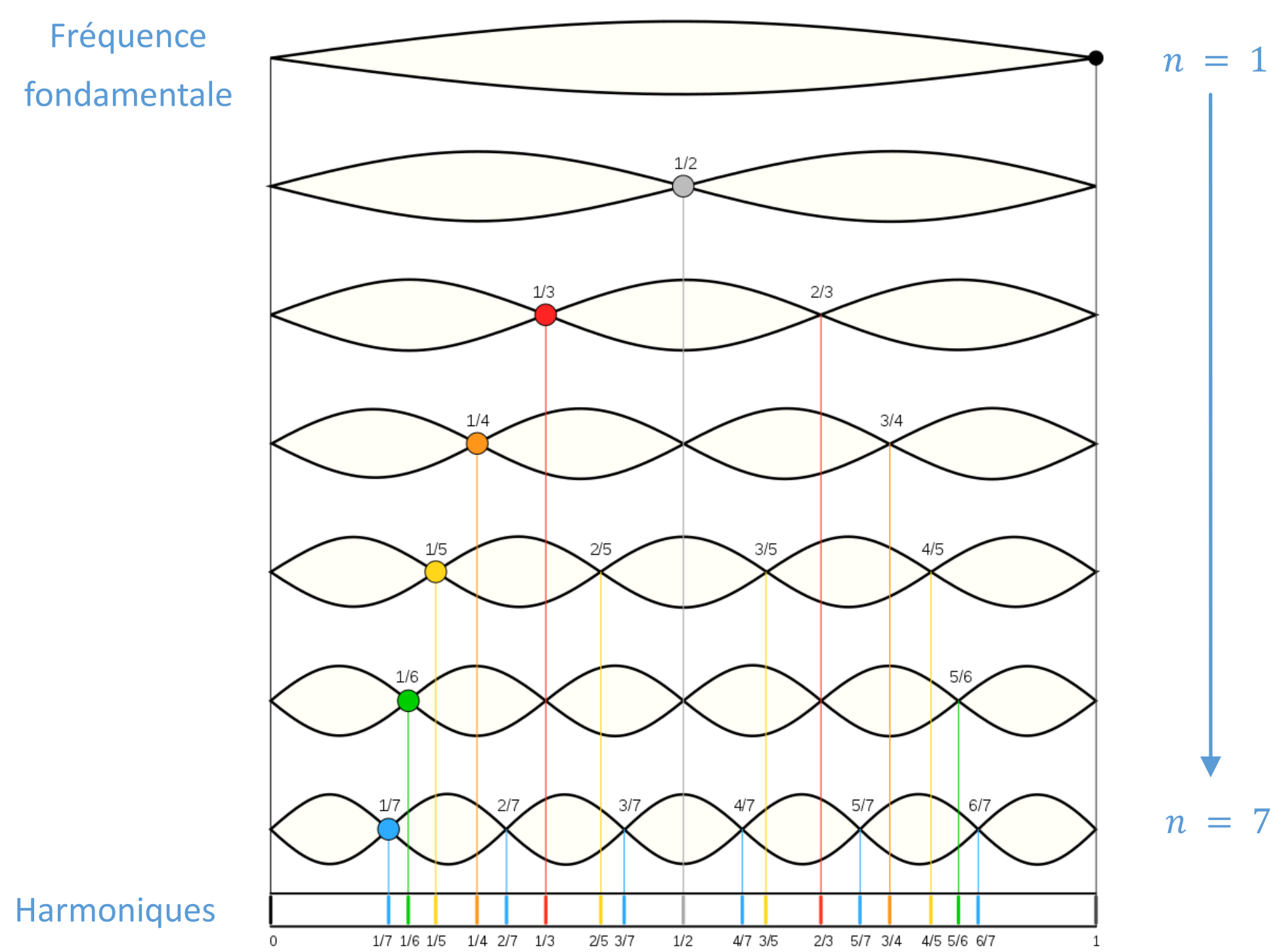

Fig. 2.4 : Ondes stationnaires (corde vibrante)

Nous voyons bien sur le schéma que les vibrations des ondes stationnaires sont « quantifiées » (ne prennent que certaines valeurs), ce qui explique intuitivement la quantification des orbites.

La condition de résonnance se traduit de façon générale par la formule suivante :

$$m_0 \oint V dl = nh \qquad (2.11).$$

Dans le cas des trajectoires circulaires de l'atome de Bohr (de rayon R), on a :

$$m_0 \oint V dl = 2\pi R m_0 V \qquad (2.12).$$

Nous pouvons également introduire la vitesse angulaire $\omega$ : $V = \omega R$.

Avec ces formules, De Broglie retrouve alors la formule de Bohr, selon laquelle les trajectoires circulaires stables sont associées à un moment de la quantité de mouvement qui est un multiple entier de $\hbar$ :

$$m_0 \omega R^2 = n\hbar \qquad (2.13).$$

De Broglie permet ainsi d'expliquer, avec son approche ondulatoire, pourquoi certaines orbites de l'atome sont stables et donne une justification à la quantification. Les orbites quantifiées peuvent s'interpréter par la résonnance de l'onde de phase sur la longueur d'onde des trajectoires circulaires.

Dans son rapport de thèse, Langevin apprécie « l'originalité et la profondeur des idées [et] la coordination remarquable qu'elles permettent ».

Les hypothèses révolutionnaires proposées par De Broglie incite son directeur de thèse Paul Langevin à demander l'avis d'Albert Einstein, qui lui répondra quelques mois plus tard : « Le travail de Louis De Broglie m'a fait grande impression. Il a levé un coin du grand voile ».

Comme toute théorie, la théorie ondulatoire des particules de De Broglie doit être confirmée expérimentalement. Ce dernier proposa la diffraction des électrons.

En 1927, Davisson et Germer découvrent expérimentalement ce phénomène de diffraction des électrons par des cristaux de nickel. De Broglie se verra attribué le prix Nobel en 1929.

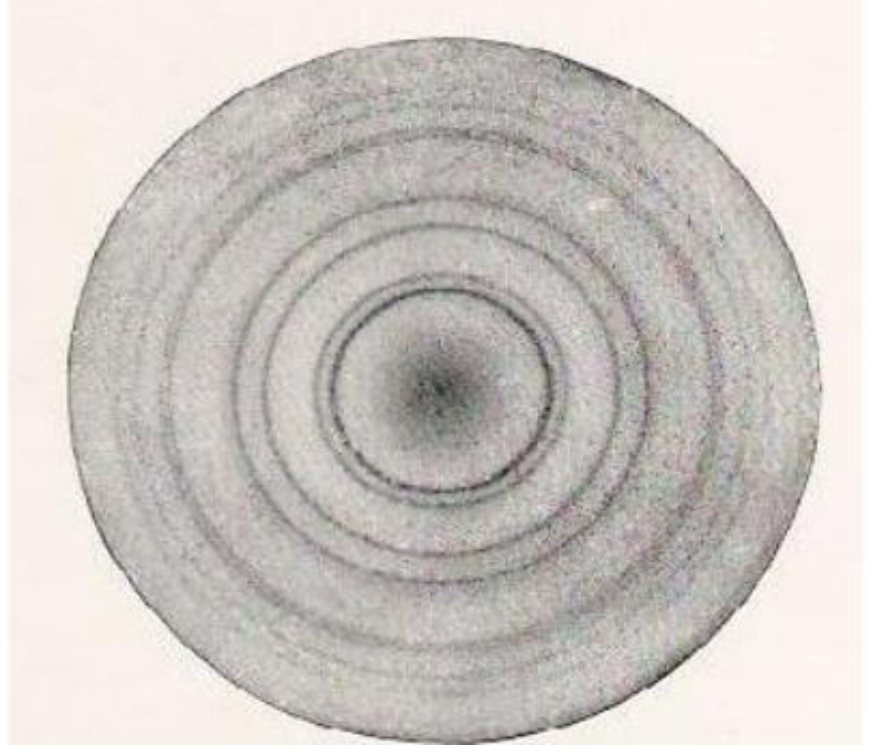

Fig. 2.5 : Figure de diffraction des électrons (*Davisson & Germer, 1927*)

Ces découvertes valurent la notoriété de De Broglie. Il est invité à participer au 5e Conseil Solvay en 1927. Il est ensuite nommé maître de conférences à l'Université de Paris en 1928, plus précisément à l'Institut Henri Poincaré; puis élu à l'Académie des Sciences en 1933 et à l'Académie française en 1944.

« Son apport à la connaissance du monde où nous vivons est si important, si bouleversant, si porteur d'images nouvelles » indiquait Jean Hamburger, directeur de l'Académie française, dans son hommage à Louis de Broglie.

Mais De Broglie reconnait également dans sa thèse la nécessité d'étudier et de compléter l'idée que le mouvement d'un point matériel dissimule la propagation d'une onde, ce qui représenterait « une synthèse d'une grande beauté rationnelle ».

### Théorie de la double solution : *The double solution theory, a new interpretation of wave mechanics. Journal de Physique, 1927*

Après la découverte des formules de dualité onde-corpuscule, De Broglie en propose une interprétation. La particule, considérée comme une petite horloge, est associée à une onde qui la guide; c'est l'onde pilote, onde physique réelle de faible amplitude. Il fait également intervenir une autre onde $\psi$, fictive, de nature statistique / probabiliste et normalisée.

En notant $v$ l'onde physique réelle, on a : $\psi = Cv$. Les solutions $v$ et $\psi$ ont la forme d'une onde, la phase étant la même mais l'amplitude étant différente.

$C$ est un facteur de normalisation tel que $\int_V |\psi|^2 d\tau = 1$, $V$ représentant le volume occupé par l'onde $v$. $|\psi|^2 d\tau$ donne la probabilité de trouver la particule dans l'élément de volume $d\tau$.

De Broglie qualifie cette théorie de « double solution » car $v$ et $\psi$ sont deux solutions de la même équation d'onde, l'équation de Schrödinger (que nous verrons après).

« J'ai introduit comme théorie de la double solution l'idée qu'il était nécessaire de distinguer deux solutions différentes qui sont toutes les deux reliées à l'équation d'onde. L'une que j'ai appelé l'onde $v$ qui était une onde physique réelle représentée par une singularité; l'autre comme l'onde $\psi$ de Schrödinger, qui est une représentation de probabilité sans singularité »

La particule forme une petite région de grande concentration d'énergie, qui peut être approximée par une singularité en mouvement.

$$v = a(\vec{x},t)exp\left(\frac{i}{\hbar}\phi(\vec{x},t)\right)$$

$$E = \frac{\partial \phi}{\partial t}, \;\; \vec{p} = -\vec{\nabla}\phi$$

$$E = \frac{M_0 c^2}{\sqrt{1-\beta^2}}, \;\; \vec{p} = \frac{M_0 \vec{V}}{\sqrt{1-\beta^2}}$$

$$\rightarrow \vec{V} = \frac{c^2\vec{p}}{E} = -c^2 \frac{\overrightarrow{grad}\phi}{\partial\phi/\partial t}. \quad \text{Formule de guidage} \quad (2.14)$$

Dans le cas d'une onde monochromatique, la vibration interne de la particule est en phase avec l'onde sur laquelle elle est portée.

Appliquons l'équation de Schrödinger à l'onde v :

$$i\hbar\frac{\partial v}{\partial t} = -\frac{\hbar^2}{2m}\Delta v + U.v.$$

De Broglie pose que $v = a.exp\left(\frac{i\phi}{\hbar}\right)$, où $a$ est l'amplitude de l'onde et $\phi$ sa phase.

Il obtient alors deux équations :

$$\frac{\partial\phi}{\partial t} - U - \frac{(\nabla\phi)^2}{2m} = -\frac{\hbar^2}{2m}\frac{\Delta a}{a} \quad \text{Équation de Jacobi généralisée}$$

$$\frac{\partial(a^2)}{\partial t} - \frac{1}{m}div(a^2\nabla\phi) = 0 \quad \text{Équation de continuité} \quad (2.15)$$

avec $U$ le potentiel classique.

En partant de l'équation de Jacobi généralisée, avec $\phi = S$ et en négligeant le terme de droite ($h \rightarrow 0$), on obtient $\frac{\partial S}{\partial t} - U = \frac{(\nabla S)^2}{2m}$. C'est l'équation de Jacobi de la mécanique classique.

Le terme de droite de l'équation de Jacobi généralisée s'interprète comme un potentiel quantique $Q$ : $Q = -\frac{\hbar^2}{2m}\frac{\Delta a}{a}$.

Le potentiel quantique est l'expression de la réaction de l'onde sur le corpuscule, et la force quantique est : $\vec{F} = -\vec{\nabla}Q$. En prenant en compte cette force, le mouvement défini par la formule de guidage n'est plus rectiligne. Les obstacles agissent sur la particule à travers le potentiel quantique, produisant une déflection.

De plus, on a :

$$\frac{\partial S}{\partial t} = E,\ \ \vec{p} = -\vec{\nabla} S\ \rightarrow\ \frac{\partial \phi}{\partial t} = E,\ \ \vec{p} = -\vec{\nabla}\phi\ ;$$

$$\vec{p} = m\vec{V}\ \rightarrow\ \vec{V} = -\frac{\vec{\nabla}\phi}{m}.\quad \text{Formule de guidage} \qquad (2.16)$$

La vitesse de la particule (à la position $x, y, z$ et au temps $t$) est fonction de la variation de la phase locale à ce point.

On peut calculer la variation de phase de l'onde associée à la particule :

$$d\phi = \frac{\partial\phi}{\partial t}dt + \frac{\partial\phi}{\partial l}dl = \left(\frac{\partial\phi}{\partial t} + \vec{V}.\overrightarrow{grad}\phi\right)dt = \left(\frac{m_0 c^2}{\sqrt{1-\beta^2}} - \frac{m_0 V^2}{\sqrt{1-\beta^2}}\right)dt = m_0 c^2\sqrt{1-\beta^2}dt \quad (2.17).$$

La variation de la phase interne de la particule est : $d\phi_i = m_0 c^2\sqrt{1-\beta^2}dt = d\phi$.

Ainsi, la vibration de la particule est constamment en phase avec celle de l'onde.

L'onde physique réelle doit inclure une petite région de très grande amplitude, singulière, qui est la particule. On la note $u_0$.

L'onde physique ($u$) est donc composée de cette région singulière, et de l'onde $v$ d'amplitude très faible :

$$u = u_0 + v \qquad (2.18).$$

Le mouvement de la singularité subirait l'influence de tous les obstacles qui influeraient sur la propagation du phénomène ondulatoire dont elle est solidaire, et ainsi expliquerait l'existence des interférences et de la diffraction.

Mais De Broglie n'a pas voulu essayer de décrire la structure interne de cette région singulière, c'est à dire la particule. Cette description impliquerait certainement des équations non-linéaires selon lui.

Pour résumé, la théorie de la double solution de De Broglie indique que la particule microscopique est guidée par son onde. Deux ondes entrent en jeu : l'onde pilote réelle centrée sur la particule et l'onde statistique prédite par la théorique quantique usuelle.

Il a proposé que l'onde pilote venait des oscillations internes de la particule à la fréquence de Compton $\nu_c = \frac{mc^2}{h}$, et qu'elle évolue selon l'équation de Klein-Gordon.

Il a souligné l'importance de l'harmonie des phases, par laquelle la vibration interne de la particule, vue comme une horloge, reste en phase avec celle de l'onde pilote. Selon sa conception, l'onde et la particule maintiennent un état de résonnance.

De Broglie présenta au 5$^{e}$ congrès de Solvay de 1927 une autre version : la théorie de l'onde pilote. À la place de la solution à singularité est introduit un point matériel qui se déplace selon la loi de guidage et dont la fonction d'onde fournit la probabilité de présence. La théorie de l'onde pilote réalise les mêmes prédictions que la théorie de la double solution, mais elle est plus simple sur le plan formel.
Son exposé au congrès de Solvay trouva cependant peu d'audience. Pauli y trouva de sérieuses objections, et les autres scientifiques (mis à part Einstein et Schrödinger) étaient convaincus de l'interprétation orthodoxe de Copenhague. Face à tant d'oppositions, De Broglie se rallie à l'interprétation probabiliste de Bohr et Heisenberg. Pendant 25 ans, il l'a enseigné et l'a exposé dans ses livres.

### Apport de De Broglie à la physique théorique et expérimentale (Vila-Valls, 2012)

Le travail théorique de De Broglie constitue une réinterprétation des lois déjà connues à l'aide de nouvelles bases conceptuelles, et en une synthèse magistrale des concepts existants (synthèse entre la dynamique du point matériel et la théorie des ondes). Selon lui, la résolution des problèmes les plus graves de physique ne peut intervenir qu'après avoir remis la théorie sur de meilleures bases. Ses idées influenceront les travaux de nombreux physiciens, comme ceux de Schrödinger comme nous le verrons après.
Ses travaux ont suscité davantage l'intérêt de physiciens théoriciens français portés vers l'abstraction plutôt que de théoriciens plus proches de l'expérience, voire des expérimentateurs.
Il donna de nombreux cours à la Sorbonne et au Collège de France, sur la mécanique ondulatoire et la mécanique quantique, adaptés cependant à des chercheurs et des étudiants avancés.

Doté de qualités littéraires et philosophiques indéniables, il vulgarisa la mécanique ondulatoire et la physique quantique dans de nombreux écrits destinés au grand public. Ces écrits ont eu une influence sur les étudiants, ont suscité des vocations.

Il a également encadré des thèses de physique théorique, et ses idées influencent les travaux de ses élèves.

Dans les années 1930 et 1940, De Broglie était connu dans la communauté scientifique, et contribua au rayonnement de la physique théorique française.

Dans une optique de favoriser les idées nouvelles, la fondation Louis de Broglie a été créée en 1973. Elle a pour vocation de « participer au rayonnement de la pensée scientifique et humaniste de Louis de Broglie, de soutenir, de développer et de favoriser la recherche, notamment en physique théorique, en mathématiques, en histoire et philosophie des sciences ».

## Critique de la théorie de De Broglie

Les idées de De Broglie ont cependant fait l'objet de nombreuses critiques de la part des autres physiciens, en France et à l'étranger.

Tout d'abord, De Broglie donne une trajectoire et une vitesse précise aux photons et aux particules, alors que la mécanique quantique indique que ces entités n'ont pas de trajectoire bien définie.

Beaucoup de physiciens mathématiciens, formalistes, comme Dirac, ne virent même pas l'intérêt de ses travaux. Pour eux, l'essentiel de la physique est dans les relations mathématiques et non pas dans les images intuitives.

La théorie de la double solution laisse en suspens de nombreuses questions et pose des problèmes : le lien exact entre la particule et l'onde, la nature et l'origine de l'onde, singularité au sein de l'onde,…

De plus, la fonction d'onde $\psi$ chez De Broglie a un double statut : un statut physique (onde pilote) et un statut probabiliste. Or, certains éléments viennent fragiliser la pertinence d'attribuer à une telle onde un caractère physique réel : son caractère complexe et son espace de propagation (l'espace de configuration).

Enfin, nous ne trouvons pas non plus beaucoup de volonté de la part de Louis de Broglie de chercher à appliquer immédiatement les idées nouvelles contenues dans sa thèse après sa publication. En effet, la théorie de la double solution, beaucoup plus complexe mathématiquement que la mécanique ondulatoire de Schrödinger, serait techniquement difficile à appliquer à un problème de physique atomique par exemple, et le bénéfice pour l'expérimentateur serait contestable.

Devant les difficultés théoriques de la théorie de la double solution et devant les objections adressées par les membres de « l'école de Copenhague », Louis de Broglie finira par renoncer à sa théorie et à accepter l'interprétation probabiliste de la mécanique ondulatoire.

### 2) La mécanique quantique et l'interprétation de Copenhague : notions fondamentales, paradoxes et critiques

Louis De Broglie, *La physique quantique restera-t-elle indéterministe ?* (1952)

Franck Laloë, *Comprenons-nous vraiment la mécanique quantique* (2018)

La dualité onde-corpuscule peut s'interpréter de trois manières différentes :

- L'approche de Schrödinger qui consiste à nier la réalité du dualisme en contestant l'existence des corpuscules. Seules les ondes auraient une signification physique.
- L'approche de De Broglie où le corpuscule est une sorte de singularité au sein d'un phénomène ondulatoire étendu dont il serait le centre, comme nous avons vu précédemment.
- La dernière approche, qui consiste à ne considérer que les idées de corpuscule et d'onde continue et les regarder comme des faces complémentaires de la réalité au sens de Bohr. C'est l'interprétation orthodoxe de la mécanique ondulatoire, aussi appelé interprétation de Copenhague.

Inspiré par la dualité onde-corpuscule de De Broglie, le physicien viennois Erwin

Schrödinger formula son équation d'onde, qui devint avec le temps l'essence même de la théorie quantique (*Schrödinger, 1926*). C'est l'équation de Schrödinger :

$$i\hbar \frac{\partial \psi(\vec{r},t)}{\partial t} = -\frac{\hbar^2}{2m}\Delta\psi(\vec{r},t) + V(\vec{r},t)\psi(\vec{r},t) \qquad (2.19)$$

avec $\Delta$ l'opérateur Laplacien $\Delta = \frac{\partial^2}{\partial x^2} + \frac{\partial^2}{\partial y^2} + \frac{\partial^2}{\partial z^2}$ et $V$ le potentiel.

Cette équation concorde avec la formule classique de l'énergie :

$$E = E_c + E_P = \frac{p^2}{2m} + V,\ avec\ E = i\hbar\frac{\partial}{\partial t}\ et\ \vec{p} = -i\hbar\vec{\nabla} \qquad (2.20).$$

L'interprétation physique de la fonction d'onde $\psi$ fut apportée par Born en 1926 : le carré de cette fonction d'onde $|\psi|^2$ donne la densité de probabilité de trouver une particule.

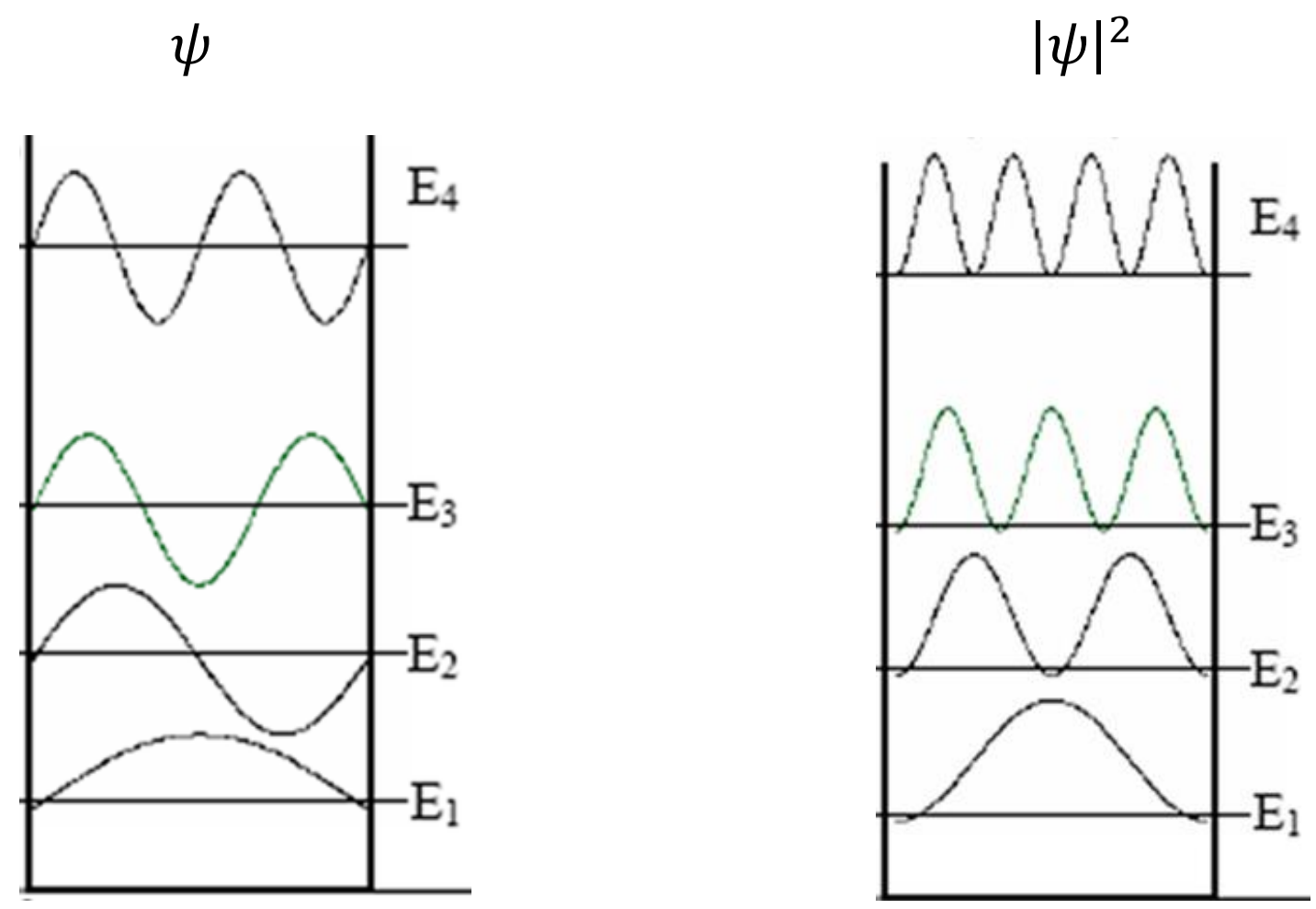

Fig. 2.6 : Exemple de fonction d'onde et de densité de probabilité pour différents niveaux d'énergie

Une autre formulation de la mécanique quantique se développa : la mécanique matricielle, construite par Heisenberg, Born et Jordan en 1925, où les grandeurs physiques classiques sont remplacées par des observables associées à des matrices. Il fut démontré par Pauli et Schrödinger que ces deux formalismes, ondulatoire et matricielle, sont équivalents.

Dans l'équation de Schrödinger, le temps apparaît comme une dérivée d'ordre 1, les coordonnées spatiales s'expriment en dérivées d'ordre 2. Cet aspect est en contradiction avec le principe inhérent à la théorie relativiste : le traitement systématique des 4 composantes (1 composante temporelle et 3 composantes spatiales) qui constituent le « quadrivecteur espace-temps ».

L'équation de Schrödinger fut généralisée au cas relativiste par Paul Dirac avec sa célèbre équation d'onde (*Dirac, 1928*) :

$$i\hbar\frac{\partial\psi}{\partial t} = \left(\alpha_0 mc^2 - i\hbar c\vec{\alpha}.\vec{\nabla}\right)\psi \Leftrightarrow \left(i\gamma^\mu\partial_\mu - mI\right)\psi = 0 \qquad (2.21)$$

avec $\psi$ : spineur de Dirac, contient les composantes de la particule et de l'antiparticule;

$\alpha_\mu$ *et* $\gamma^\mu, \mu = 0,1,2,3$ : matrices de Dirac; $I$ : matrice identité.

## Postulats et notions fondamentales de l'interprétation de Copenhague

L'interprétation orthodoxe de la mécanique quantique, appelée interprétation de Copenhague, est née au congrès Solvay de 1927.

Les postulats de l'interprétation de Copenhague sont les suivants :

- L'état d'une particule quantique est entièrement contenue dans un vecteur d'état noté $|\psi(x,t)\rangle$. L'ensemble des états d'une particule se composent linéairement sur une base, c'est le principe de superposition :

  $$|\psi\rangle = \sum_i \alpha_i |u_i\rangle, \text{ avec } \alpha_i \in \mathbb{C} \qquad (2.22).$$

  L'ensemble des états possibles d'un système définit ainsi un espace vectoriel, plus précisément un espace de Hilbert (noté $\mathcal{H}$).

- Toute grandeur physique mesurable $A$ (comme la position, la quantité de mouvement, le spin) est décrite par un opérateur linéaire $\hat{A}$, appelé observable, agissant sur les vecteurs d'un espace de Hilbert $\mathcal{H}$.

  Par exemple, l'opérateur quantité de mouvement est défini par : $\hat{P} = -i\hbar\vec{\nabla}$.

  La valeur moyenne de la mesure de l'observable $\hat{A}$ est donnée par :

  $$< \hat{A} > = \sum_k a_k P(a_k) = \langle\psi|A|\psi\rangle \qquad (2.23).$$

- L'évolution dans le temps de la fonction d'onde $\psi(x,t)$ est régie, dans le cas non-relativiste, par l'équation de Schrödinger :

  $$i\hbar\frac{d}{dt}|\psi(t)\rangle = \widehat{H}(t)|\psi(t)\rangle\,, avec\ \widehat{H} = -\frac{\hbar^2}{2m}\Delta + V \qquad (2.24).$$

Dans le cas relativiste, il faut utiliser l'équation de Dirac.

- La mesure d'une grandeur physique A donne comme résultat une des valeurs propres $a_n$ de l'observable $\hat{A}$ :

$$\hat{A}|\psi\rangle = a_n|\psi\rangle \qquad (2.25).$$

- La densité de probabilité est donnée par le carré de l'amplitude, $|\psi|^2$;
la probabilité de trouver à l'instant t une particule dans le volume $V$ est quant à elle donnée par : $P = \int_V |\psi(r,t)|^2 d^3r$.
- Réduction du paquet d'ondes : si la mesure de la grandeur physique A donne le résultat $a_n$, l'état du système immédiatement après la mesure est la projection normée de $\psi(x,t)$ sur le sous-espace propre associé à $a_n$.

Dans cette conception, l'onde et le corpuscule ne peuvent se représenter de manière classique. On ne peut attribuer ni position, ni vitesse, ni trajectoire. La position et la vitesse se révèlent uniquement au moment de la mesure, avec des certaines probabilités. Le corpuscule n'est plus un objet bien défini dans le cadre de l'espace et du temps.

La fonction d'onde $\psi$ caractérise complètement la particule quantique. Elle ne permet de prédire que des probabilités concernant les résultats des différentes mesures et évolue de façon régulière et prédictible selon l'équation de Schrödinger; dès que l'on effectue une mesure, elle effectue des sauts imprévisibles, non-déterministes, selon le postulat de réduction du paquet d'onde.

Pour un système de $N$ particules, la fonction d'onde $\psi$ se propage dans un espace des configurations à $3N$ dimensions, très différent de l'espace habituel.

La transition quantique-classique s'effectue par la procédure mathématique appelée « quantification canonique ». Les crochets de Poisson $\{p_i, q_i\} = \delta_{ij}$, $\{p_i, p_i\} = 0$, $\{q_i, q_i\} = 0$ sont remplacés par les commutateurs des opérateurs quantiques :

$$[p_i, q_i] = i\hbar, \; [p_i, p_i] = 0, \; [q_i, q_i] = 0 \qquad (2.26).$$

Le formalisme quantique a pour conséquence les relations d'incertitude de Heisenberg. On ne peut pas connaître de manière précise à la fois la position et la quantité de mouvement d'un corpuscule, comme illustré par la formule suivante :

$$\Delta x.\Delta p_x \geq \frac{\hbar}{2} \qquad (2.27).$$

La mécanique quantique déjoue notre sens commun et ne ressemble à rien de ce que l'on connaît dans la vie de tous les jours. Selon Bohr, il est impossible de transposer notre vision macroscopique dans le monde microscopique. La tâche de la physique n'est pas de comprendre la nature, mais de trouver une théorie qui colle avec l'expérience.

### Critiques et paradoxes de la mécanique quantique

L'interprétation actuelle de la mécanique quantique fait jouer une part fondamentale au point de vue probabiliste. S'effondrent alors la réalité objective du monde et la compréhension intuitive de la nature...

Bien que la mécanique quantique soit parfaitement en accord avec l'expérience, elle a fait l'objet de débats animés entre philosophes et scientifiques au cours du XX$^{e}$s. La physique quantique décrit-elle la réalité ou bien uniquement la perception que nous avons du monde ? Les principes probabilistes sont-ils intrinsèques ou bien sont-ils des conséquences de fondements que nous ne comprenons pas ? Existe-t-il une réalité parfaitement déterminée et descriptible dans le cadre de l'espace-temps par des variables cachées ?

Les grands savants de l'époque classique, de Laplace à Poincaré, ont toujours proclamé que les phénomènes naturels sont déterminés et que les probabilités résultent de notre ignorance ou notre incapacité à suivre un déterminisme trop compliqué. Dans la théorie cinétique des gaz, les lois de probabilités étaient considérées comme résultant de notre ignorance de mouvements déterminés, mais désordonnés et complexes, des molécules du gaz. En mécanique quantique, il s'agirait de «probabilité pure» qui ne résulterait pas d'un déterminisme caché.

Certains scientifiques contemporains, comme Einstein et Schrödinger, n'ont pas

accepté l'interprétation probabiliste de Copenhague. Selon Einstein, elle est incompatible avec la théorie de l'espace-temps (relativité) et avec une propagation de proche en proche à vitesse finie.

L'interprétation de Copenhague est fondée sur deux postulats peu compatibles :

- La mesure de Von Neumann, qui correspond à une réduction du vecteur d'état. C'est un processus discontinu et aléatoire. Cette interprétation fait jouer un rôle spécial à l'observateur.
- L'évolution de la fonction d'onde $\psi$ est basée sur l'équation de Schrödinger. Cette équation est continue et déterministe.

John Von Neumann a démontré un théorème, portant le nom de théorème d'incompatibilité, qui montre que la forme des lois de probabilité de la mécanique quantique est incompatible avec l'existence de paramètres cachés. Il serait donc impossible d'interpréter le formalisme de la mécanique quantique à l'aide de variables cachées. Il s'avéra par la suite que cette preuve n'était en fait relative qu'à une classe restreinte de telles variables. Le raisonnement de Von Neumann n'est pas applicable à la théorie de l'onde pilote de De Broglie par exemple.

Le principe de superposition est à l'origine de ce qu'on appelle le problème de la mesure quantique, avec le paradoxe du chat (ni mort, ni vivant) imaginé en 1935 par Schrödinger. Dans cette expérience de pensée, un chat est enfermé dans une boite avec un dispositif qui tue l'animal dès qu'il détecte la désintégration d'un atome d'un corps radioactif. Si les probabilités indiquent qu'une désintégration a une chance sur deux d'avoir eu lieu au bout d'une minute, la mécanique quantique indique que, tant que l'observation n'est pas faite (ou plus précisément qu'il n'y a pas eu de réduction du paquet d'onde), l'atome est simultanément dans deux états : intact et désintégré. Ainsi, le chat serait simultanément dans deux états (mort et vivant), jusqu'à ce que l'ouverture de la boîte (observation) déclenche le choix entre les deux états.
Cette situation d'un chat en état de superposition peut nous sembler grotesque. De plus, un animal a-t-il un niveau de conscience lui permettant de savoir s'il est vivant et de réduire le paquet d'ondes ? Est-il capable à lui seul de forcer l'émergence classique d'un résultat unique, ou peut-il être mis dans un état quantique où il serait à la fois mort et vivant ?

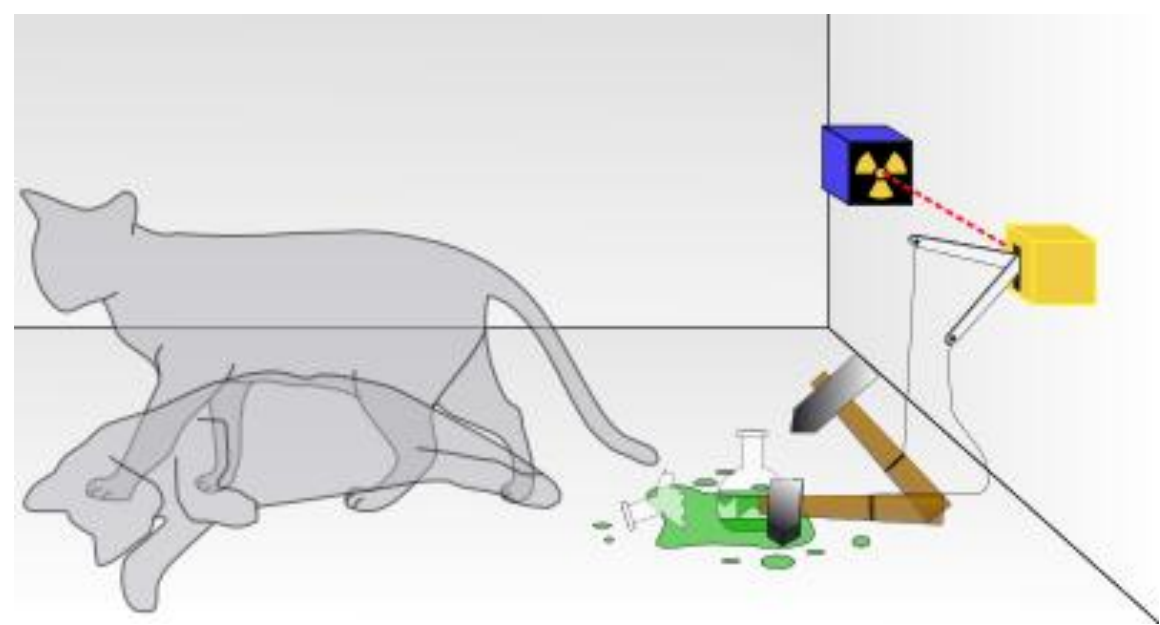

Fig. 2.7 : Illustration de l'expérience du chat de Schrödinger

Un autre problème de l'interprétation usuelle est que si l'on trouve des expériences en contradiction avec elle, on peut toujours introduire des opérateurs et champs afin que la théorie soit en accord avec l'expérience, sans requérir de changement fondamental dans l'interprétation physique. Les expériences ne peuvent pas vraiment la contredire, on a une sorte de «trap».

Enfin, évoquons des critiques faites à l'égard de théories développées à partir de la mécanique quantique : la théorie quantique des champs et l'électrodynamique quantique. Dans la théorie quantique, l'image d'un électron à extension spatiale est une particule élémentaire, sans structure interne.

L'équation de Dirac relativiste a conduit au développement de la théorie quantique des champs, unification de la mécanique quantique et de la relativité restreinte.

Une partie essentielle est l'électrodynamique quantique, qui régit les interactions électromagnétiques entre lumière et les particules chargées (comme l'électron). Elle a été développé par Richard Feynman et d'autres (Dyson, Schwinger,…), à partir de la mécanique quantique et de l'idée de particule ponctuelle (*Schweber, 1994*). Elle a été appelée par Feynman comme « perle de la physique » (lumière et matière) pour ses prédictions très précises dans la détermination théorique de quantités telles que le facteur de Landé $g$.

Feynman a également développé les diagrammes de Feynman, dont le but est représenter de façon plus intuitivement et imagée les interactions lumière – matière dans l'espace-temps (*Feynman, 1987*) :

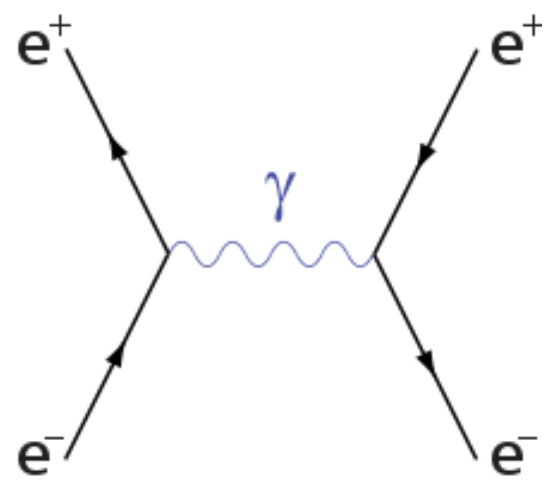

Fig. 2.8 : Exemple de diagramme de Feynman

La théorie quantique des champs indique que des particules / antiparticules se créent en permanence à partir des fluctuations électromagnétiques du vide. L'électron est ainsi entouré en permanence de ces particules /antiparticules virtuelles, on parle de polarisation du vide. (*Tong, 2009*)
De plus, un électron peut émettre un photon qui est ensuite réabsorbé par l'électron lui-même; on parle de self-énergie de l'électron.

L'évaluation des processus de la polarisation du vide et de la self-énergie de l'électron, par les calculs électrodynamiques perturbatifs, conduit à des intégrales divergentes.
Toutes les théories des champs connues dans les années 1960 ont comme propriété que les interactions deviennent infinies à des échelles de distance suffisamment petites. Là constitue une limite importante à ces théories des champs basées sur la mécanique quantique traditionnelle et sur l'idée de particule ponctuelle.
Ces infinis des calculs ont cependant pu être éradiqués par une procédure appelée renormalisation, mais est considérée par certains physiciens comme une véritable « escroquerie » (aux dires même de Feynman). En effet, cette théorie implique de négliger des infinis qui apparaissent dans les équations, ce qui ne peut avoir de sens mathématiquement. Comme l'indique Dirac lui-même, initiateur de l'électrodynamique quantique : « La renormalisation est juste une procédure d'interruption. Il devrait y avoir un changement fondamental dans nos idées ».

Alors certes, cette théorie est totalement en accord avec l'expérience, cela prouve-t-il réellement que la théorie soit correcte ?

La mécanique quantique a triomphé de bien des difficultés. Cependant, selon Franck Laloë, les responsables de ce succès ont parfois été trop loin, tant dans leur désir de convaincre que dans leurs convictions propres, affirmant que leur point de vue était le seul compatible avec l'expérience. Selon eux, aucune description plus fine ne serait jamais possible; en particulier, il était prouvé que la Nature était fondamentalement indéterministe. Bohr ne semble pas avoir réalisé l'impact de sa position, et que la frontière qu'il voulait établir entre macroscopique et microscopique n'est pas facile à maintenir. Les arguments avancés ne sont pas solides, principalement car ils imposent aux théories visant à compléter la mécanique quantique de trop lui ressembler.

Selon mathématicien français René Thom, « la mécanique quantique est incontestablement le scandale intellectuel du siècle. La science a renoncé à l'intelligibilité du monde. », il est persuadé qu' « il y a une dynamique sous-jacente à la mécanique quantique » (*Thom, 1991*).

Selon De Broglie, la science a toujours cherché à établir des liens de causalité entre les phénomènes, et cette causalité n'a plus lieu d'être dans l'interprétation de Copenhague. Il faut selon lui revenir à une théorie qui n'est pas fondamentalement probabiliste, mais où les probabilités émergent de phénomènes de causalité. Dès lors, les probabilités refléteraient notre ignorance partielle de l'état du monde physique.

De Broglie questionnait également le rôle prépondérant joué par les mathématiques dans la physique quantique et la physique moderne en général (« Shut up and calculate » prône le physicien Feynman), et le danger associé. Faire des développements mathématiques sans questionner la physique, sans esprit critique, peut souvent mener à des impasses. Seules l'intuition et l'imagination permettent de briser le cercle dans lequel s'enferme naturellement toute pensée purement déductive.

### 3) Interprétation alternative de De Broglie-Bohm

*David Bohm, A suggested interpretation of the quantum theory in terms of hidden variables, Phys. Rev. 85:166–79, 1952*

*Michel Gondran & Alexandre Gondran, Mécanique quantique : Et si Einstein et de Broglie avaient raison ?, Editions Matériologiques, 2014*

La théorie de l'onde pilote fut reprise par David Bohm, qui proposa une version améliorée de cette théorie, plus complète et plus simple (*Bohm, 1952*).

Bohm considère que l'onde et la particule coexistent et interagissent. Il ne considère cependant pas le champ $v$ de De Broglie, mais les entités suivantes :

- La fonction d'onde $\psi$, champ réel et classique, pas directement observable, contrôlable expérimentalement. C'est une représentation mathématique d'un champ de force réel. Elle évolue dans un espace à $3N$ dimensions, avec $N$ le nombre de particules.
- La position de la particule, observable, mais pas contrôlée expérimentalement. Elle est décrite dans un espace à 3 dimensions.

La position est la variable «cachée» (ou supplémentaire) de la théorie de De Broglie-Bohm. Cependant, la position est mesurée par l'expérience, il n'est pas logique de la qualifier de «variables cachée» : dans une chambre à brouillard (détecteur de particules créé par Wilson en 1911), on peut voir les trajectoires des particules, la position a bien une réalité concrète. En revanche, l'onde $\psi$ n'est jamais mesurée directement mais reconstruite statistiquement. La position semble donc être une variable moins « cachée » que la fonction d'onde $\psi$.

Nous détaillons ici le raisonnement suivant par Bohm dans son article de 1952.

Il introduit dans son article une transformation qui consiste en une formulation alternative de l'équation de Schrödinger en termes de variables hydrodynamiques

(*Madelung, 1926*) :

$$i\hbar\frac{\partial\psi}{\partial t} = -\frac{\hbar^2}{2m}\Delta\psi + V(x)\psi\ ;$$

posons $\psi = R.exp\left(\frac{iS}{\hbar}\right),\ (R,S) \in \mathbb{R}^2 \rightarrow \begin{cases} \frac{\partial R}{\partial t} = -\frac{1}{2m}(R.\Delta S + 2\nabla R.\nabla S) \\ \frac{\partial S}{\partial t} = -\frac{(\nabla S)^2}{2m} - V(x) + \frac{\hbar^2}{2m}\frac{\Delta R}{R} \end{cases};$

avec $P(x) = R^2(x),\ \begin{cases} \frac{\partial P}{\partial t} + \nabla\left(P\frac{\nabla S}{m}\right) = 0 \\ \frac{\partial S}{\partial t} + \frac{(\nabla S)^2}{2m} + V(x) - \frac{\hbar^2}{2m}\left(\frac{\Delta P}{P} - \frac{1}{2}\frac{(\nabla P)^2}{P^2}\right) = 0 \end{cases}$ Équation de continuité / Équation de Jacobi généralisée

Équations de Madelung (2.28)

Dans la limite $h \longrightarrow 0$, on retrouve l'équation de Hamilton-Jacobi classique. $P(x)$ s'interprète comme une probabilité : $P(x) = R^2(x) = |\psi|^2$

L'onde pilote gouverne le mouvement de la particule en suivant l'équation de Schrödinger.

La formule de guidage, dans la théorie de Bohm, est donnée par la formule suivante :

$v(x) = \frac{\nabla S(x)}{m} = \frac{\hbar}{m}Im\left(\frac{\psi^*\nabla\psi}{\psi^*\psi}\right)$, avec $v(x)$ la vitesse quantique (2.29).

Ce qui conduit à l'équation de continuité : $\frac{\partial P}{\partial t} + \nabla(P.v) = 0.$

L'influence de l'onde pilote se caractérise sous la forme d'un potentiel quantique, dérivé de la fonction d'onde, agissant sur la particule de façon analogue à un champ électrique.

$$Q(x) = -\frac{\hbar^2}{2m}\left(\frac{\Delta P}{P} - \frac{1}{2}\frac{(\nabla P)^2}{P}\right) = -\frac{\hbar^2}{2m}\frac{\Delta R}{R} = -\frac{\hbar^2}{2m}\frac{\Delta|\psi(\vec{r},t)|}{|\psi(\vec{r},t)|} \quad (2.30).$$

Les formules précédentes peuvent se généraliser dans le cas où l'on considère un système à $n$ particules. Le système est décrit par une onde de Schrödinger à 3n dimensions et une trajectoire à $3n$ dimensions. On a les équations suivantes :

$$i\hbar\frac{\partial\psi}{\partial t} = -\frac{\hbar^2}{2m}(\Delta_1\psi + \cdots + \Delta_n\psi) + V(\vec{x}_1, \dots, \vec{x}_n)\psi$$

$$\psi = R(\vec{x}_1, \dots, \vec{x}_n)exp\left(\frac{i}{\hbar}S(\vec{x}_1, \dots, \vec{x}_n)\right)$$

(2.31)

$$\vec{v}_i = \frac{\vec{\nabla}_i S(\vec{x}_1, \dots, \vec{x}_n)}{m}$$

$$R^2 = P(\vec{x}_1, \dots, \vec{x}_n) \to \frac{\partial P}{\partial t} + \frac{1}{m}(\nabla_1 . P\nabla_1 S + \cdots + \nabla_n . P\nabla_n S) = 0$$

$$\frac{\partial S}{\partial t} + \frac{(\nabla_1 S)^2 + \cdots + (\nabla_n S)^2}{m} + V(\vec{x}_1, \dots, \vec{x}_n) - \frac{\hbar^2}{2mR}(\Delta_1 R + \cdots + \Delta_n R) = 0$$

$$Q(\vec{x}_1, \dots, \vec{x}_n) = -\frac{\hbar^2}{2mR}(\Delta_1 R + \cdots + \Delta_n R).$$

Pour les systèmes composés de plusieurs particules, le potentiel quantique Q introduit un phénomène de non-localité dans l'équation du mouvement.

En effet, $Q$ dépend de l'amplitude $R$ de l'onde; lorsque la distance entre les particules devient grande, $R \to 0$, mais $Q$ ne s'annule pas.

Dans cette vision, chaque particule est à l'origine d'un potentiel / champ non-local. Dans un système compose de plusieurs particules, la modification de l'état d'une particule a un effet sur le reste du système instantanément. C'est l'intrication, que nous détaillerons après.

A très courtes distances (au delà de $10^{-13}$ cm), la fonction d'onde $\psi$ ne satisfait plus l'équation de Schrödinger et $v(x) \neq \frac{\nabla S(x)}{m}$. Donc $|\psi|^2$ ne satisfait plus l'équation de conservation, et ne représente plus la densité de probabilité des particules.

Trois affirmations consistantes sont nécessaires pour obtenir les mêmes résultats que l'interprétation usuelle de Copenhague :

- La fonction d'onde $\psi$ satisfait l'équation de Schrödinger;
- La vitesse quantique est donnée par : $v(x) = \frac{\nabla S(x)}{m}$;
- La densité de probabilité s'exprime comme suit : $P(x) = |\psi|^2$.

La formulation habituelle de Copenhague mène à des difficultés insolubles à des distances de l'ordre de $10^{-13}$ cm ou moins. L'interprétation proposée ici par Bohm est peut-être une solution pour résoudre ces difficultés.

Dans cette vision, les résultats des mesures dépendent des positions initiales aléatoires du système dans l'espace des configurations. L'onde guide la particule d'une façon qui force sa position à reproduire les caractéristiques d'une figure d'interférence. Les trajectoires bohmiennes peuvent être courbées, comme par exemple dans une expérience d'interférences. La théorie a un caractère non-locale. Selon Bell, « c'est un des mérites de De Broglie-Bohm de mettre en avant la non-localité ».

La théorie de Bohm possède de nombreuses ressemblances avec l'onde pilote de De Broglie, elle est appelée de nos jours théorie de De Broglie-Bohm.

La théorie de De Broglie-Bohm mène aux mêmes résultats que l'interprétation habituelle (*Dürr et al., 1993*). Elle permet une description précise et continue des processus, même au niveau quantique.

Nous résumons ici les postulats de la théorie de De Broglie-Bohm :

- La fonction d'onde $\psi$ est considérée comme un champ réel et objectif, et non comme une entité purement mathématique
- Il existe des particules qui ont des coordonnées bien définies et qui évoluent de façon déterministe.
- La vitesse des particules est $v = \frac{\nabla S}{m}$, et la fonction d'onde est $\psi = Re^{iS/\hbar}$.
- La particule réagit au potentiel classique $V(x)$ et également à un potentiel quantique additionnel : $Q = -\frac{\hbar^2}{2m}\frac{\Delta R}{R}$.

Dans la deuxième partie de son article, Bohm cherche à montrer que sa théorie mène aux mêmes prédictions que celles de l'interprétation usuelle.

Le principe d'incertitude est vu comme une limite pratique à la précision à laquelle position et quantité de mouvement peuvent être simultanément mesurées, émanant de perturbations du système. En effet, le potentiel quantique subit de violentes et rapides fluctuations, qui tendent à rendre la trajectoire de la particule aléatoire. Même si les variables initiales de la particule sont bien définies, nous devrions en pratique perdre toute possibilité de suivre le mouvement de la particule et serions obligés d'avoir recours à une sorte de théorie statistique. Les forces chaotiques compliquées amènent à un ensemble avec une densité de probabilité $|\psi|^2$. Ce phénomène est semblable au problème de la cinétique des gaz en thermodynamique, où l'ensemble statistique repose sur un chaos moléculaire. Ainsi les fluctuations quantiques et classiques (comme le mouvement brownien) auraient la même origine.

Bohm s'intéresse également au théorème de Von Neumann indiqué précédemment, selon lequel les probabilités quantiques ne peuvent être comprises en termes de distribution sur des paramètres cachés. Selon Bohm, les conclusions de ce théorème sont critiquables, car la démonstration se restreint à une petite classe de paramètres cachés et exclut les paramètres proposés par sa théorie (position notamment).

En 1954, Bohm publie avec le physicien français Jean-Pierre Vigier une version stochastique où est introduit un 5$^e$ axiome : le champ $\psi$ est dans un état de fluctuations aléatoires et chaotiques telle que la valeur $\psi$ constitue une moyenne de ces fluctuations; les fluctuations proviennent d'un niveau sous-jacent, de la même manière que les fluctuations du mouvement brownien proviennent d'un niveau atomique plus profond. (*Bohm & Vigier, 1954*). En 1952, Bohm démontre que pour certains cas, l'action de ces forces chaotiques font tendre le système vers la distribution de Born, $|\psi|^2$, sans parvenir cependant à généraliser.

Vigier, assistant de De Broglie, perçoit dans la théorie de la double solution un moyen de rapprocher la mécanique ondulatoire et la théorie de la relativité générale. L'ambition de De Broglie était alors de renouveler pour la physique quantique et la relativité générale ce qu'il avait entrepris jadis pour l'optique et la mécanique. Mais il n'y parvint pas…

L'enjeu et l'intérêt de la théorie de De Broglie-Bohm est de faire reposer la physique microscopique sur de meilleures bases, qui expliquent tout aussi bien les prédictions quantiques que la théorie usuelle. Cette approche est analogue à l'hypothèse atomique postulée afin d'expliquer des résultats macrophysiques, qui pouvaient cependant être expliqués directement en termes de concepts macroscopiques.

Bohm s'oppose profondément à Bohr : la réalité n'est pas uniquement ce que l'on peut observer. Le monde est objectivement réel et peut être regardé comme ayant une structure analysable d'une immense complexité.

L'interprétation de De Broglie-Bohm est cependant ignorée par la communauté des physiciens de l'époque, comme l'indique John Bell :

« Mais alors pourquoi Born ne m'avait pas parlé de cette « onde-pilote » ? Ne serait-ce que pour signaler ce qui n'allait pas avec elle ? Pourquoi von Neumann ne l'a pas envisagée ? Plus extraordinaire encore, pourquoi des gens ont-ils continué à produire des preuves d'impossibilité, après 1952, et aussi récemment qu'en 1978 ? Alors que même Pauli, Rosenfeld, et Heisenberg, ne pouvaient guère produire de critique plus dévastatrice de la théorie de Bohm que de la dénoncer comme étant « métaphysique » et « idéologique » ? Pourquoi l'image de l'onde-pilote est-elle ignorée dans les cours ? Ne devrait-elle pas être enseignée, non pas comme l'unique solution, mais comme un antidote à l'auto-satisfaction dominante ? Pour montrer que le flou, la subjectivité, et l'indéterminisme, ne nous sont pas imposés de force par les faits expérimentaux, mais proviennent d'un choix théorique délibéré ? » (*Bell, 1987*).

Nous allons étudier maintenant la théorie de De Broglie-Bohm à travers trois expériences «cruciales» de la mécanique quantique : les fentes de Young, l'expérience de Stern et Gerlach, et l'expérience EPR.

Nous ferons cette étude à l'aide du livre de *M.&A. Gondran*, *Mécanique quantique, Et si Einstein et De Broglie avaient aussi raison ? (2014)*.

## Les fentes de Young

L'expérience des fentes de Yong a pendant longtemps été l'expérience cruciale dans l'interprétation de la dualité onde-corpuscule et de la mécanique quantique.

Selon Feynman, « [Elle aborde] le point fondamental du comportement mystérieux [des objets quantiques] sous son aspect le plus étrange. C'est un phénomène qu'il est impossible à expliquer de façon classique et qui contient le cœur de la mécanique quantique. En réalité, il en contient même l'unique mystère » (*Feynman, 1964*).

L'expérience de la double fente a été réalisée pour la lumière (expérience de Young), puis avec des électrons (Jönsson). Dans l'expérience de Jönsson de 1961, un canon à électrons émet un à un dans le plan horizontal, à travers un trou de quelques millimètres, des électrons à une vitesse de $1,8*10^8$ m/s. Une plaque percée de 2 fentes horizontales larges de 0,2 µm est installée à 35 cm de la source d'électrons. L'impact de chaque électron apparaît sur l'écran au fur et à mesure que l'expérience se déroule. Au bout de quelques milliers d'impacts, la répartition des impacts des électrons fait apparaître une figure d'interférence.

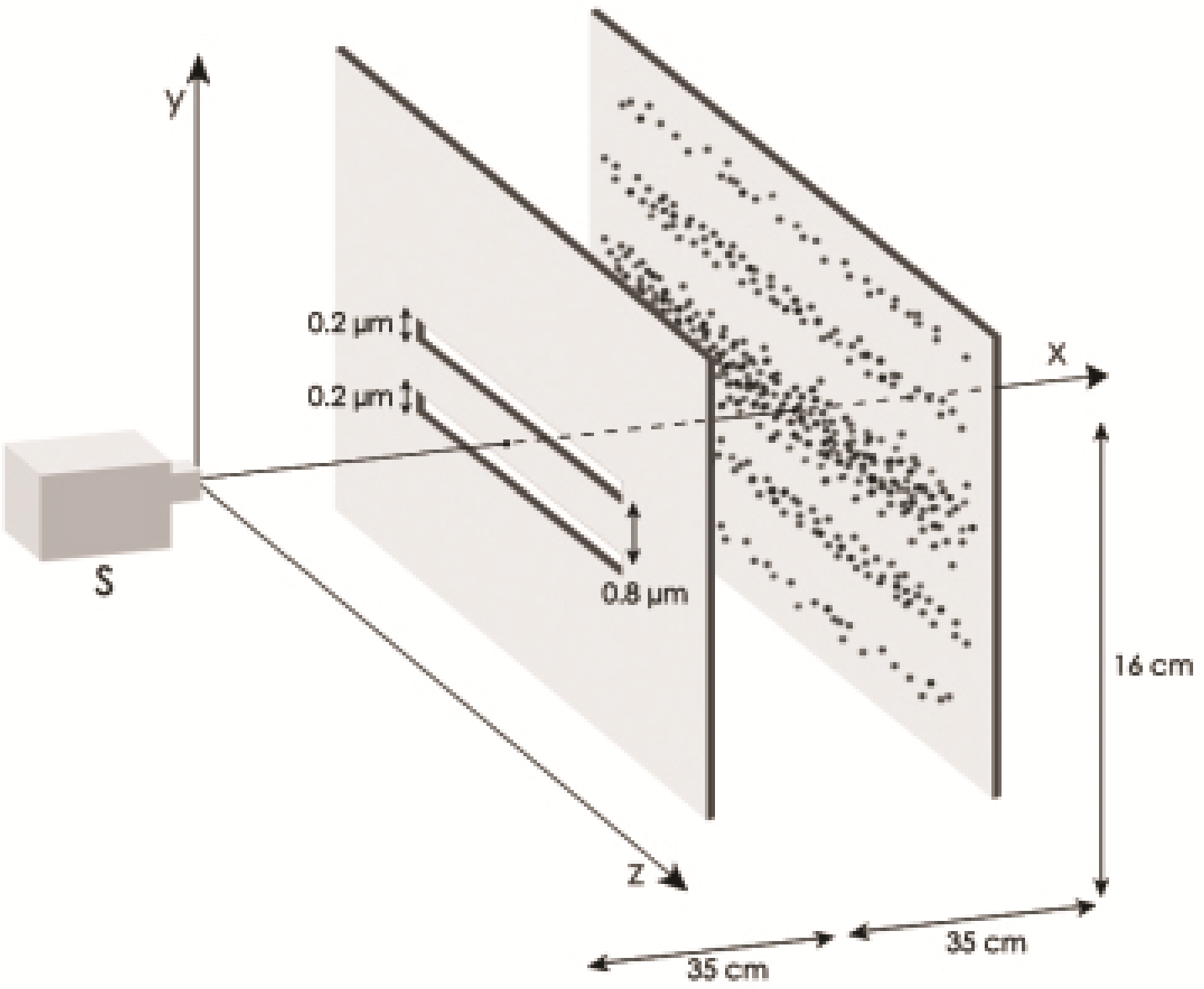

Fig. 2.9 : Expérience de Jönsson

L'électron se comporte comme un corpuscule à son arrivée sur l'écran, de par son impact. Le caractère ondulatoire d'un électron se révèle statistiquement :

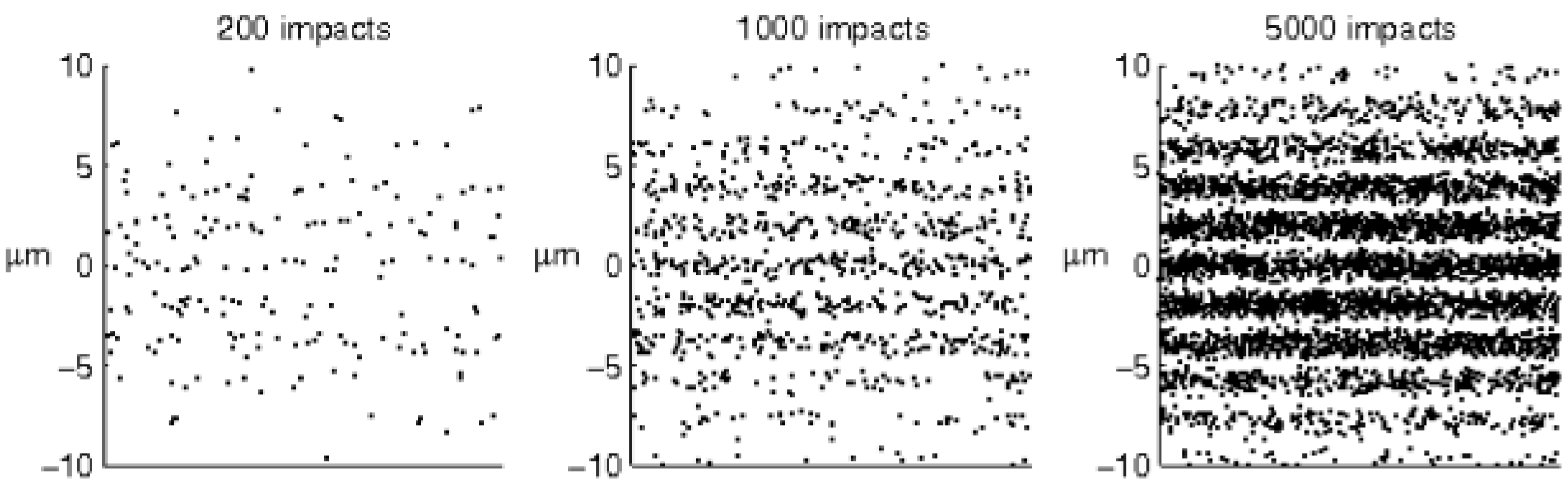

Fig. 2.10 : Apparition d'une figure d'interférence après l'envoi d'un grand nombre d'électrons

Dans l'interprétation de Copenhague, l'électron est soit une onde soit un corpuscule (Ou exclusif). Durant l'expérience, l'électron est une onde, et est un corpuscule seulement lors de son impact sur l'écran. L'électron-onde passe par les 2 fentes à la fois.

Dans l'interprétation de De Broglie-Bohm, l'électron est à la fois onde et corpuscule, l'onde guidant le corpuscule. Ici, l'électron-onde passe ici également par les deux fentes à la fois, mais l'électron-corpuscule ne passe que par l'une des fentes. Les particules ont une position initiale et suivent une trajectoire dont la vitesse est donnée par $v(x,t) = \frac{\nabla S(x,t)}{m}$.

M.&A. Gondran ont réalisé des simulations des trajectoires bohmiennes des électrons, et ont obtenu des résultats en accord avec l'expérience, dans les cas de la diffraction et des fentes de Young. Dans le cas des fentes de Young, le principe est de tirer au hasard les positions initiales des électrons dans le paquet d'ondes initial. Sont représentées ci-dessous 100 trajectoires quantiques d'électrons traversant une des deux fentes. Ne sont pas représentées les trajectoires des électrons arrêtés par le 1er écran.

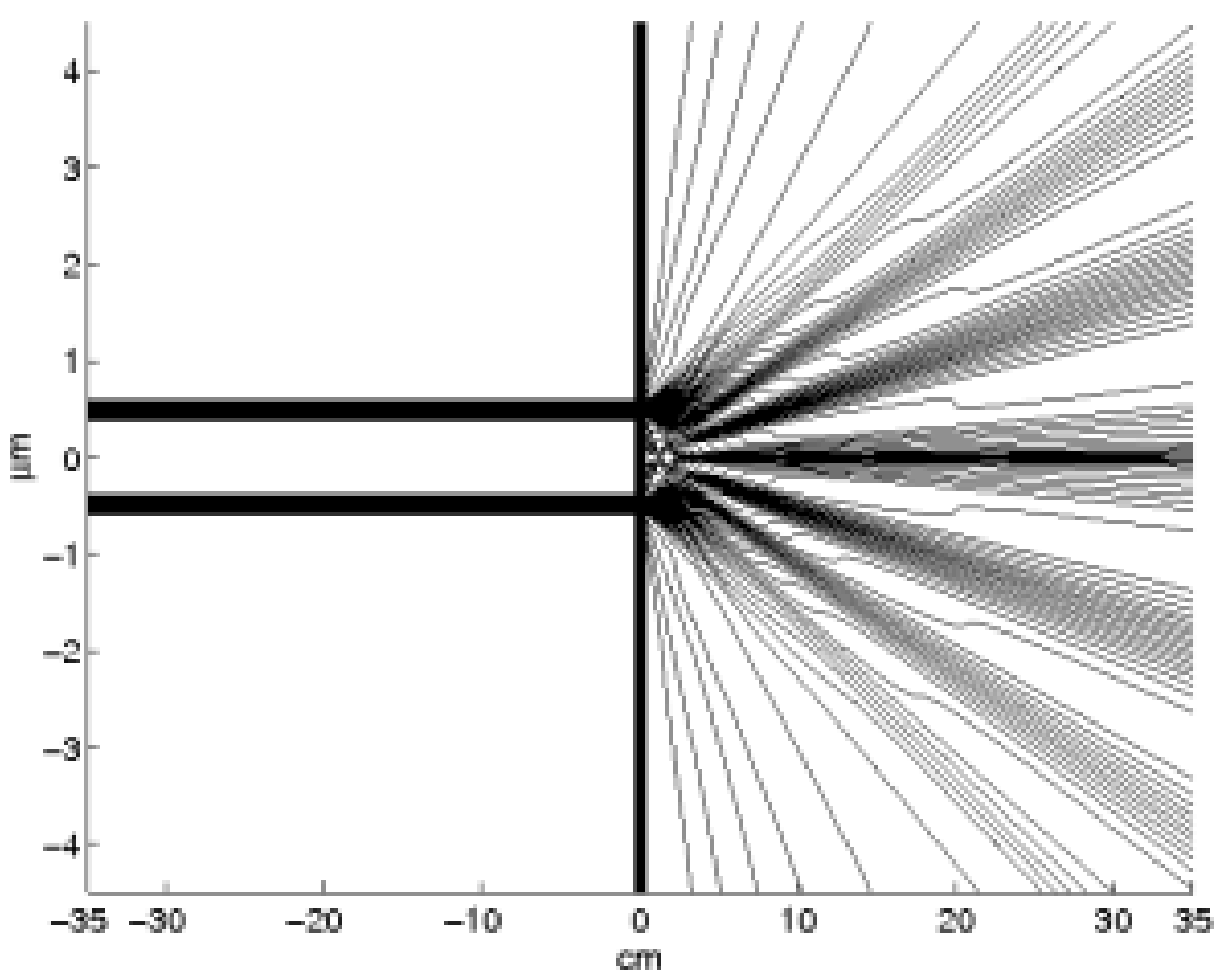

Fig. 2.11 : Trajectoires bohmiennes des électrons - expérience des fentes de Young

Les simulations réalisées par M.&A. Gondran montrent également que lorsque l'on fait tendre $h$ vers $0$, les franges d'interférences disparaissent; on est dans le cas classique.

Le phénomène d'interférences a été mis en évidence avec des objets quantiques mésoscopiques individuels comme les fullerènes ($C_{60}$, $C_{70}$), par des chercheurs de l'université de Vienne en 1999.

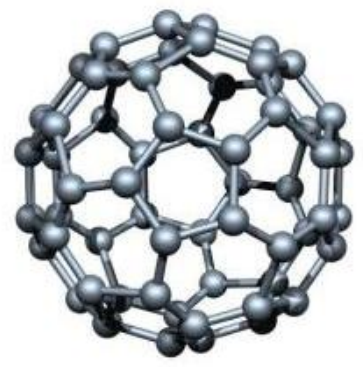

Fig. 2.12 : Molécule de fullerène $C_{60}$

Il est difficile de refuser une réalité physique à la molécule de fullerène, plus précisément de lui refuser une position dans l'espace et une trajectoire passant par une seule fente. La découverte de la dualité onde-corpuscule et du phénomène d'interférences pour ces macromolécules est donc un bon argument en faveur de la théorie de De Broglie-Bohm, et en défaveur de l'interprétation de Copenhague.

## Le spin

La notion de spin a été introduite à la suite de l'effet Zeeman anomal et de l'expérience de Stern et Gerlach (1922).

Elle consiste à faire passer des atomes d'argent dans un champ magnétique non homogène de direction verticale. Les atomes d'argent dans leur état fondamental ayant un moment cinétique nul, le faisceau ne devrait classiquement pas subir l'influence du champ magnétique. Cependant, l'expérience montre que le faisceau se sépare en deux :

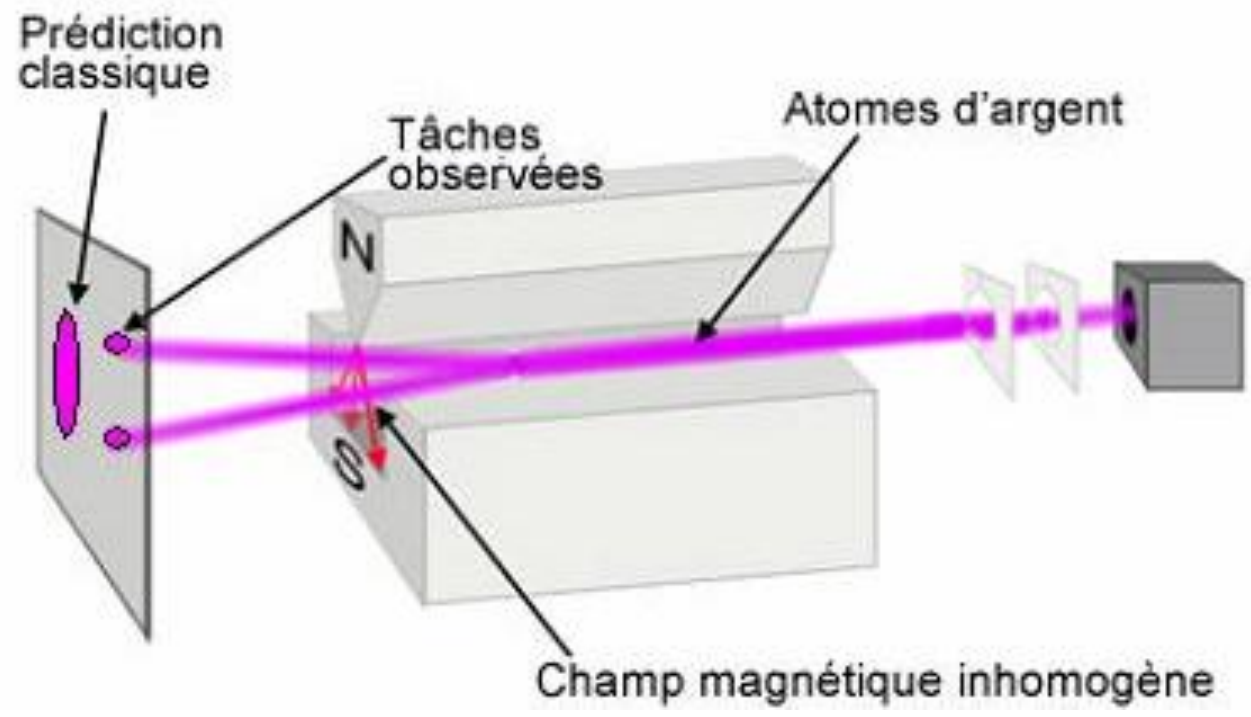

Fig. 2.13 : Expérience de Stern et Gerlach

On explique ce phénomène en introduisant une observable de nature essentiellement quantique : le moment cinétique de spin, ou plus simplement spin. La séparation en deux faisceaux révèle qu'il existe deux états possibles pour le spin de l'atome : $+1/2$ et $-1/2$.

On peut théoriser ce phénomène à l'aide de la mécanique quantique standard.

Exprimons le bispineur de Dirac en prenant en compte la dimension spatiale. La résolution en espace et en temps de l'équation de Pauli permet de décrire complètement les résultats statistiques de l'expérience de Stern et Gerlach. A l'arrivée dans l'électro-aimant, à l'instant $t = 0$, chaque atome préparé peut être décrit par un spineur gaussien :

$$\psi^0(x,z) = A.exp\left(-\frac{x^2+z^2}{4{\sigma_0}^2}\right)\begin{pmatrix} \cos\left(\frac{\theta_0}{2}\right)e^{i\varphi_0/2} \\ \text{isin}\left(\frac{\theta_0}{2}\right)e^{-i\varphi_0/2} \end{pmatrix} \quad (2.32).$$

Écrivons l'équation de Pauli, qui correspond à l'équation de Schrödinger pour les particules de spin 1/2 dans un champ électromagnétique :

$$i\hbar\frac{\partial\psi}{\partial t} = -\frac{\hbar^2}{2m}\Delta\psi + \mu_B\vec{B}.\vec{\sigma}\psi \quad (2.33)$$

avec $\psi = \begin{pmatrix}\psi_+\\ \psi_-\end{pmatrix}, \vec{B} = (B_x, B_y, B_z), \vec{\sigma} = (\sigma_x, \sigma_y, \sigma_z)$.

Exprimons le bispineur : $\psi(x,z,t) = \begin{pmatrix} R_+ exp\left(i\frac{S_+}{\hbar}\right) \\ R_- exp\left(i\frac{S_-}{\hbar}\right)\end{pmatrix}$ (2.34).

Après calcul, on peut obtenir la densité de présence en $z$ des atomes d'argent à l'instant $t$ :

$$\rho(z, t+\Delta t) = A.\left\{exp\left(-\frac{(z-z_\Delta-ut)^2}{2{\sigma_0}^2}\right) + exp\left(-\frac{(z+z_\Delta+ut)^2}{2{\sigma_0}^2}\right)\right\} \quad (2.35).$$

Où $\Delta t$ est le temps après le passage par le champ magnétique et $z_\Delta$ la distance en $z$ au champ magnétique.

Après calcul numérique, on obtient que la séparation se fait à partir de 10 cm après la sortie du champ magnétique :

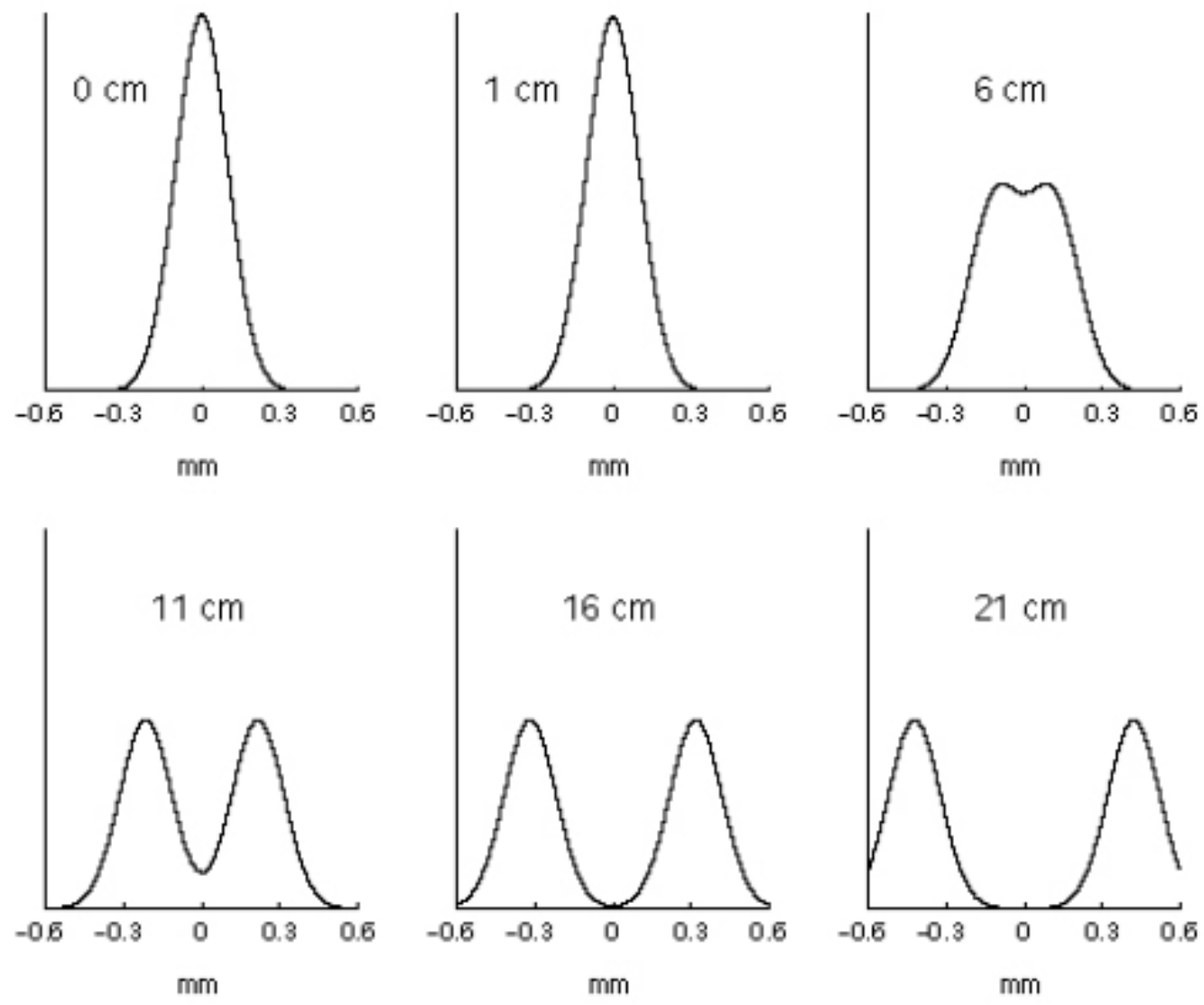

Fig. 2.14 : Évolution de la probabilité de présence des atomes d'argent (*Gondran*)

Pour expliquer les impacts, *M.&A. Gondran* ont simulé les trajectoires des atomes d'argent dans l'interprétation de De Broglie-Bohm. Les centres de masse des atomes issus du jet atomique ont des positions qui sont distribuées suivant la densité donnée par la fonction d'onde et suivent des trajectoires compatibles avec l'équation de continuité (hypothèse de De Broglie-Bohm). La particule quantique possède une position locale comme une particule classique, mais possède également un comportement non-local dû à la fonction d'onde.

Cette approche, qui utilise uniquement la résolution de l'équation de Pauli, donne les mêmes résultats statistiques que l'interprétation de Copenhague, mais elle détermine également les impacts individuels.

On peut donner une interprétation intuitive de la séparation du faisceau en deux dans l'expérience de Stern et Gerlach, comme illustré ci-dessous :

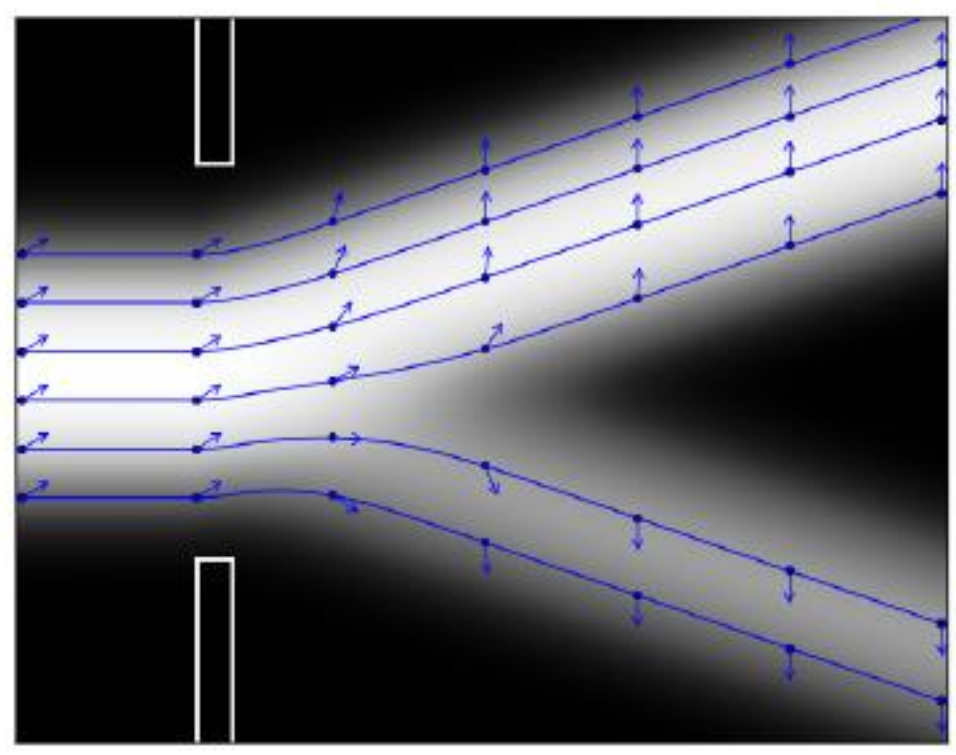

Fig. 2.15 : Trajectoires d'atomes d'argent; les flèches bleues représentent l'orientation du spin.

projection Selon cette vision, l'expérience de Stern et Gerlach n'est pas la mesure de la projection du spin selon l'axe $O_z$, mais le redressement de l'orientation du spin soit dans la direction du gradient du champ magnétique, soit dans la direction opposée. La valeur mesurée (spin) n'est pas une valeur préexistante comme la masse et la charge de la particule, mais une valeur contextuelle.

Le spin n'a pas forcément une direction constante le long d'une trajectoire. La direction du spin change au fur et à mesure que la trajectoire pénètre dans le gradient de champ magnétique, afin de devenir parallèle (spin $+$) ou antiparallèle (spin $-$) à l'axe $O_x$.

*M.&A. Gondran* font un parallèle entre l'expérience de Stern et Gerlach et le lancer d'une pièce de monnaie. L'expérience de la pièce de monnaie consiste à jouer pile ou face dans l'espace (sans gravité, donc sans direction privilégiée), et l'on mesure pile ou face au claquement des mains. L'orientation de la pièce est une variable à valeurs non prédéterminées (avant la mesure). On parle aussi de mesures contextuelles, c'est à dire dont le résultat dépend du contexte.

Il en est de même pour le spin en mécanique quantique, avec deux résultats possibles : spin up (déviation de la particule vers le haut par rapport à l'appareil de Stern et Gerlach) et spin down (déviation de la particule vers le bas).

Dans les deux cas, on peut dire que l'orientation de la pièce (resp. spin) avant la mesure est dans une superposition de l'état pile et face (resp. état up et down).

Résultats possibles pour une mesure (*source : Gondran*) :

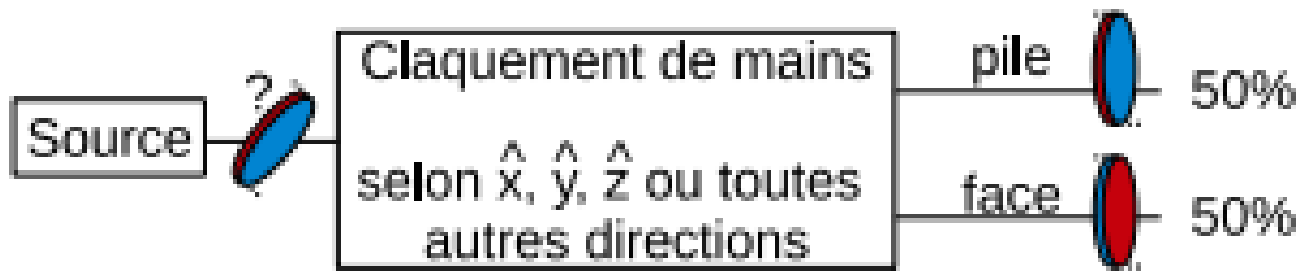

Résultats possibles pour deux mesures selon le même axe :

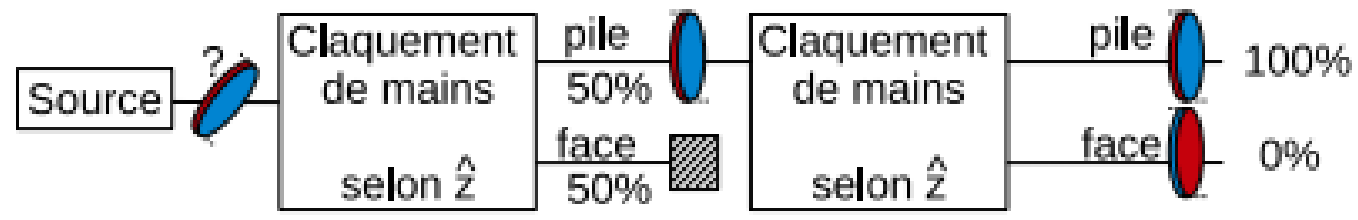

Résultats possibles pour deux mesures selon des axes différents :

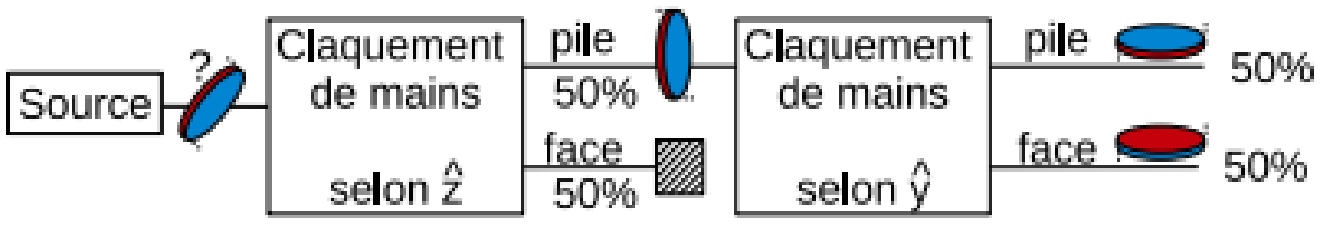

En résumé :

- si l'orientation des 2 appareils est inférieure à 90°, on a 100% de chance d'obtenir le même résultat;
- si elle est supérieure à 90°, 100% de chance d'obtenir le résultat opposé;
- si elle est égale à 90°, on a une chance sur 2 d'obtenir pile ou face.

Dans le cas quantique, l'appareil de mesure modifie l'état du système. La mesure est associée à un bombardement de photons qui modifie la trajectoire d'un atome ou d'un électron.

## Expérience EPR

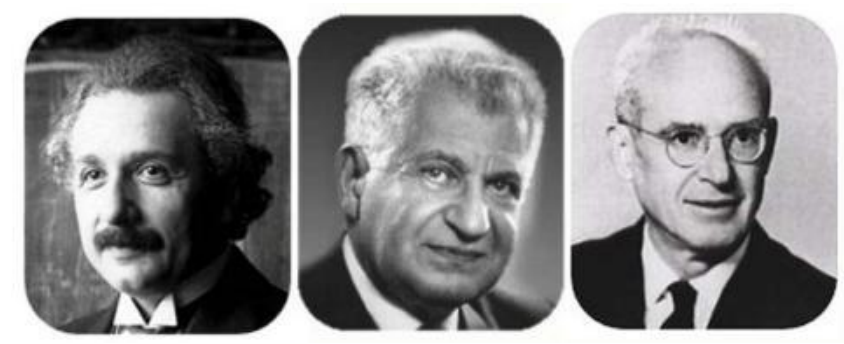

Fig. 2.16 : Einstein, Podolsky et Rosen

Einstein, Podolsky et Rosen proposèrent une expérience de pensée dont le but était de réfuter l'interprétation de Copenhague de la physique quantique. Cette expérience de pensée est appelée «paradoxe EPR» (*Einstein, Podolsky & Rosen, 1935*).

Cette expérience de pensée concerne l'intrication quantique de 2 particules, c'est à dire elle consiste à observer les paramètres ($q$, $p$, spin) d'un couple d'électrons éloignés l'un de l'autre, mis en corrélation au départ. Une source S crée une paire d'atomes identiques $A$ et $B$, intriqués par leur spin, avec des spins opposés. Les atomes $A$ et $B$ se séparent suivant l'axe $O_x$ dans des directions opposées ($A$ avec la vitesse $+v_x$ et $B$ avec la vitesse $-v_x$), et rencontrent 2 appareils identiques pour la mesure du spin.

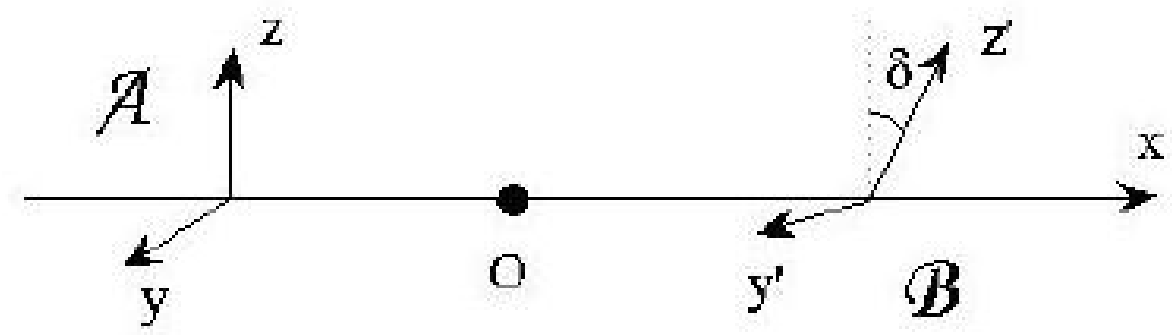

Fig. 2.17 : Schéma de l'expérience EPR-B

Selon la mécanique quantique, le second électron présente toujours un spin opposé au premier, les particules demeurent en corrélation / intrication. Et cela en dehors de toute relation causale, c'est-à-dire sans qu'il y ait eu d'interaction entre les deux.

En notant $+$ pour le spin up et – pour le spin down, on a donc les états intriqués superposés suivants :

$$| + - \rangle + | - + \rangle \quad (2.36).$$

Selon Einstein, Polosky et Rosen, il s'agit là d'une contradiction de la mécanique quantique en ce qu'elle viole les principes suivants :

- Causalité : impossibilité pour un signal de dépasser la vitesse de la lumière
- Localité : deux objets distants ne peuvent avoir une influence instantanée l'un sur l'autre
- Réalisme : il existe une réalité indépendante de l'observateur, des variables cachées, et la mécanique quantique est incomplète.

Selon eux, la corrélation consiste en une propriété commune que chacune des particules « emporte avec elle ». Ce raisonnement démontre alors le caractère incomplet de la mécanique quantique « orthodoxe », il doit exister des variables supplémentaires, « cachées ». C'est une approche déterministe.

En se fondant sur ces hypothèses, John Bell en a déduit les fameuses inégalités qui portent son nom. Nous détaillons ici son raisonnement.

Le raisonnement de Bell, comme prolongation du théorème EPR, est fondé sur les probabilités classiques résultant de fluctuations d'une cause inconnue. On peut aussi voir les choses différemment et considérer que le théorème de Bell part de l'existence de variables supplémentaires (ou «variables cachées») $\lambda$ dont on ne cherche pas à préciser l'origine, et qui vont influencer le résultat des mesures.

On considère une paire de particules de spin $1/2$ soumises à la mesure de la composante de leur spin, selon le vecteur $a$ pour la 1ère et $b$ pour la 2ème.

$\theta$ désigne l'angle entre les directions $a$ et $b$; $\lambda$ désigne les éléments de réalité EPR.

On note $+1$ le résultat de mesure pour une déviation vers le haut; $-1$ vers le bas.

Probabilités de résultats : $$P_{+,+} = P_{-,-} = sin^2\theta \quad (2.37)$$

$$P_{+,-} = P_{-,+} = cos^2\theta$$

On considère deux directions d'analyse pour chaque mesure : $a$ et $a'$ pour la 1ère; $b$ et $b'$ pour la 2ème.

Les résultats de mesure dépendent de $\lambda$ et des conditions de mesure, on pose :

$$A(a,\lambda) \equiv A \ ; \ A(a',\lambda) \equiv A' \quad (2.38)$$

$$B(b,\lambda) \equiv B \ ; \ B(b',\lambda) \equiv B'.$$

On pose également : $M = AB + AB' - A'B' + A'B' = (A - A').B + (A + A').B'$.

Cela conduit à l'inégalité suivante, qui est la forme BCHSH du théorème de Bell :

$$-2 \leq < M > \leq 2 \quad (2.39).$$

Pour certains choix de directions, l'inégalité précédente est violée. Ainsi, les raisonnements EPR-Bell conduisent à une contradiction avec les prédictions de la mécanique quantique. Cette dernière n'est pas une théorie réaliste et locale au sens de EPR et Bell.

Cependant, les résultats des expériences confirment les prédictions de la mécanique quantique et la violation des inégalités de Bell. Dans les années 1980, le chercheur français Alain Aspect a réalisé l'expérience EPR (*Aspect et al., 1982*). Le second électron présentait toujours un spin opposé au premier, les particules demeuraient en corrélation. Et cela en dehors de toute relation causale, c'est-à-dire sans qu'il y ait eu d'interaction entre les deux.

Cette expérience d'Aspect a montré les inégalités de Bell étaient violées, invalidant ainsi la vision d'Einstein. La corrélation quantique est bien une réalité.

Cependant, Bell lui-même indiquait qu'il ne fallait pas tirer trop de conclusions hâtives de ses inégalités. En effet, la violation de ses inégalités montre juste qu'une des hypothèses énoncées par Einstein, Podolski et Rosen ne tient pas dans le monde microscopique. Selon Bell, la violation de ses inégalités vient de la non-localité des phénomènes à l'échelle microscopique.

La théorie de De Broglie-Bohm, mise en avant précédemment, est non-locale. Elle fait intervenir un potentiel quantique (différent des potentiels classiques), qui ne dépend pas de la force ou de la « grandeur » du potentiel mais uniquement de sa forme. C'est pour cette raison que des objets lointains peuvent exercer une influence forte sur le mouvement d'autres objets.

Dans cette approche, si l'on mesure la position de la première particule, on introduit alors des fluctuations dans l'impulsion de chaque particule. De même, si l'on mesure l'impulsion de la première particule, les fluctuations dans la fonction d'onde du système mène, à travers les forces quantiques, à des changements dans la position de chaque particule. Les forces quantiques-mécaniques peuvent transmettre des perturbations instantanément d'une particule à une autre à travers le milieu du champ $\psi$. Ce n'est pas inconsistant avec la relativité car aucun signal n'est transmis de cette manière instantanément.

La non-localité entre deux particules se traduit par une impossibilité de décomposer la fonction d'onde du système en les fonctions d'onde individuelles :

$$\psi(\vec{r}_1, \vec{r}_2; t) \neq \varphi(\vec{r}_1, t) . \chi(\vec{r}_2, t) \quad (2.40).$$

On peut donner une interprétation intuitive de l'intrication, issue de M.&A. Gondran :

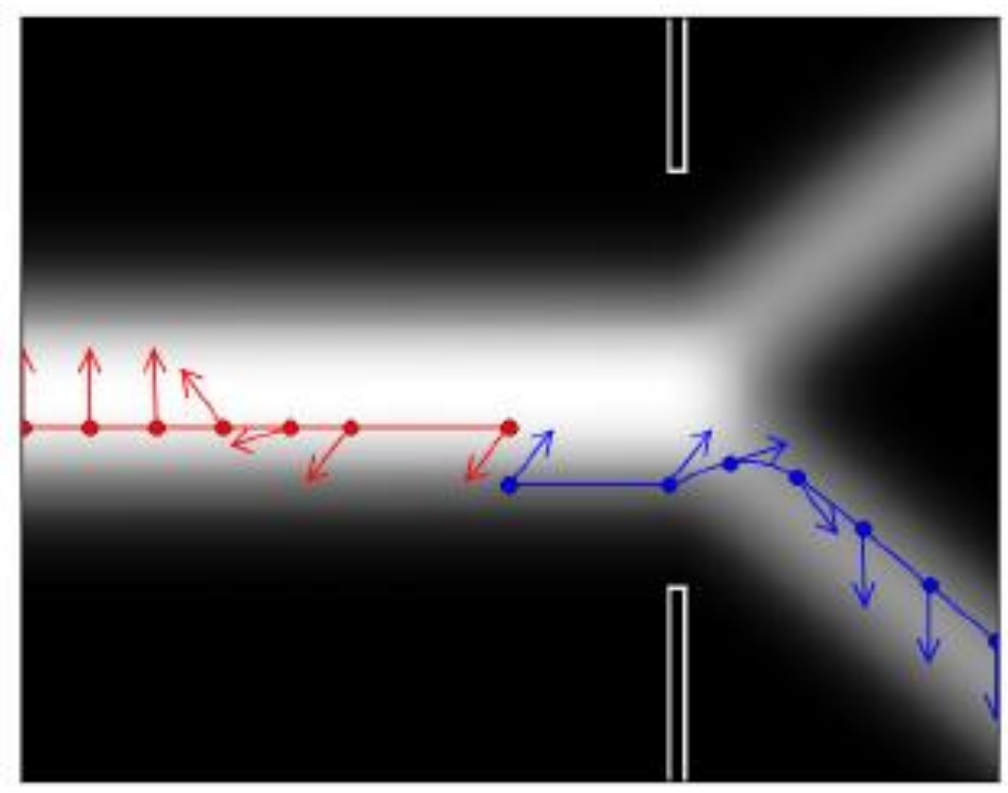

Fig. 2.18 : Illustration de l'intrication des spins (*Gondran*)

Les deux atomes intriqués A (bleu) et B (rouge) sont créés avec des spins opposés.

Puis, la particule A traverse l'aimant; pendant la mesure de A, le spin de B se redresse pour rester en opposition avec celui de A

## Développement ultérieur de la théorie de De Broglie-Bohm

La théorie de De Broglie-Bohm est un sujet toujours très actif de nos jours. Un problème fréquemment entendu à propos de cette théorie est qu'elle n'est pas relativiste. En effet, la formule de guidage a la vitesse de chaque particule définie en termes de la fonction d'onde et de ses dérivées spatiales évaluées au point de configuration du système entier à N-particules à cet instant. La vitesse d'une particule dépend donc des positions instantanées des autres particules.

Cependant, des travaux récents ont montré que cette théorie peut être compatible avec l'invariance de Lorentz (*Dürr et al., 2004*), et même peut être étendue à la théorie quantique des champs (*Dürr et al., 2014*).

## 4) Perspectives modernes de l'approche ondulatoire - analogies hydrodynamiques

*Yves Couder & Emmanuel Fort, Single-particle diffraction and interference at a macroscopic scale, Physical Review letters, 2006*

*John W.M. Bush, Pilot-Wave Hydrodynamics, Annu. Rev. Fluid Mech. 47:269-92, 2015*

### Expérience des gouttelettes

Tout d'abord, nous allons étudier une expérience menée en premier lieu par des chercheurs français (*Couder & Fort, 2006)*.

Ils ont découvert qu'une gouttelette millimétrique suspendue sur une surface d'un fluide vibrant peut avoir une auto-propulsion à travers une interaction résonnante avec sa propre fonction d'onde.

L'expérience est la suivante : prenons un fluide de densité ρ, tension superficielle σ dans un bain de hauteur $H$, subissant un mouvement de vibration vertical d'amplitude $A_0$ et de pulsation $\omega$. Cette vibration produit des ondes, appelées ondes de Faraday (*Faraday, 1831*), ce sont des ondes stationnaires non-linéaires.

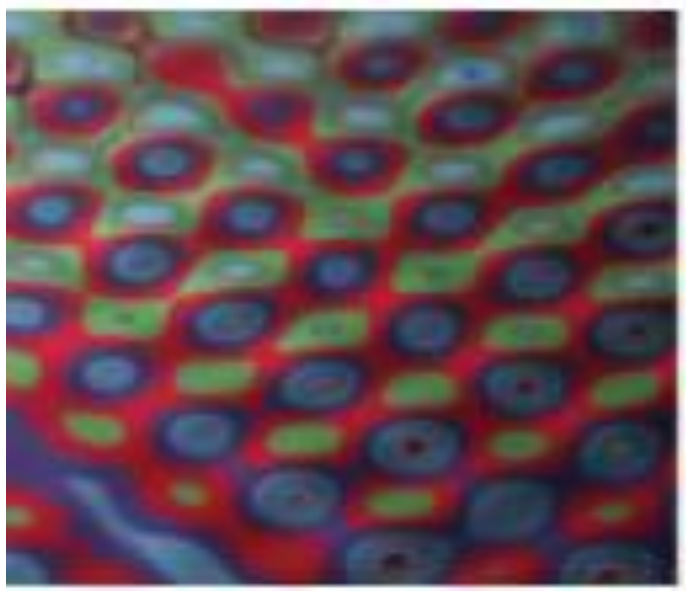

Fig. 2.19 : Ondes de Faraday (*Bush, 2015)*

Ils ont découvert qu'une gouttelette millimétrique suspendue sur cette surface de fluide vibrant peut avoir une auto-propulsion à travers une interaction résonnante avec sa propre fonction d'onde. On parle de «marcheur», guidé par sa fonction d'onde. Plus précisément, en considère le coefficient $\gamma = A_0\omega^2$, les expériences du marcheur fonctionnent pour $\gamma < \gamma_F$ (seuil de Faraday); au-delà, le système est instable.

La particule oscille à une certaine fréquence notée $\nu_F$, cette fréquence dépend de la fréquence de vibration du fluide et également de la tension superficielle $\sigma$ de ce fluide.

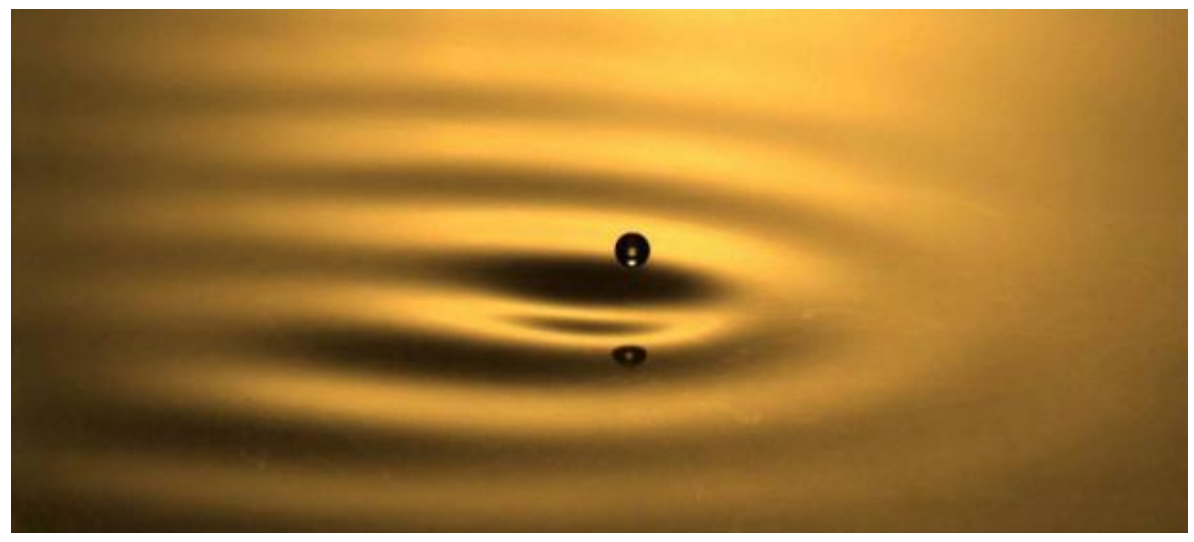

Fig. 2.20 : «Marcheur» (*Bush, 2015)*

Les expériences menées ont montré que ce système gouttelette + fluide vibrant permet de reproduire les résultats de la mécanique quantique : diffraction d'une simple particule (*Couder & Fort 2006*), effet tunnel, orbites quantifiés (*Fort et al. 2010*), états de spin,... La gouttelette est associée à une onde, et cette onde va guider la particule (onde pilote). Par exemple, nous illustrons ci-dessous les résultats pour les expériences de diffraction.

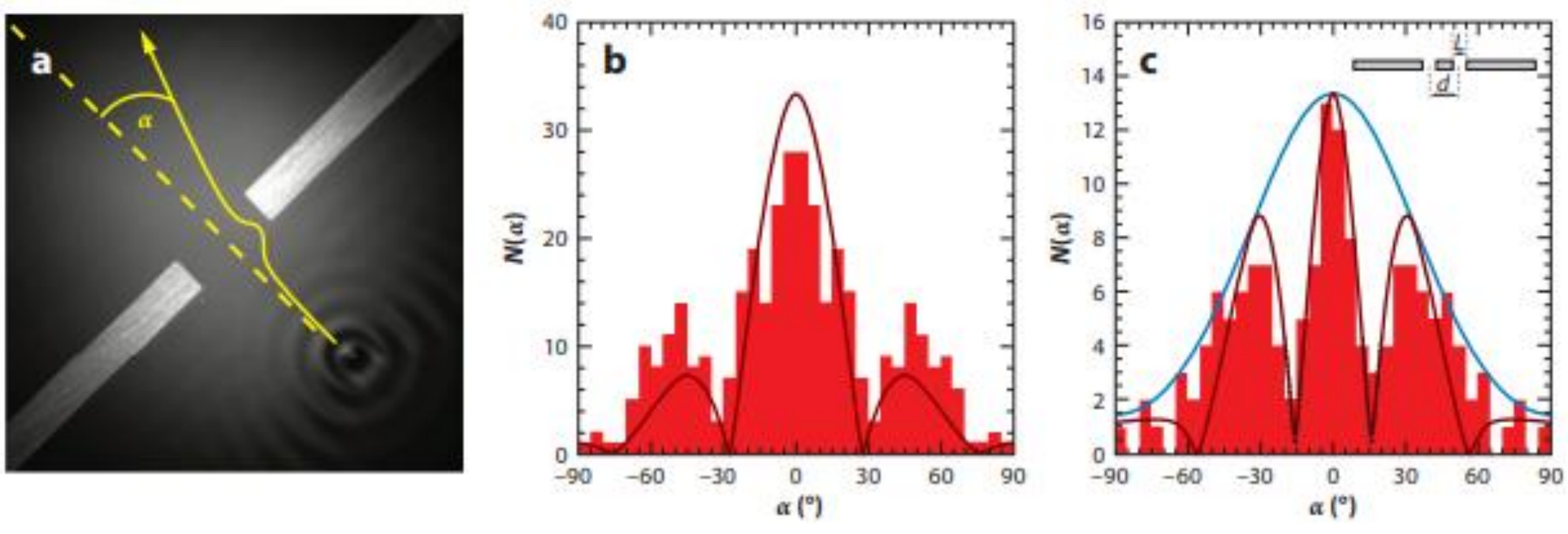

Fig. 2.21 : Diffraction d'une particule (*Bush, 2015*).
(a) Particule et son onde passant à travers une fente;
(b) Histogramme de répartition des particules en fonction de l'angle α - simple fente (diffraction);
(c) Histogramme - double fente (interférences)

Le processus de la double fente est le suivant : tandis que le marcheur passe par l'une ou l'autre fente, son onde pilote passe par les deux; donc le marcheur «sent » effectivement la seconde fente par l'intermédiaire de son onde pilote. Si l'on ne peut observer directement le marcheur, on ne peut prédire la trajectoire finale du marcheur.

Ces expériences de Couder et Fort suggèrent de plus que la dynamique de l'onde pilote est suffisamment complexe pour être chaotique, avec un angle de déflection α sensible au point de passage de la fente et à la phase de rebond.

Tout cela montre que les résultats des expériences quantiques n'ont rien d'étrange, étant donné que l'on peut les reproduire à l'échelle macroscopique. Ce sont des phénomènes essentiellement ondulatoires.

En examinant le comportement d'une gouttelette «libre» sans influence extérieure, on observe qu'elle a un comportement très aléatoire, analogue à un mouvement brownien. Mais à cause de l'onde dont elle est associée, on observe au bout d'un certain temps que la particule va plus aller au niveau de certaines orbites : on retrouve alors une distribution similaire à la fonction d'onde de la mécanique quantique.

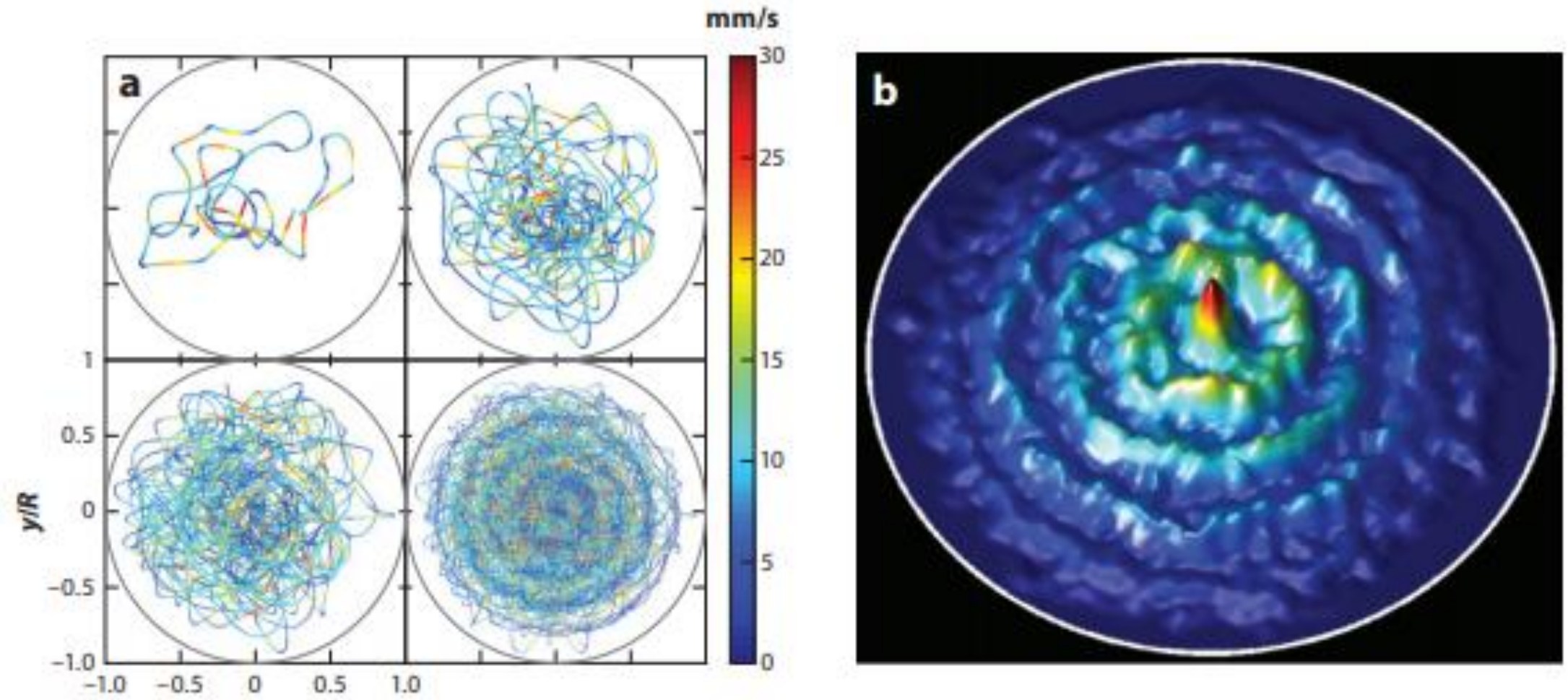

Fig. 2.22 : Trajectoire de la particule formant une fonction d'onde statistique

Source : *Bush, 2015.* Expérience « Coral »

L'analogie entre phénomènes quantiques et hydrodynamique a également été mises en évidence à travers d'autres phénomènes : effet Casimir (*Denardo et al., 2009*), superfluides, condensats de Bose-Einstein.

L'intérêt du système de l'onde pilote hydrodynamique est son accessibilité : les propriétés du marcheur, de la gouttelette et de son onde pilote sont visibles.

Le système du marcheur ressemble à la conception de l'onde pilote De Broglie et Bohm, selon laquelle la particule microscopique bouge en résonnance avec sa propre onde. Il est cependant plus proche de la théorie de la double solution de De Broglie : dans cette théorie, comme dans le système hydrodynamique, on a une onde pilote réelle et une onde statistique; dans la mécanique bohmienne, l'onde de guidage est une onde statistique.

De Broglie n'a jamais spécifié l'origine physique ni la forme géométrique de l'onde pilote. Il a proposé qu'elle était liée linéairement à l'onde statistique dans le champ lointain, mais était non-linéaire dans les environs de la particule. Il a évoqué une singularité au niveau de la position de la particule.

Trois facteurs clés du système du marcheur sont absents de la conception De Broglie :

- Le marcheur interagit avec un champ existant (l'interface) et il n'y a nul besoin d'une singularité;
- La relation entre l'onde réelle et l'onde statistique émerge à travers la dynamique chaotique de l'onde pilote.

### Approche stochastique

La mécanique bohmienne a été étendue pour incorporer l'influence d'un domaine sub-quantique stochastique, tout d'abord avec les travaux de Bohm et Vigier (1954) que nous avons décrit précédemment. Ceci mena à une littérature considérable.

Puis, Edward Nelson a montré qu'en considérant un système classique stochastique (processus de Wiener) de masse $m$ avec un coefficient de diffusion $\hbar$/m, on obtient l’équation de Schrödinger (*Nelson, 1966*).

Ici, l’évolution de la fonction d’onde n’est plus donnée par un postulat, mais déduite d’autres postulats plus fondamentaux (mouvement brownien). Comme la théorie de Bohm, la mécanique de Nelson reproduit exactement les résultats de la mécanique quantique. Elle donne de plus un rôle important aux équations hydrodynamiques de Madelung.

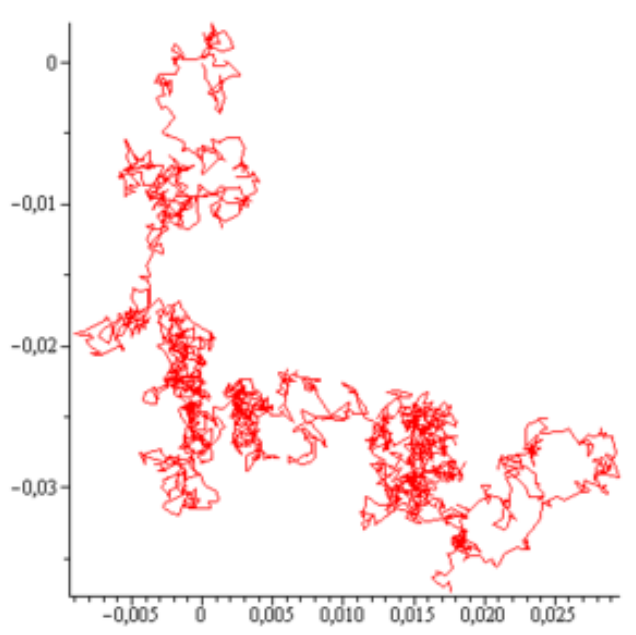

Fig. 2.23 : Mouvement brownien (*Brown, 1828*)

25 ans après avoir abandonné sa théorie de la double solution, De Broglie prend connaissance des travaux de Bohm avec enthousiasme.

Selon De Broglie, l’élément aléatoire d’origine cachée doit être admis. Le mouvement de la particule est une combinaison de mouvement régulier défini par la formule de guidage, avec un mouvement aléatoire de caractère brownien. On a ici une analogie avec la thermodynamique, où l’élément aléatoire résulte des collisions des molécules entre elles dans le gaz («chaos moléculaire»); les molécules invisibles jouent le rôle de thermostat caché.

La particule échange en permanence de l’énergie et de l’impulsion avec un tel thermostat caché. Ce milieu ne serait pas un milieu de référence universel, car il violerait la relativité, mais plutôt un ensemble de thermostat.

Un argument important va en faveur de cette approche stochastique est l’analogie entre la physique statistique et la mécanique quantique (*Mallick, 2012*). Cette analogie, ou dictionnaire, a été mise en valeur au cours de la 2e moitié du XX$^{e}$s., et est résumé dans le tableau ci-dessous :

Tab. 2.1 : Dictionnaire physique statistique - mécanique quantique

| **Physique statistique** | **Mécanique quantique** |
|---|---|
| Energie $E$ | Action $S$ |
| Poids de Boltzmann : $P_n = \frac{N_n}{N} = \frac{\exp(-\beta E_n)}{Z}$ | Amplitude de Feynman : $< X_b, t_b \| X_a, t_a >= \sum_{chemins} exp\left(\frac{iS}{\hbar}\right)$ |
| Température $T$ | Constante de Planck $h$ |
| Temps imaginaire $\tau$ | Temps réel $t$ |

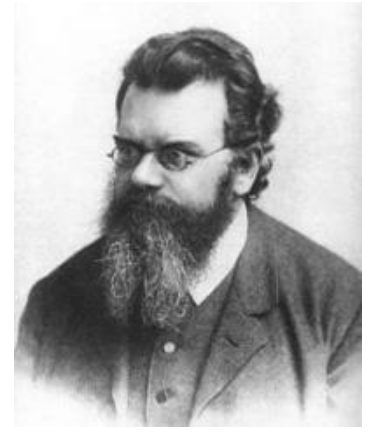

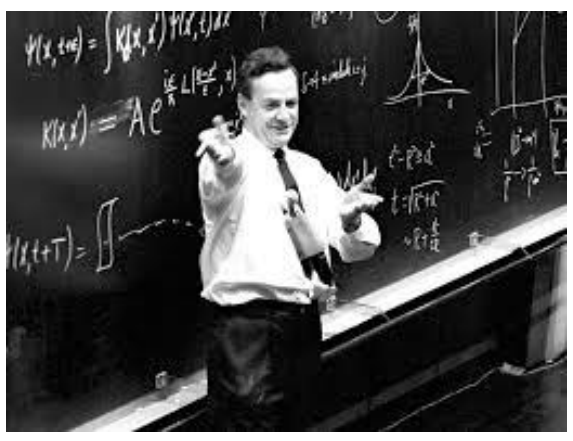

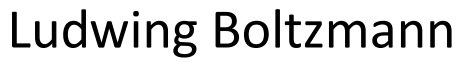

Ludwing Boltzmann

Richard Feynman

Selon l'électrodynamique stochastique (SED), la stochasticité dans le domaine sub-quantique a pour origine les fluctuations électromagnétique du vide, ou champ du point zéro *(De la Pena & Cetto, 1986)*. C'est l'énergie qui subsiste lorsque toute autre forme d'énergie a été enlevée (particules massives, photons,…), elle a été prédite par la théorie quantique des champs. Chaque mode de ce champ a un niveau d'énergie minimum moyen de : $E(\omega) = \frac{1}{2}\hbar\omega$.

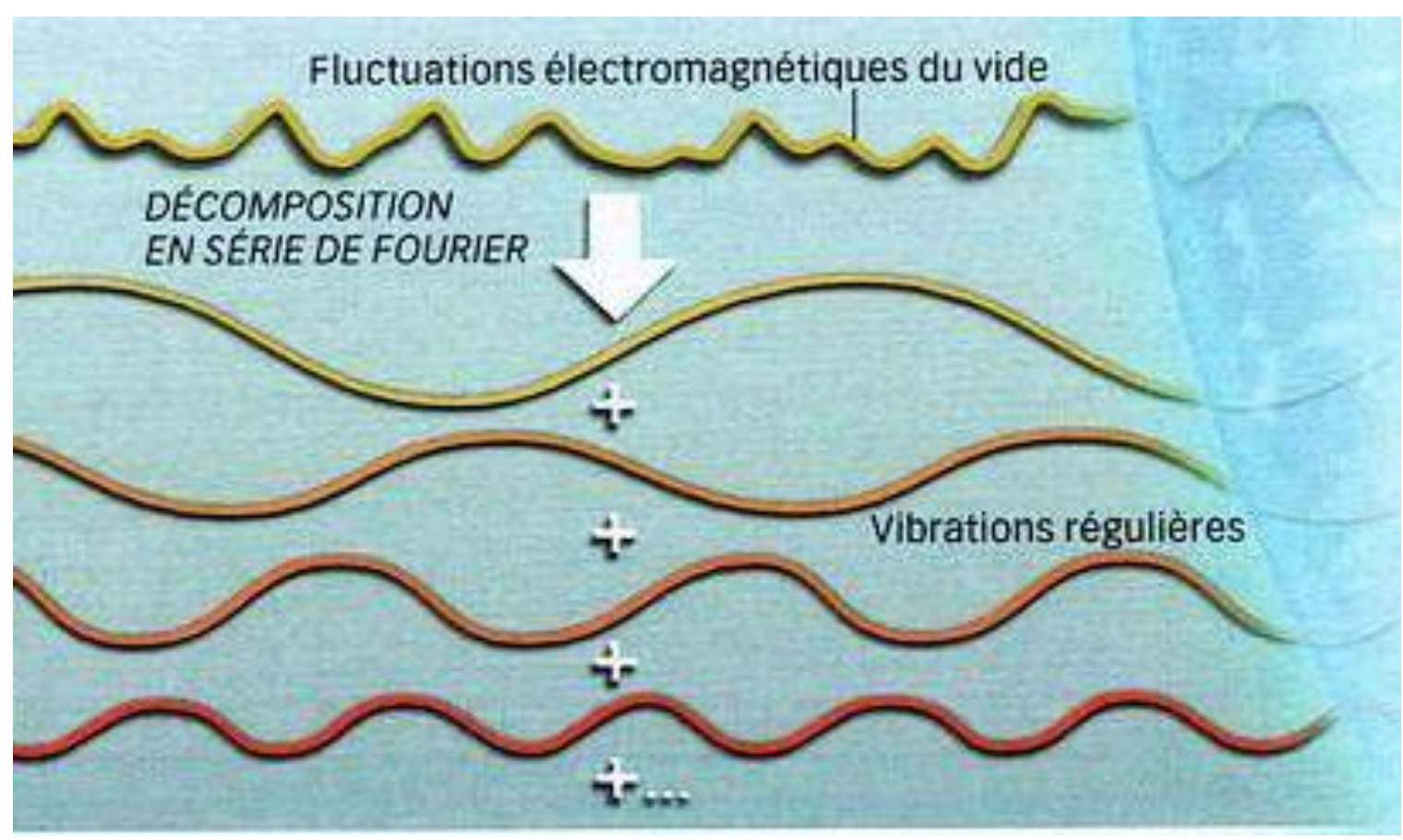

Fig. 2.21 : Fluctuations électromagnétiques du vide

L'onde pilote est alors, dans cette vision, de nature électromagnétique et en résonnance avec la particule.

On peut représenter dans un tableau les analogies et différences entre l'onde pilote de De Broglie, l'électrodynamique stochastique, et le système hydrodynamique.

Tab. 2.2 : Analogies hydrodynamique – onde pilote (*Bush, 2015*)

| **Hydrodynamique** | **Onde pilote** |
| --- | --- |
| Fluide | Espace |
| Tension superficielle du fluide $\sigma$ | Constante de Planck $h$ |
| Ondes de Faraday | Fluctuations du vide (électrodynamique stochastique) |
| Gouttelette | Particule |
| Rebond | Vibration - Zitterbewegung |
| Fréquence de rebond $\nu_c$ | Fréquence de vibration $\nu_p = \frac{2mc^2}{h}$ |
| Onde hydrodynamique | Onde pilote |
| Résultats obtenus : Diffraction, interférences, effet tunnel, effet Casimir | Résultats obtenus : Diffraction, interférences, effet tunnel, effet Casimir + intrication |

## 5) Questionnements sur l'éther, l'espace et le temps

Différents éthers ont été considérés au fil des siècles, à travers les effets qu'ils produisent. Ces effets concernaient la transmission de la force gravitationnelle (Isaac Newton), le transport de la lumière (depuis Descartes, Robert Hooke, Newton et bien d'autres jusqu'au début du XX$^{e}$ siècle), le transport de la force électrique, magnétique, et ensuite du courant électromagnétique (Maxwell),... Les ondes ont besoin d'un milieu pour se propager, comme c'est le cas du son (dans l'air) et des vagues (dans l'eau).

Comme nous avons vu précédemment, l'expérience de Michelson et Morley a remis en cause l’existence de l’éther, Einstein a développé la relativité restreinte sans besoin de le faire intervenir, et l’éther a par la suite été abandonné par les physiciens, jusqu’à nos jours.

Cependant, en 1916, après la publication de la théorie de la relativité générale, Einstein reconnaît la possibilité d’introduire un nouveau concept d’éther. Il rejette toutefois énergiquement le caractère stationnaire de l'éther défendu par Lorentz, c'est-à-dire la conception d'un milieu rigide qui possède son propre référentiel dans lequel il est au repos, car cela est contraire au principe de relativité. Il admet la possibilité d'un éther qui ne serait pas un medium doté d'un état de mouvement et donc qui ne violerait pas le principe de relativité. Ce « nouvel éther » serait doté d'un état qui déterminerait le mouvement des objets physiques, dont le comportement serait décrit par le tenseur métrique.

La distinction fondamentale entre l’ancien éther et celui d’Einstein est résumée par Hermann Weyl : l'éther de la relativité générale dote l'espace d'un « champ d'états », interagissant avec la matière et étant influencée par elle.

Einstein explique, dans la troisième partie du discours *Éther et la théorie de la relativité (1920)*, que l'idée d'un éther peut revenir afin d'attribuer des propriétés physiques (autres que mécaniques ou cinématiques) à l'espace, et que l'espace - même dépourvu de matière - ne peut être considéré comme réellement vide. Inspiré par les idées de Mach, il énonce le besoin d'un milieu pour transmettre l'interaction gravitationnelle de ces masses distantes, cette vision est reliée à la relativité générale.

« D’après la théorie de la relativité générale, l'espace est pourvu de propriétés physiques, et dans ce sens il existe un éther. Un espace sans éther est inconcevable, car la propagation de la lumière y serait impossible, et car il n’y aurait même aucune possibilité d’existence pour les règles et horloges et par conséquence aussi pour les distances spatio-temporelles dans le sens de la physique » (*Einstein, 1920*).

Cette approche de l’espace doté de propriétés physiques est d’autant plus d’actualité de nos jours, avec la découverte des fluctuations du vide, de la notion de champ quantique, de la création / annihilation de particules virtuelles.

Certains physiciens considèrent l'éther comme un référentiel privilégié, ce qui serait consistant avec la non-localité. Cette vision n'est cependant pas compatible avec la relativité, il est tout à fait possible que l'éther opère à un niveau sous-jacent où les lois de la relativité ne s'appliquent plus. C'est cette vision de l'éther qui a été développée par Lorentz, Poincaré, puis soutenue par Bell et Gondran de nos jours. Réhabiliter un référentiel privilégié, comme celui de Lorentz-Poincaré, est peut-être le chemin pour réconcilier les deux théories de la relativité générale et de la mécanique quantique.

Le réel serait alors doublement voilé :

- En mécanique quantique, la fonction d'onde est une variable cachée non locale;
- En relativité, le temps et l'espace que nous mesurons ne seraient qu'un temps et un espace apparents, l'espace et le temps vrais du référentiel privilégié nous étant cachés.

Citons ici la vision de Bell en *1986* : « Un retour à la relativité comme avant Einstein, quand des gens comme Lorentz et Poincaré pensaient qu'il y avait un éther - un référentiel de référence privilégié - mais que nos instruments de mesure étaient déformés par le mouvement de telle sorte que nous ne pouvions pas détecter le mouvement à travers l'éther. Derrière l'invariance apparente des phénomènes de Lorentz, il existe un niveau plus profond, qui n'est pas invariant de Lorentz. La raison pour laquelle je veux revenir ici à l'idée d'un éther est que dans ces expériences EPR, il y a suggestion que dans les coulisses quelque chose va plus vite que la lumière. »

### Conclusion de la partie II :

Nous pouvons nous demander l'intérêt de modifier l'interprétation actuelle de Copenhague si elle suffit à rendre compte de tous les phénomènes observables ? Le retour à des conceptions claires, causales, respectant la validité du cadre de l'espace-temps, satisferait beaucoup d'esprits et permettrait de lever les objections d'Einstein et de Schrödinger.

Si l'interprétation actuelle suffit à la prévision des phénomènes à l'échelle atomique (jusqu'à $10^{-11}$ cm), il pourrait ne pas en être de même à l'échelle inférieure (en dessous de $10^{-13}$ cm).

De nos jours, l'attitude qui prévaut dans la communauté des physiciens est beaucoup plus prudente, peut-être car les théorèmes d'impossibilité énoncés par les tenants de Copenhague (ceux de Von Neumann en particulier) ne sont plus vraiment pertinents.

Laloë indique : « Nous savons maintenant que l'interprétation orthodoxe de Copenhague n'est pas la seule possibilité logique. D'autres interprétations restent parfaitement possibles, le déterminisme n'étant pas du tout éliminé ».

Selon De Broglie, l'extension, la modification de l'interprétation de Copenhague mènerait à une meilleure compréhension de la coexistence des ondes et particules, alors que la mécanique quantique ne donne qu'une information statistique, souvent correcte, mais incomplète. Il indique : « Il est certainement utile de reprendre le problème très difficile de l'interprétation de la mécanique ondulatoire, afin de voir si celle qui est actuellement orthodoxe est vraiment la seule que l'on puisse adopter ».

De plus, selon De Broglie, la physique a un besoin urgent de pouvoir définir une structure des particules et notamment de pouvoir introduire un «rayon» de l'électron comme dans l'ancienne théorie de Lorentz. L'onde statistique $\psi$ empêche d'employer une image structurale de ces particules.

Il indique que la représentation de la particule comme simple singularité se déplaçant sur l'onde n'est certainement pas une bonne image de la structure de la particule. C'est un fait que la physique n'a pas réussi, même de nos jours, à représenter théoriquement des corpuscules par des champs sans singularités.

C'est dans cette optique que nous proposerons dans la partie suivante des idées nouvelles sur la structure de la particule, et son lien avec la relativité.

« Long may Louis de Broglie continue to inspire those who suspect that what is proved by impossibility proofs is lack of imagination » John Bell

# III. Ouverture : modèle ondulatoire des particules

« Vous savez, il suffirait de comprendre l'électron » Einstein

Nous allons proposer ici un modèle ondulatoire des particules élémentaires, fermions et bosons, et voir que l'on retrouve les lois de la physique actuelle (mécanique quantique, relativité).

Les modèles proposés reposent sur les motivations suivantes :

- Chaque particule élémentaire est associée à une fréquence et a une longueur d'onde. C'est la dualité onde-particule.
- La relation intime entre lumière et matière, comme dans $E = mc^2$.
- La notion d'hélicité L/R des particules massives élémentaires.
- La remise en question du caractère ponctuel des particules, vision qui est incompatible avec la contraction/dilatation des longueurs en relativité restreinte.

Cas des particules non-massives (photon) :

Mouvement rectiligne de vitesse $c$, ce qui correspond à la vision de la physique actuelle.

Paramètres : longueur d'onde $\lambda_\gamma$, periode $T_\gamma$.

Cas des particules massives (fermions, bosons W/Z) :

Particule de vitesse $c$ (photon) dotée d'une trajectoire hélicoïdale.

Paramètres : rayon de l'hélice $r$, longueur d'une spire $l$, temps pour parcourir une spire $t$, et pas de l'hélice $L$.

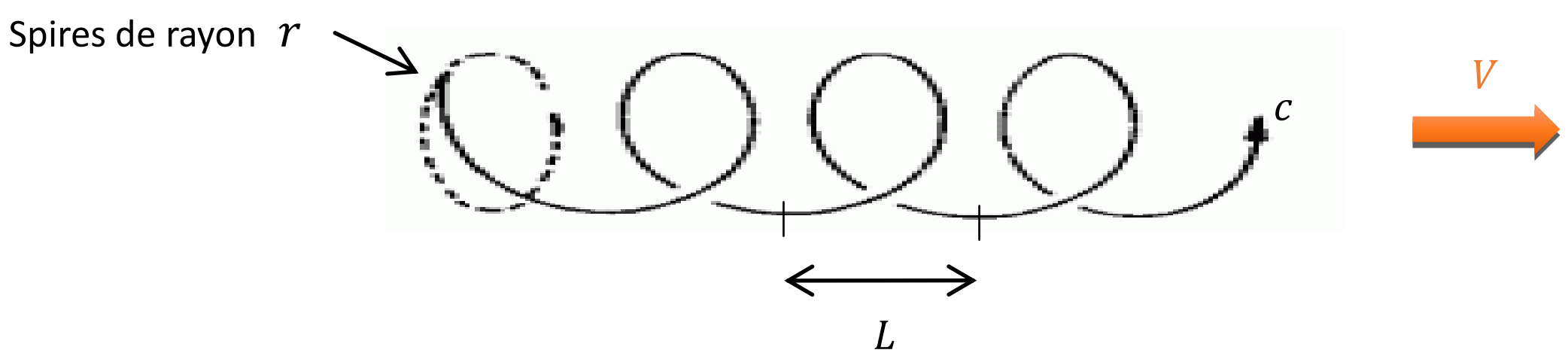

Fig. 3.1 : Mouvement hélicoïdal – particule en mouvement (vue de côté)

La vitesse V vaut : $V = \frac{L}{t}$ ; de plus, $c = \frac{l}{t}$.

Lorsque $r \to 0$, nous obtenons le cas d'une particule non-massive, se déplaçant rectilignement a la vitesse c.

Au repos, $V = 0 \Leftrightarrow L = 0 \to$ mouvement cyclique de rayon $r_0$, de périmètre $l_0 = 2\pi r_0$, et de période $t_0$.

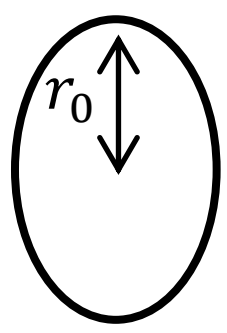

Fig. 3.2 : particule au repos (vue de côté)

Le modèle en hélice a été pensé également par d'autres physiciens (*Vigier, 1956; Hestenes, 2008*).

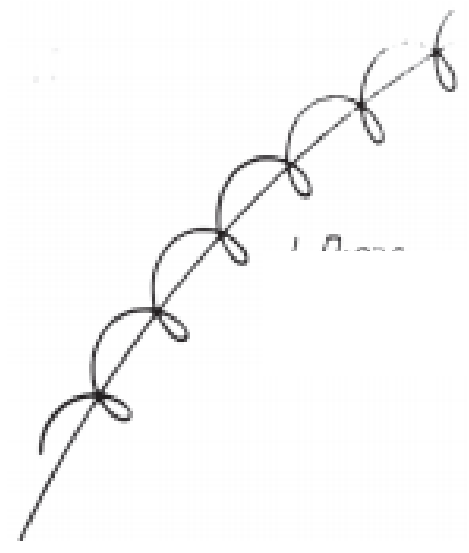

Fig. 3.3 : Trajectoire hélicoïdale des particules à spin. Source : Vigier, 1956.

## 1) Les lois de la mécanique quantique

Selon notre modèle ondulatoire, une particule au repos a un mouvement cyclique de rayon $r_0$. Comment introduire la notion de spin ?

Géométriquement, le spin ½ correspond à un ruban de Möbius (*Möbius, 1886*)., tandis que le spin 1 correspond à un cercle. Nous devons considérer plus précisément les champs électrique et magnétique associés au photon constituant la particule.

Illustrons cela avec le cas des fermions, par exemple l'électron.

## Modèle ondulatoire des fermions de spin 1/2

La structure en ruban de Möbius peut paraitre surprenante à première vue.

Il est la manifestation géométrique du spineur; en effet, le spineur $\psi$ change de signe suite à une rotation de $2\pi$ et retourne à sa position initiale après une rotation de $4\pi$.

Les spineurs ont été introduits pour la première fois par le mathématicien Elie Cartan en 1913 (*Cartan, 1966*), et a ensuite été réutilisé par Dirac en physique pour décrire la fonction d'onde d'un fermion.

Le modèle de fermion en ruban de Möbius a été pensé par d'autres théoriciens (*Williamson & Van Der Mark, 1997; Q. Hu, 2004*).

Le chercheur Isaac Freund a démontré théoriquement que lorsque deux faisceaux de lumière à polarisation circulaire opposée interfèrent, il est possible que la lumière créée suive une trajectoire en ruban de Möbius. Le champ électrique oscillant tourne autour d'une bande d'un seul côté, la polarisation de la lumière est « tordue », formant un ruban de Möbius (*Freund, 2010).*

Ce fait a récemment été observé expérimentalement (*Banzer et al.*, 2015).

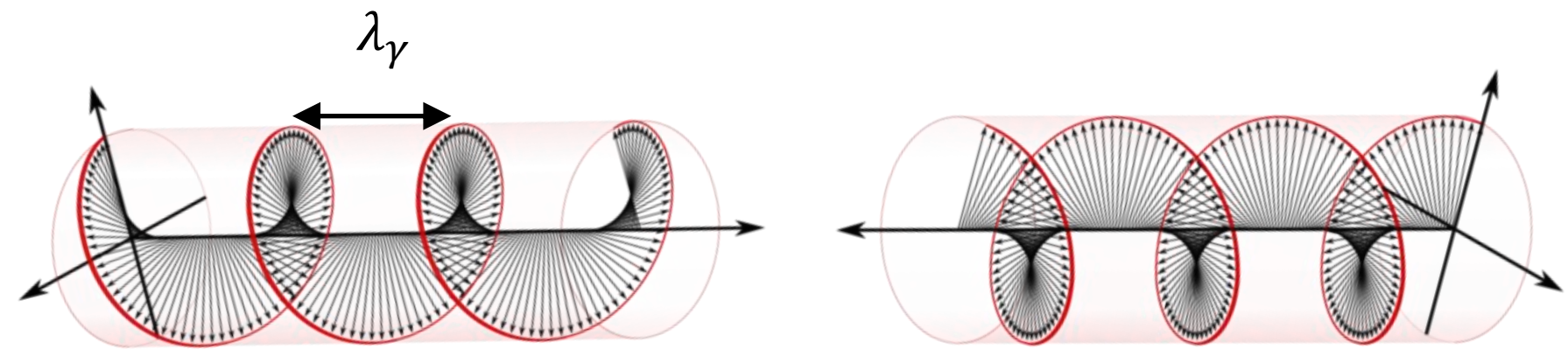

2 faisceaux lumineux de polarisation circulaire opposée

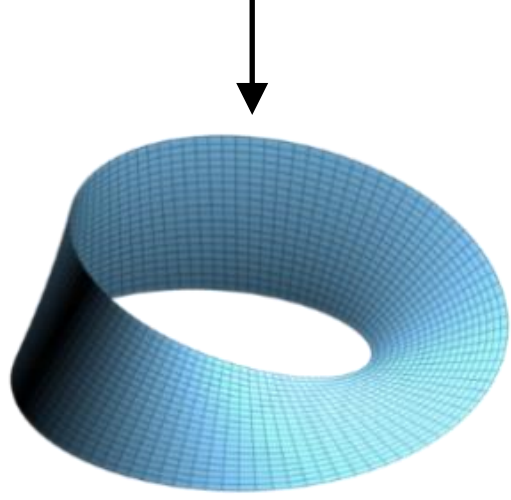

Fig. 3.4 : Ruban de Möbius

Bien qu'il ne soit pas certain que ce processus observé expérimentalement soit exactement le même que celui à l'œuvre dans la création des particules-antiparticules, il a le mérite de montrer qu'une trajectoire d'un photon en ruban de Möbius n'est pas aberrante.

Ce phénomène met de plus en valeur la richesse topologique cachée dans les lois de l'électromagnétisme.

La démonstration faite par Einstein de la relation $E = mc^2$ ne repose pas sur les hypothèses de la relativité mais sur les lois de la physique classique ($p = mV, E = \frac{p}{c}$, conservation de l'impulsion). Elle s'applique donc également avec notre modèle.

Le couple électron/positron peut être créé à partir de deux photons virtuels, ainsi que décrit par la théorie quantique des champs.

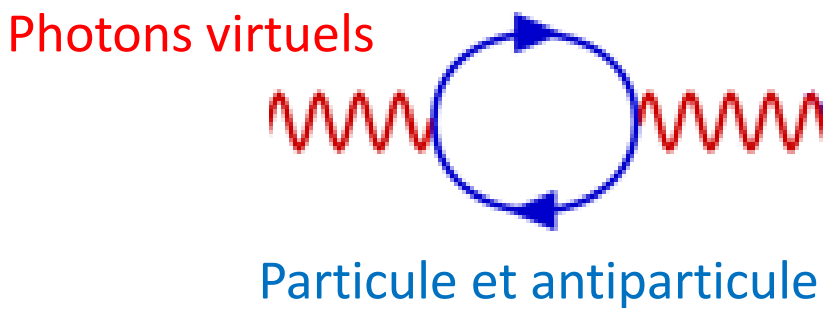

Fig. 3.5 : Polarisation du vide

L'énergie du photon est $E_\gamma = h\nu_\gamma$; celle de la particule est $E = mc^2 = h\nu_c = \frac{hc}{\lambda_c}$.

Par conservation de l'énergie lors du processus de polarisation du vide, on a :

$$E_\gamma = E \rightarrow \nu_\gamma = \nu_c.$$

La longueur d'onde du photon correspond donc à la longueur d'onde de Compton :

$$\lambda_\gamma = \lambda_c \quad (3.1).$$

En notant $r$ le rayon du ruban de Möbius, on a : $4\pi r = \lambda_c$ (on parcourt une longueur d'onde en faisant 2 tours). Donc $r = \frac{\lambda_c}{4\pi} = \frac{\hbar}{2mc}$ $(3.2)$.

L'électron a lui-même un mouvement d'oscillation. Nous devons donc introduire une autre fréquence, en plus de celle de Compton.

Le photon accomplit une longueur d'onde en parcourant 2 tours $(\lambda_c = 4\pi r)$. Chaque tour correspond à une longueur d'onde d'oscillation de l'électron ($\lambda_p$). On a :

1 longueur d'onde $\lambda_c$ du photon $\leftrightarrow$ 2 tours du ruban de Möbius

1 longueur d'onde $\lambda_p$ de l'électron $\leftrightarrow$ 1 tour ruban de Möbius

$$\rightarrow \lambda_c = 2.\lambda_p \rightarrow \nu_p = 2.\nu_c \quad (3.3)$$

où l'on a apposé l'indice *p* pour particule (ici l'électron).

$$\rightarrow \nu_p = 2.\nu_c = \frac{2mc^2}{h} \quad (3.4).$$

On retrouve ici la fréquence de vibration $\nu_p$ prédite par Schrödinger à partir des équations de Dirac (*Schrödinger,* 1930).

Comme l'indique Dirac, « il a été trouvé que l'électron doit avoir un mouvement oscillant de haute fréquence, de faible amplitude, superposé sur le mouvement régulier ». Ce mouvement oscillatoire frénétique à très haute fréquence s'ajoute à celui du mouvement moyen de déplacement de la particule.

Schrödinger l'a appelé le «zitterbewegung» (mouvement de tremblement en allemand).

Le zitterbewegung a été simulé récemment avec l'ion $Ca^{2+}$, les condensats de Bose-Einstein, les isolants topolotiques,… (Stepanov et al., 2016).

Nous avons ainsi les relations suivantes pour l'énergie de la particule :

$$E = mc^2 = h\nu_c = \frac{h}{2}\nu_p = \frac{h}{T_c} = \frac{h}{2.T_p} = \frac{hc}{\lambda_c} = \frac{hc}{2.\lambda_p} \quad (3.5).$$

Au repos, on appose un indice 0 :

$$E = m_0c^2 = h\nu_{c_0} = \frac{h}{2}\nu_{p_0} = \frac{h}{T_{c_0}} = \frac{h}{2.T_{p_0}} = \frac{hc}{\lambda_{c_0}} = \frac{hc}{2.\lambda_{p_0}} \quad (3.6).$$

Les termes $m_0$ / $\nu_{p_0}/\lambda_{p_0}/T_{p_0}$ sont appelées alors masse/fréquence/longueur d'onde/période *propres* de la particule.

Le ruban de Möbius a une propriété remarquable : en considérant les deux faces du ruban, chaque face sera orientée vers l'extérieur alternativement.

Lorsque le photon se déplace, le champ électrique positif représente une face du ruban, le champ électrique négatif l'autre face. Ainsi, avec un ruban de Möbius, le champ électrique sera soit toujours négatif (la charge est alors négative), soit toujours positif (la charge est alors positive).

Le champ électrique du photon est à l'origine de la charge de l'électron, tandis le champ magnétique du photon génère le moment magnétique de spin $\vec{\mu}$ de l'électron.

Nous pouvons ainsi donner une représentation de l'électron où nous indiquons par $\vec{\mu}$ so moment magnétique de spin, et $\vec{\pi} = \frac{\vec{E} \wedge \vec{B}}{\mu_0}$ le vecteur de Poynting du photon.

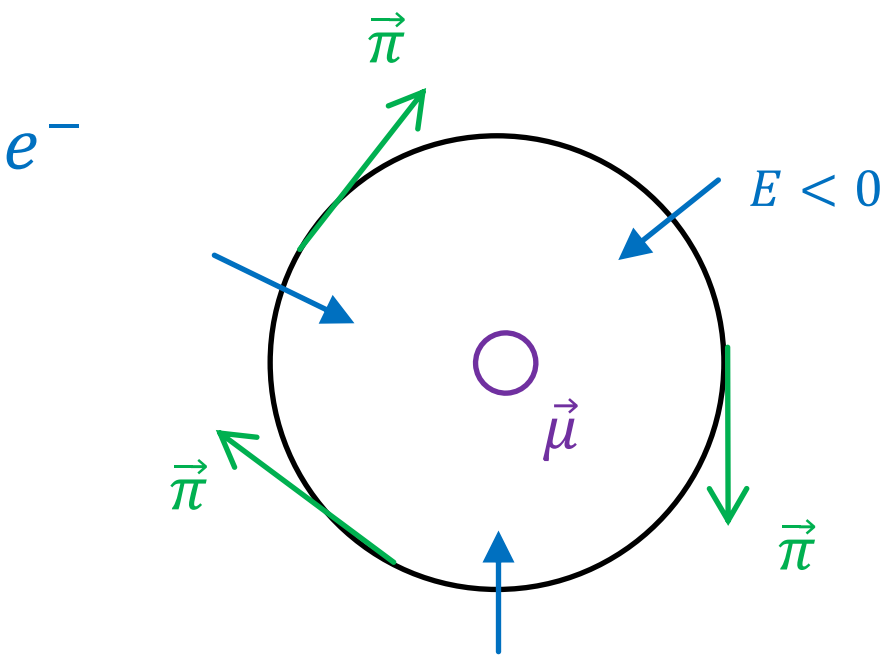

Fig. 3.6 : Représentation d'un électron (au repos, vue de dessus)

Le spin :

Après la découverte du spin par Pauli, deux jeunes étudiants de l'université de Leyde (Pays-Bas), George Uhlenbeck et Samuel Goudsmit, associèrent le nouveau nombre quantique au moment cinétique correspondant au mouvement de rotation de l'électron sur lui-même (*Uhlenbeck & Goudsmit, 1926*). Ce moment cinétique avait une valeur de $\frac{\hbar}{2}$ et ne pouvait prendre que deux orientations possibles en présence d'un champ magnétique externe. Ce nombre quantique correspondrait alors à un degré de liberté supplémentaire de l'électron, sa rotation intrinsèque.

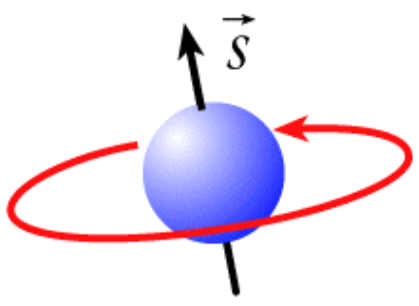

Fig. 3.7 : Illustration « naïve » du spin

Cependant, les calculs réalisés par Pauli montrèrent que, dans le cas d'un électron de taille finie avec une charge répartie de façon homogène, la vitesse angulaire à la surface de l’électron devait être considérablement supérieure à la vitesse de la lumière, ce qui entrait en contradiction avec la relativité. Les physiciens déduisirent alors que l’électron est une particule élémentaire / ponctuelle, elle n’a pas une taille finie. Le spin est considéré par la physique moderne comme n’ayant pas d’équivalent dans le monde classique, il est purement quantique (*Bekaert, 2015*).

Dans notre approche, le spin correspond au mouvement de rotation du photon, ce mouvement étant à l’origine de la masse et de la charge de la particule. La charge n’est pas ici distribuée de façon homogène sur une sphère, donc la réfutation d’une vitesse angulaire supérieure à $c$ ne s’applique pas ici.

De plus, notons que notre approche ondulatoire permet de retrouver la valeur de spin du fermion :

$$S = |\vec{r} \wedge \vec{p}| = r.p = \frac{\lambda_c}{4\pi}.\frac{h\nu_c}{c} = \frac{\lambda_c}{4\pi}.\frac{h\nu_c}{\lambda_c \nu_c} = \frac{\hbar}{2} \quad (3.7).$$

Nous pouvons de plus établir la relation entre le moment magnétique de spin et le spin.

En appliquant le même raisonnement qu’avec la fréquence, le moment magnétique de spin sur un ruban de Möbius est 2 fois celui sur un cercle.

$$\vec{\mu}_{Mobius} = 2.\vec{\mu}_{cercle} = 2.I.A\,\vec{n}, \text{ avec } \vec{n} \text{ vecteur unitaire}$$

$$= 2.\frac{q}{T}A\,\vec{n} = 2\,\frac{q.c}{2\pi r}A\,\vec{n} = 2\,\frac{q.c}{2\pi r}\pi r^2\,\vec{n} = 2.\frac{q.c}{2}.r\,\vec{n}$$

$$= 2.\frac{q.c}{2}.\frac{\hbar}{2mc}\,\vec{n} = 2.\frac{q}{2m}.\frac{\hbar}{2}\,\vec{n} = 2.\frac{q}{2m}.S\,\vec{n} = -2.\frac{q}{2m}.S(-\vec{n})$$

$$= g.\frac{q}{2m}.\vec{S}, \text{ avec } g = -2 \text{ et } \vec{S} = -\frac{\hbar}{2}\vec{n} \quad (3.8).$$

On retrouve la relation entre le moment magnétique et le spin telle que prédite par la mécanique quantique, avec le facteur de Landé de l’électron $g = -2$.

De plus, $\vec{\mu}$ et $\vec{S}$ sont en sens opposé.

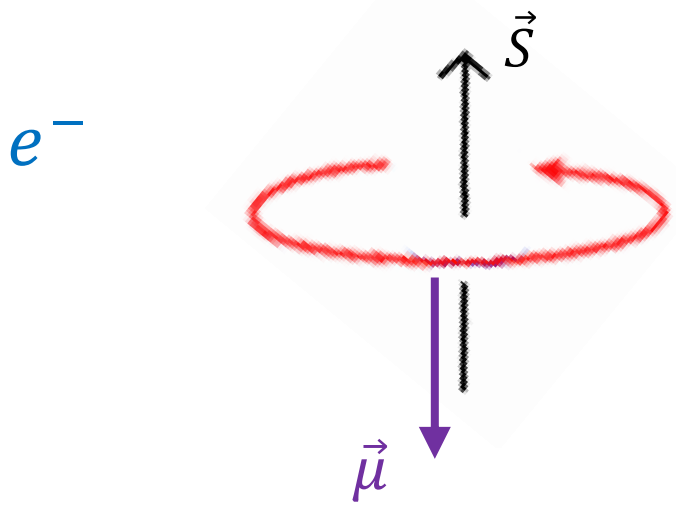

Fig. 3.8 : Représentation d'un électron (au repos, vue de

Le mouvement de rotation du photon créé un moment magnétique de spin, orienté dans la direction de l'axe de rotation. Mais il est possible que cette direction de rotation change en permanence, influencée par les interactions de la particule avec l'environnement (notamment les fluctuations du vide). En présence d'un champ magnétique externe, le moment magnétique de spin de la particule s'aligne dans la direction du champ magnétique ou dans le sens inverse.

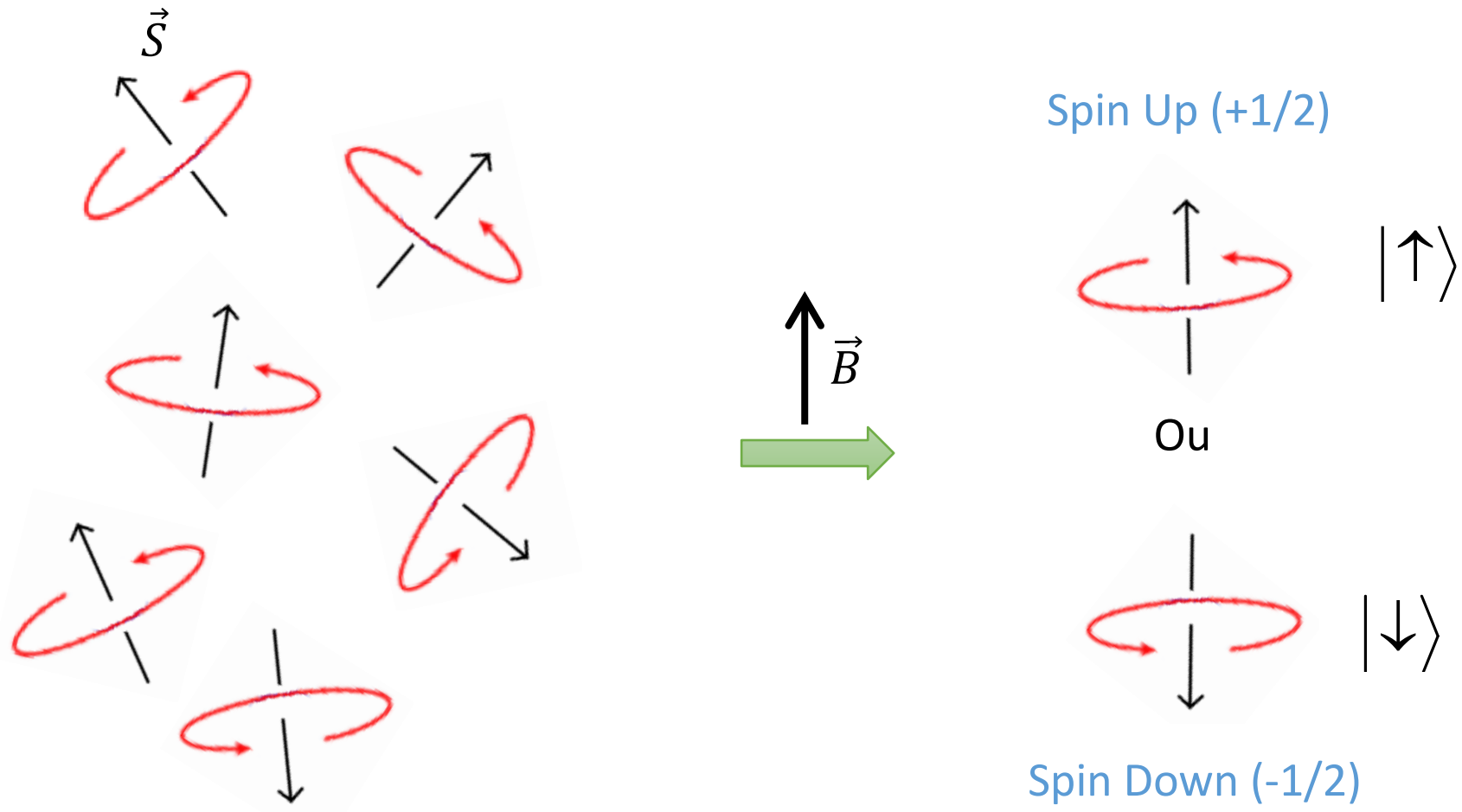

Fig. 3.9 : Orientation du spin de la particule sans et avec champ magnétique externe

La particule prendra alors nécessairement deux valeurs (+1/2 ou -1/2), comme illustré sur le schéma. Les deux valeurs de spin possibles ne semblent donc pas réellement intrinsèques à la particule, mais dépendent de l'expérience. Cela rejoint d'ailleurs la vision de M.&A. Gondran sur le spin.

Nous avons vu ici que l'axe de rotation du mouvement cyclique du photon (ou Zitterbewegung) détermine la direction du spin. Ce lien a été mis en évidence par de nombreux auteurs (*Schrödinger, 1930*; *Hestenes, 2008*).

### Modèle ondulatoire des particules massives de spin 1

Le cas des particules massives de spin 1 (bosons W et Z) est analogue au cas précédent. Dans ce cas, les formules suivantes s'appliquent :

$$\lambda_c = \lambda_p \; ; \; E = h\nu_p = \frac{hc}{\lambda_p} \; ; \; \nu_p = \frac{mc^2}{h} \; ; \; \lambda_c = 2\pi r \; ; s = \hbar \; ; |g| = 1 \quad (3.9).$$

### Notion d'hélicité

En physique des particules, on peut définir l'hélicité d'une particule : $H = \frac{\vec{S}.\vec{p}}{|p|}$, avec S le spin.

L'hélicité est positive (ou droite) quand le spin est orienté dans le même sens que le mouvement; elle est négative (ou gauche) quand le spin est orienté dans le sens contraire du mouvement.

Notre modèle hélicoïdal permet de bien comprendre cette notion d'hélicité, alors qu'il n'est pas possible de le faire en considérant la particule comme ponctuelle. C'est même une conséquence naturelle de notre modèle.

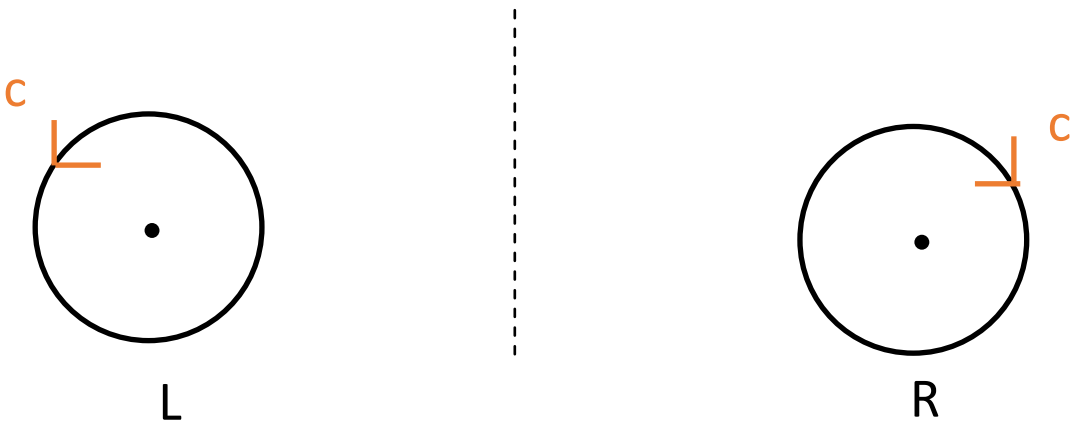

Fig. 3.10 : Illustration de l'hélicité pour une particule arrivant vers l'observateur

### Equations de Schrödinger et de Dirac

Le mouvement hélicoïdal du photon peut être mathématisé par la fonction suivante :

$$\psi(x) = A.exp\left(i\frac{p.x}{\hbar}\right) = A.\left\{\cos\left(\frac{p.x}{\hbar}\right) + isin\left(\frac{p.x}{\hbar}\right)\right\} \quad (3.10)$$

avec $A$ l'amplitude du mouvement.

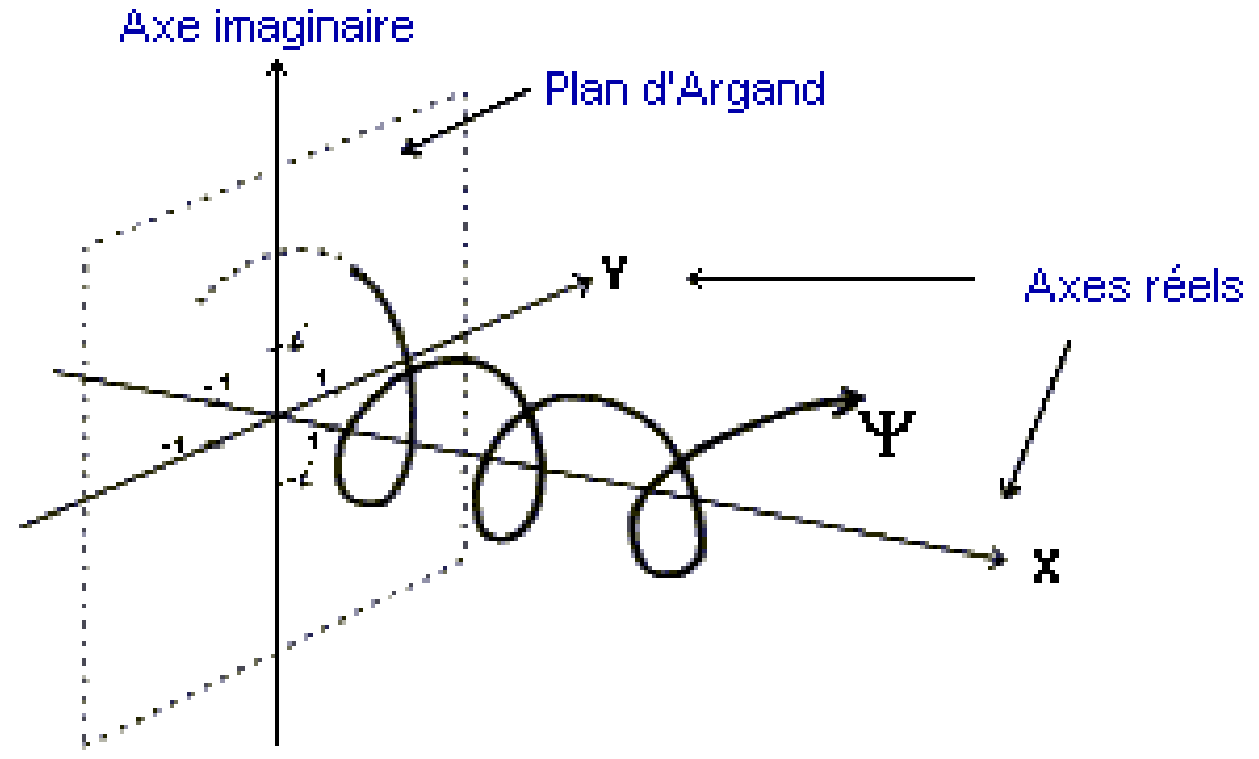

Fig. 3.11 : Mouvement hélicoïdal dans un repère complexe

Deux cas peuvent se présenter :

- Faible vitesse :

  Le rayon $r$ est proche de $r_0$, donc l'amplitude est quasiment constante.

  On a alors : $\psi(x) = A.exp\left(i\frac{p.x}{\hbar}\right)$, avec $A = const.$

  On constate que cette fonction satisfait l'équation d'onde de Schrödinger pour une particule libre :

$$i\hbar\frac{\partial\psi(\vec{r},t)}{\partial t} = -\frac{\hbar^2}{2m}\Delta\psi(\vec{r},t) \quad (3.11).$$

- Vitesse importante / relativiste :

  Le rayon $r$ dépend de la vitesse / de l'impulsion $\left(r = \frac{r_0}{\gamma}\right)$, donc l'amplitude en dépend également.

  On a alors : $\psi(x) = A(p).exp\left(i\frac{p.x}{\hbar}\right)$, ce qui correspond aux solutions générales de l'équation de Dirac :

$$\left(i\hbar\gamma^\mu\partial_\mu - mcI\right)\psi = 0 \quad (3.12).$$

  D'autres auteurs ont montré par des voix différentes que le modèle de l'électron correspondant à une onde électromagnétique en mouvement circulaire conduit à l'équation de Dirac (*Khan & Asif, 2015*).

Lorsqu'il y a création de particule et antiparticule à partir de photons virtuels, particule et antiparticule vont en sens opposé (conversation de l'impulsion), l'une va suivant l'axe $x$, l'autre suivant l'axe $-x$. Ainsi, la fonction d'onde s'exprime alors

comme :

$$\psi(x) \propto b \exp\left(-\mathrm{i}\frac{p.x}{\hbar}\right) + c \exp\left(i\frac{p.x}{\hbar}\right) \quad (3.13).$$

Cela rejoint une formule fondamentale de théorie quantique des champs (*Tong, 2009*) :

$$\psi(x) = \int \frac{d^3p}{(2\pi)^3}\frac{1}{\sqrt{p^0}}\sum_{s=1}^{2}\left\{b_s(p)u_s(p)\exp\left(-\mathrm{i}\frac{p.x}{\hbar}\right) + c^{\dagger}(p)v_s(p)\exp\left(i\frac{p.x}{\hbar}\right)\right\} \quad (3.14).$$

Où l'on l'intègre sur les moments, on somme sur les spins (up et down), et on met des coefficients $u(p)$ pour la particule et $v(p)$ pour l'antiparticule.

Quel lien peut-on faire entre notre modèle et l'électrodynamique quantique (où la particule est ponctuelle) ? Si l'électron est observé à une échelle spatiale bien plus grande que sa longueur d'onde de Compton et à une échelle temporelle bien plus grande que la très petite période du Zitterbewegung, l'électron peut être approximé par une particule ponctuelle qui bouge le long de l'axe de l'hélice. Il peut alors être décrit par l'électrodynamique quantique.

Notre modèle de particule avec une structure interne non ponctuelle pourrait permettre d'éviter les divergences qui apparaissent en électrodynamique quantique…

### Lien avec la théorie de De Broglie-Bohm

Nous avons vu que l'électron a un mouvement oscillant, de fréquence $\nu_p = 2.\nu_c = \frac{2mc^2}{h}$. A l'image d'un objet oscillant sur la surface de l'eau, la vibration de l'électron dans l'éther va produire des ondes (dans différentes directions de l'espace), de même fréquence $\nu_p$ et toujours en phase avec la particule.

Cette onde va possiblement « guider » la particule, et comme mis en évidence par la théorie de De Broglie-Bohm, être à l'origine des effets quantiques (fentes de Young, effet tunnel,…).

## 2) Les lois de la relativité restreinte

### Formules de l'énergie

D'autre part, nous allons démontrer la formule relativiste de l'énergie en nous plaçant par rapport à un référentiel, et donner une expression ondulatoire au facteur de Lorentz. Nous supposons que dans ce référentiel, la longueur de la spire reste constante quel que soit la vitesse de la particule ($l = l_0$).

Les formules dérivées ici s'appliquent à la fois aux particules massives ($r > 0$) et non-massives ($r = 0$).

Nous allons dans un premier temps mathématiser ce mouvement hélicoïdal.

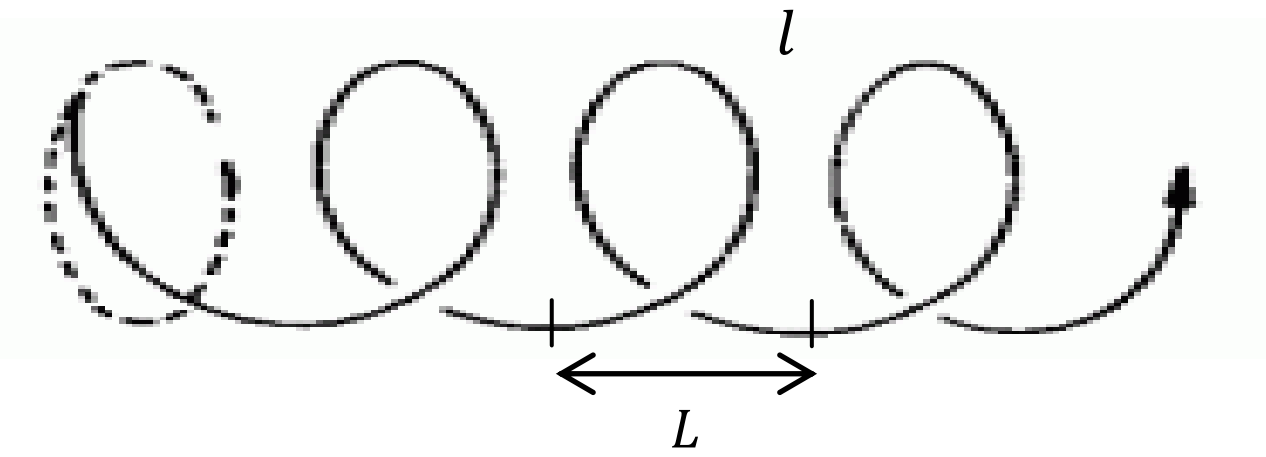

Les mathématiques donnent la relation suivante :

$$l^2 = L^2 + (2\pi r)^2 \quad (3.15).$$

Ce qui conduit aux relations :

$$l^2 = L^2 + (2\pi r)^2 \Leftrightarrow \left(\frac{l}{t}\right)^2 = \left(\frac{L}{t}\right)^2 + \left(\frac{2\pi r}{t}\right)^2 \Leftrightarrow c^2 = V^2 + \left(\frac{2\pi r}{t}\right)^2 \quad (3.16).$$

Nous pouvons prendre deux cas limites pour vérifier que ces relations sont vraies :

- Pour une particule au repos : $L = 0,\ r = r_0 \rightarrow l_0 = 2\pi r_0.\ Ok$ !

- Pour un photon : $V = c \rightarrow V^2 = c^2 \rightarrow \left(\frac{2\pi r}{t}\right)^2 = 0 \rightarrow r = 0.\ Ok$ !

Nous introduisons le facteur :

$$\gamma = \frac{t}{2\pi r} c \quad (3.17).$$

En partant de $(3.16)$, on obtient que $\gamma = \frac{1}{\sqrt{1-\frac{V^2}{c^2}}}$.

Notre facteur $\gamma$ correspond donc au facteur de Lorentz de la relativité restreinte.

Avec $c = \frac{l}{t}$, on a :

$$\gamma = \frac{t}{2\pi r}.\frac{l}{t} = \frac{l}{2\pi r} = \frac{\sqrt{L^2 + (2\pi r)^2}}{2\pi r} \geq 1$$

$$= \frac{l}{2\pi r}.\frac{l_0}{l_0} = \frac{l}{l_0}.\frac{2\pi r_0}{2\pi r} = \underbrace{\frac{l}{l_0}}_{=1}.\frac{r_0}{r} = \frac{r_0}{r} \quad (3.18).$$

Avec les formules des paragraphes précédents, on a :

$$\frac{E}{E_0} = \frac{m}{m_0} = \frac{\nu_p}{\nu_{p_0}} = \frac{T_{p_0}}{T_p} = \frac{\lambda_{p_0}}{\lambda_p} = \frac{r_0}{r} \quad (3.19).$$

Ainsi, $\gamma = \frac{E}{E_0} = \frac{m}{m_0} = \frac{\nu_p}{\nu_{p0}} = \frac{T_{p0}}{T_p} = \frac{\lambda_{p0}}{\lambda_p} = \frac{r_0}{r}$ (3.20).

Nous retrouvons ainsi le résultat de la relativité restreinte en ce qui concerne l'énergie et la masse.

Revenons à la relation $c^2 = V^2 + \left(\frac{2\pi r}{t}\right)^2$. Sachant que $\gamma = \frac{t}{2\pi r}c$, on obtient :

$$c^2 = V^2 + \left(\frac{c}{\gamma}\right)^2 \quad (3.21).$$

On multiplie de chaque côté par $(mc)^2$:

$$(mc)^2c^2 = (mc)^2V^2 + (mc)^2\left(\frac{c}{\gamma}\right)^2 \rightarrow \underbrace{(mc^2)}_{E}{}^2 = \underbrace{(mV)}_{p}{}^2 c^2 + \underbrace{\left(\frac{m}{\gamma}\right)}_{m_0}{}^2 c^4$$

$$\rightarrow E^2 = p^2c^2 + {m_0}^2c^4 \quad (3.22).$$

C'est là la formule fondamentale de l'énergie en relativité restreinte.

### c, vitesse limite

$$V = \frac{L}{t} = \frac{L}{l}.\frac{l}{t} = \frac{L}{l}.c = \frac{L}{\sqrt{L^2 + (2\pi r)^2}}.c < c \quad (3.23).$$

On en déduit que la vitesse de la lumière est une vitesse limite, les particules ne peuvent aller plus vite que $c$. Ce fait est d'ailleurs très intuitif avec notre modèle hélicoïdal.

### Intervalle d'espace-temps

Revenons à la formule $l^2 = L^2 + (2\pi r)^2$. Elle conduit à :

$$c^2.t^2 = L^2 + (2\pi r)^2$$

$$\rightarrow (2\pi r)^2 = c^2.t^2 - L^2 \quad (3.24).$$

Cela est à mettre en correspondance avec l'intervalle d'espace-temps de la relativité d'Einstein donne : $ds^2 = c^2.dt^2 - dl^2$.

On identifie alors l'intervalle d'espace-temps au rayon de l'hélice de la particule :

$$ds = 2\pi r \quad (3.25).$$

De plus, $L \leftrightarrow dl$, $t \leftrightarrow dt$ et $V = \frac{L}{t} = \frac{dl}{dt}$.

Prenons le cas limite du photon : $r = 0 \rightarrow ds = 0$, ce qui est cohérent avec la relativité restreinte.

## Contraction des longueurs, dilatation des temps

Nous considérons maintenant la même particule dans deux référentiels différents : un référentiel $(R)$ dans lequel l'objet est immobile, et un autre référentiel $(R')$ en translation a la vitesse V.

C'est la même particule qui est considérée, donc le rayon $r$ est le même entre les deux référentiels, ce qui est à mettre en correspondance avec l'invariance de l'intervalle d'espace-temps $ds$.

De plus, étant donné que l'on considère des référentiels différents, il n'y maintenant plus de raison que la longueur de la spire soit la même dans ces référentiels.

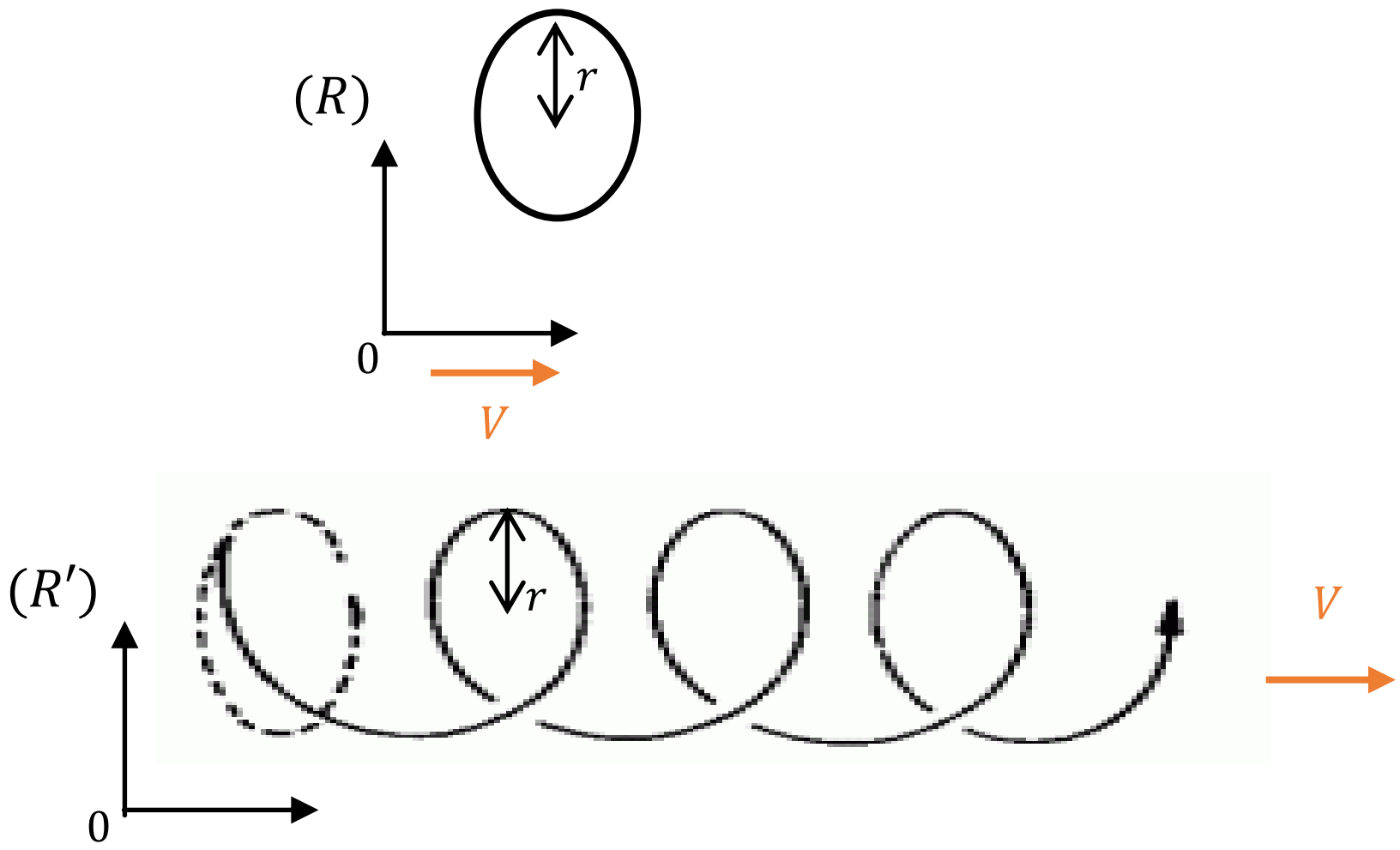

Fig 3.12 : représentation de la même particule vue dans des référentiels différents

Concernant le temps :

$$r = r' \rightarrow (2\pi r)^2 = (2\pi r')^2 \rightarrow c^2.{t_0}^2 = c^2.t^2 - L^2 \rightarrow c^2.{t_0}^2 = c^2.t^2 - (V.t)^2$$

$$\rightarrow t = \frac{1}{\sqrt{1 - \frac{V^2}{c^2}}}.t_0 = \gamma.t_0 \geq t_0 \quad (3.26).$$

C'est la dilatation du temps. Chaque particule possède une période propre, qui est son horloge interne.

## Petit résumé

Précédemment, nous avons distingué deux cas :

- Observation depuis un même référentiel de la même particule dans deux états différents (rayon $r$ différents). C'est le cas 1;
- Observation depuis deux référentiels différents de la même particule dans un même état (même rayon $r$). C'est le cas 2.

Nous avons les formules suivantes, ainsi que démontré précédemment :

$$\gamma = \frac{E}{E_0} = \frac{m}{m_0} = \frac{\nu_p}{\nu_{p0}} = \frac{T_{p0}}{T_p} = \frac{\lambda_{p0}}{\lambda_p}$$

$$= \frac{l}{l_0}.\frac{r_0}{r} = \begin{cases} \frac{r_0}{r} \;\; si\ cas\ 1\ (invariance\ l) \\ \frac{l}{l_0} \;\; si\ cas\ 2\ (invariance\ r\ ou\ ds) \end{cases} \quad (3.27).$$

De plus, nous avons pu comprendre également la contradiction dans les périodes/temps, pointée par De Broglie :

- dans le cas 1 et 2, $T_p = \frac{T_{p_0}}{\gamma}$ (période de la particule);
- dans le cas 2, $t = \gamma . t_0$ (temps pour parcourir une spire).

## 3) De la gravitation et la tentative d'unification

### La gravitation et la relativité générale

La relativité générale prédit que la force gravitationnelle se transmet par l'intermédiaire d'une particule sans masse appelée graviton de spin 2.

Nous allons ici proposer une représentation ondulatoire de cette particule, en inspirant de l'onde électromagnétique.

Nous représentons ci-dessous le cas d'une propagation rectiligne, par simplification. Mais ce sont principalement les polarisations circulaires que l'on trouve dans la nature, tout comme les ondes électromagnétiques.

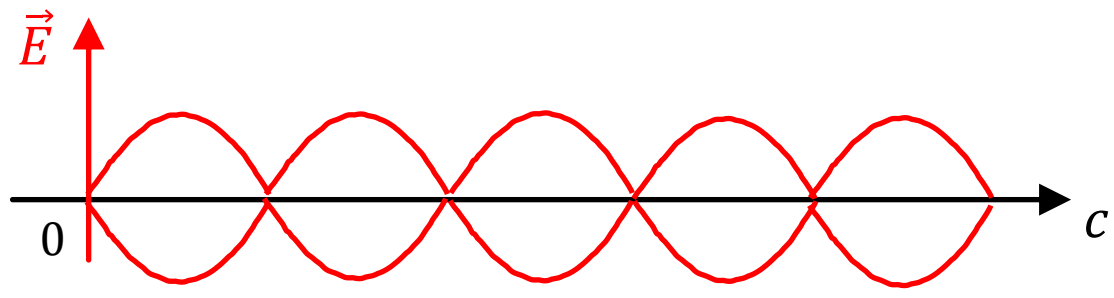

Fig. 3.13 : Représentation ondulatoire d'une onde gravitationnelle

La particule ici représentée va à la vitesse de la lumière, n'est pas chargée (pas de champ électrique résultant) et a un spin 2 (invariance par rotation de π).

Cette perspective ondulatoire de l'onde quantique gravitationnelle pourrait permettre de mieux comprendre la gravitation.

De plus, ce modèle de la particule dotée d'une structure interne bien définie avec un rayon, pourrait permettre d'éviter l'apparition des singularités en relativité générale.

## Le boson de Higgs

Nous avons occulté un fait précédemment : comment un photon peut-il avoir une trajectoire circulaire, ou plus précisément en ruban de Möbius ?

Dans le cadre de la physique des particules, les théoriciens François Englert, Robert Brout et Peter Higgs ont prédit l'existence d'un boson qui donne la masse aux bosons de l'interaction faible (*Higgs, 1964*). Il porte le nom de boson de Higgs, il a été observé au CERN en 2012. C'est un boson scalaire, c'est à dire qu'il a un spin 0.

Ce boson est également considéré comme expliquant la masse des fermions.

Quel est le mécanisme en jeu ? Un champ vectoriel photonique se combine avec le boson de Higgs pour produire un champ massif; on parle de brisure de symétrie. Lorsque le photon, de trajectoire rectiligne, interagit avec le champ de Higgs, il acquière une trajectoire cyclique. Plus la fréquence du photon est importante, plus le rayon circulaire est faible, et donc plus la masse de la particule est importante. En effet, on a : $m \propto \nu \propto \frac{1}{r}$.

On peut ainsi faire le schéma suivant :

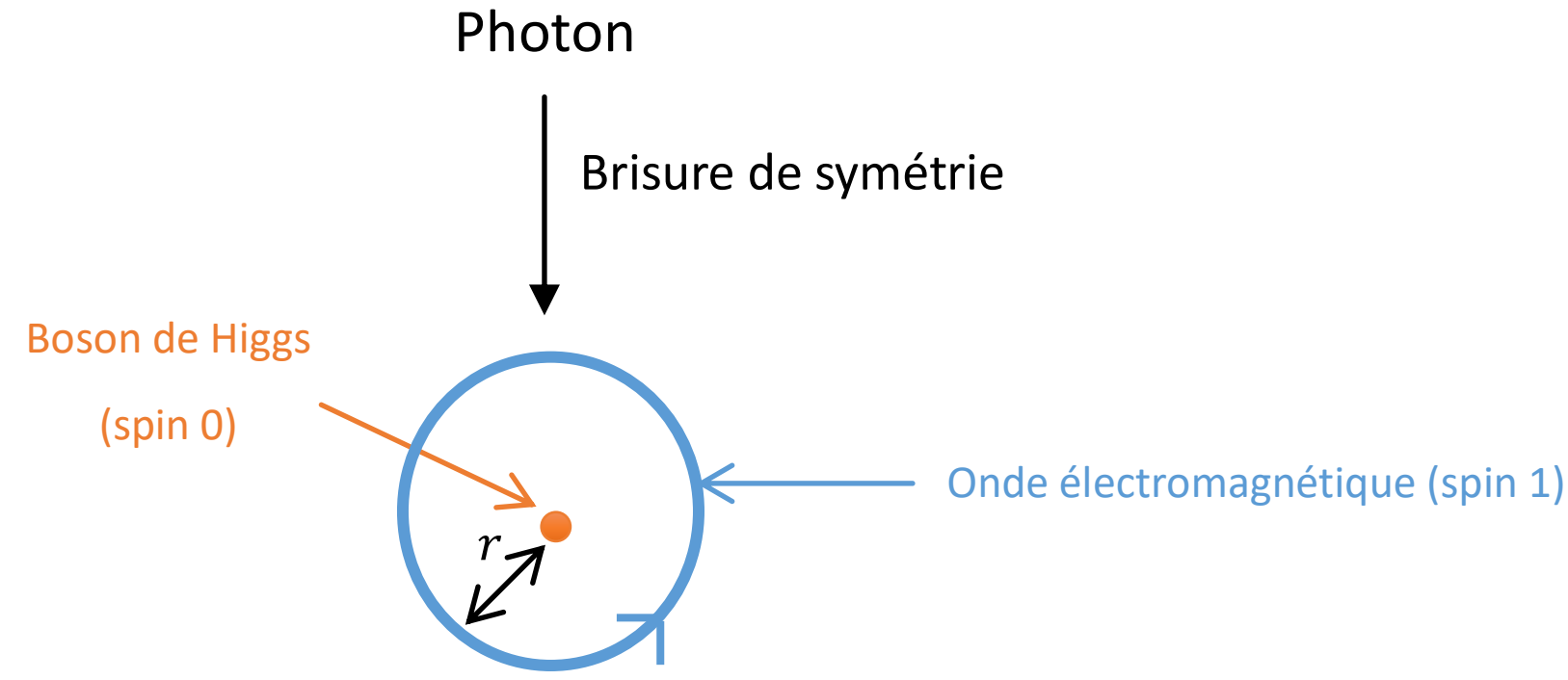

Fig. 3.14 : Représentation d'une particule massive (spin ½)

## Lien avec les théories modernes d'unification

Cette vision de la nature ondulatoire des particules fait grandement penser à la théorie des cordes. Dans cette théorie, les particules massives sont des cordes vibrantes, dont la masse est déterminée par sa longueur d'onde. Les lignes d'univers des particules ne sont pas des lignes droites, mais des surfaces d'univers :

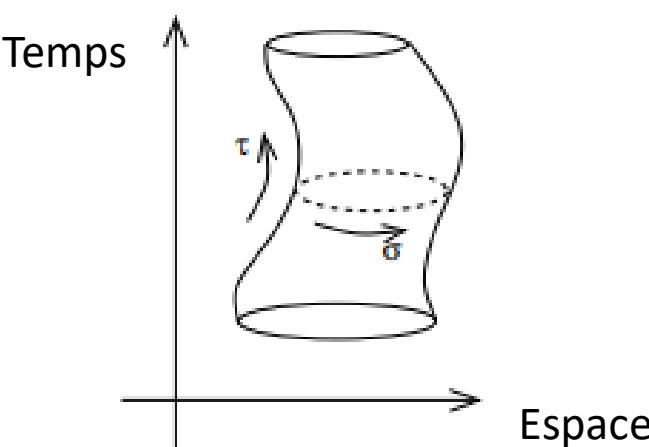

Cependant, notre théorie en diffère grandement en ce qu'elle n'introduit pas de dimensions supplémentaires, de supersymétrie,...

## 4) Résumé des propriétés ondulatoires des particules

| Spin | Non massive (vitesse c) | Massive |
|---|---|---|
| 0 | X | Boson de Higgs |
| ½ | X | c<br>Mouvement hélicoïdal avec champ électromagnétique suivant ruban de Möbius |
| 1 | $\vec{E}$ 0 c | c<br>Mouvement hélicoïdal avec champ électromagnétique suivant cercle |
| 2 | $\vec{E}$ 0 c | X |

De nombreuses questions restent encore en suspens dans notre théorie : description ondulatoire des quarks (de charge non-entière); quantification de la charge ; lien entre l'approche ondulatoire et la gravitation; compréhension du Higgs,...

## Conclusion et questionnements philosophiques

Nous avons étudié dans cet essai les idées de Louis De Broglie, avec notamment la mécanique ondulatoire, qui ont eu un impact majeur sur la science du XX$^{e}$s. et celle de nos jours (mécanique quantique, fluctuations du vide, phénomènes ondulatoires,...). Mais la physique ondulatoire n'intéresse pas que les physiciens, elle est une incitation à une révision déchirante de la confiance que nous portons à notre sens commun, dans la compréhension de la nature.

Nous pouvons résumer ici les propriétés ondulatoires de la nature telles que nous les connaissons aujourd'hui, dans la continuité des idées de De Broglie, avec des analogies hydrodynamiques :

| | | |
|---|---|---|
| Espace-éther | ↔ | Océan |
| Fluctuations du vide | ↔ | Vaguelettes sur l'océan |
| Lumière | ↔ | Vagues de plus grande amplitude |
| Particules de matière | ↔ | Objets se déplaçant sur l'océan |

Une réflexion peut cependant être menée à propos de la physique théorique moderne et de son lien avec la philosophie.

La physique théorique moderne est devenue très mathématique, sans réelle intuition physique. La démarche générale actuelle est de considérer les théories existantes, en modifier certains traits, et en tirer les conséquences. Une vision couramment admise parmi les physiciens théoriciens modernes est que la philosophie et la connaissance scientifique ancienne n'est pas pertinente quand on cherche de nouvelles théories. En effet, selon eux, les grandes questions qui étaient discutées par les philosophes sont maintenant dans les mains des physiciens. *« philosophy is dead »* indique *S.*

*Hawking*.

Cependant, certains théoriciens ne vont pas dans ce sens, comme Carlo Rovelli. Selon lui, la méthodologie et la vision décrite précédemment a engendré des montagnes de travaux inutiles dans la physique et des investissements expérimentaux inutiles. La recherche progresse en s'inspirant des philosophes et de la science, il y a une continuité; le travail d'Einstein par exemple était dans la continuité de ceux de Galilée et Newton. Notre connaissance générale est le résultat des contributions de différents domaines, de la science à la philosophie, en passant par la littérature et les arts, et de notre capacité à les intégrer. Les découvertes récentes en physique expérimentales sont toutes en contradiction avec l'attitude spéculative libre de nos jours en physique théorique. Les ondes gravitationnelles, le boson de Higgs, l'absence de super-symétrie au LHC sont des confirmations de la « vieille » physique (relativité générale, modèle standard) et des réfutations des spéculations répandues actuellement en physique théorique (comme la théorie des cordes, la gravité massive,...). (*Rovelli, 2018*)

Selon De Broglie, la physique moderne ne semble pas bâtie sur de bonnes bases. Une science peut très bien avoir un fort rôle prédictif, coller à l'expérience, tout en étant basée sur des principes contestables. C'est le cas selon lui de la mécanique quantique, comme nous l'avons vu précédemment. De plus, la physique moderne peine à expliquer les notions élémentaires de masse, d'énergie, de matière, de temps, alors que tous les raisonnements physiques se basent sur ces concepts.

Sa réflexion concernait également le rôle prépondérant joué par les mathématiques dans la physique moderne (« Shut up and calculate » prône le physicien Feynman), et le danger associé. Selon lui, le langage mathématique ne suffit pas par sa nature abstraite à rendre entièrement compte du caractère concret phénomènes physiques dans le cadre de l'espace et du temps.

« Le désir de comprendre et de se représenter clairement la réalité physique a toujours été et restera sans doute toujours le but le plus élevé et l'effort finalement le plus fructueux de la recherche scientifique fondamentale » indique De Broglie.

De Broglie questionna également l'apport moral de la science et de la physique (*Lochak, 2008*). On peut aimer la science pour ses applications et pour les commodités qu'elle a apportées à la société. Cependant, toutes les applications scientifiques ne sont pas bienfaisantes, et il n'est pas certain que son développement doive assurer le progrès réel de l'humanité, car ce progrès dépend beaucoup plus de l'élévation spirituelle et morale de l'homme que des conditions matérielles de sa vie.
Cette vision nous incite vers une plus grande sensibilité, une connexion à la nature, un plus grand respect de la vie. La science moderne semble entraînée dans cette sorte de course folle de la société, guidée par l'industrialisation, la matérialité et la rationalisation, accroissant les inégalités, et semblant vouée à sa propre destruction (*Barrau, 2019*).

Seule la science fondamentale peut nous faire progresser dans la compréhension profonde du monde. Elle doit se fonder sur les valeurs de liberté et de vérité, bien que nous sommes conscients de notre propre défaillance sur un si long chemin…

« Je vois que je n'ai été qu'un petit enfant qui jouait sur le bord de la mer, tandis que le grand océan de la vérité s'étendait, inexploré, devant moi. » Isaac Newton

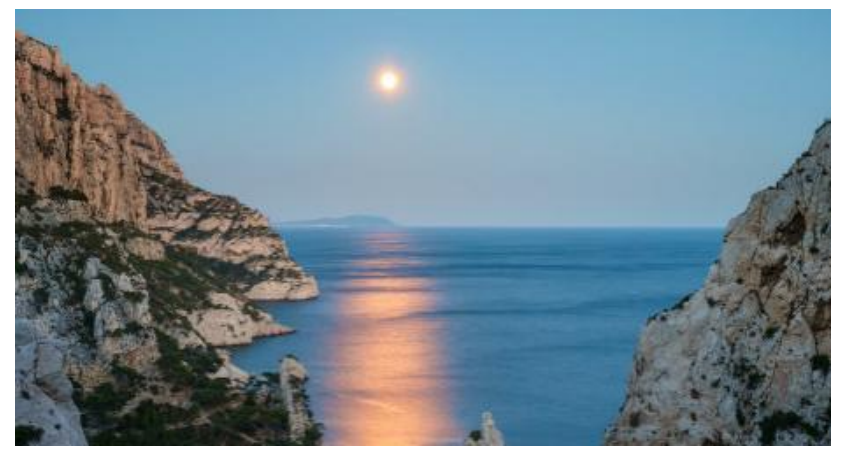

# Bibliographie